\pdfoutput=1
\documentclass[11pt,twoside,a4paper,pdftex,cmspaper,final,collab]{cms-tdr}

\begin{document}\cmsNoteHeader{CFT-09-011}
%
%
%

%
%
\hyphenation{env-iron-men-tal}
\hyphenation{had-ron-i-za-tion}
\hyphenation{cal-or-i-me-ter}
\hyphenation{de-vices}
%
%
%
%
\RCS$Revision: 1.10 $
\RCS$Date: 2010/02/22 23:12:20 $
\RCS$Name:  $
%
%
%

\newcommand {\etal}{\mbox{et al.}\xspace} 
\newcommand {\ie}{\mbox{i.e.}\xspace}     
\newcommand {\eg}{\mbox{e.g.}\xspace}     
\newcommand {\etc}{\mbox{etc.}\xspace}     
\newcommand {\vs}{\mbox{\sl vs.}\xspace}      
\newcommand {\mdash}{\ensuremath{\mathrm{-}}} 

\newcommand {\Lone}{Level-1\xspace} 
\newcommand {\Ltwo}{Level-2\xspace}
\newcommand {\Lthree}{Level-3\xspace}

\providecommand{\ACERMC} {\textsc{AcerMC}\xspace}
\providecommand{\ALPGEN} {{\textsc{alpgen}}\xspace}
\providecommand{\CHARYBDIS} {{\textsc{charybdis}}\xspace}
\providecommand{\CMKIN} {\textsc{cmkin}\xspace}
\providecommand{\CMSIM} {{\textsc{cmsim}}\xspace}
\providecommand{\CMSSW} {{\textsc{cmssw}}\xspace}
\providecommand{\COBRA} {{\textsc{cobra}}\xspace}
\providecommand{\COCOA} {{\textsc{cocoa}}\xspace}
\providecommand{\COMPHEP} {\textsc{CompHEP}\xspace}
\providecommand{\EVTGEN} {{\textsc{evtgen}}\xspace}
\providecommand{\FAMOS} {{\textsc{famos}}\xspace}
\providecommand{\GARCON} {\textsc{garcon}\xspace}
\providecommand{\GARFIELD} {{\textsc{garfield}}\xspace}
\providecommand{\GEANE} {{\textsc{geane}}\xspace}
\providecommand{\GEANTfour} {{\textsc{geant4}}\xspace}
\providecommand{\GEANTthree} {{\textsc{geant3}}\xspace}
\providecommand{\GEANT} {{\textsc{geant}}\xspace}
\providecommand{\HDECAY} {\textsc{hdecay}\xspace}
\providecommand{\HERWIG} {{\textsc{herwig}}\xspace}
\providecommand{\HIGLU} {{\textsc{higlu}}\xspace}
\providecommand{\HIJING} {{\textsc{hijing}}\xspace}
\providecommand{\IGUANA} {\textsc{iguana}\xspace}
\providecommand{\ISAJET} {{\textsc{isajet}}\xspace}
\providecommand{\ISAPYTHIA} {{\textsc{isapythia}}\xspace}
\providecommand{\ISASUGRA} {{\textsc{isasugra}}\xspace}
\providecommand{\ISASUSY} {{\textsc{isasusy}}\xspace}
\providecommand{\ISAWIG} {{\textsc{isawig}}\xspace}
\providecommand{\MADGRAPH} {\textsc{MadGraph}\xspace}
\providecommand{\MCATNLO} {\textsc{mc@nlo}\xspace}
\providecommand{\MCFM} {\textsc{mcfm}\xspace}
\providecommand{\MILLEPEDE} {{\textsc{millepede}}\xspace}
\providecommand{\ORCA} {{\textsc{orca}}\xspace}
\providecommand{\OSCAR} {{\textsc{oscar}}\xspace}
\providecommand{\PHOTOS} {\textsc{photos}\xspace}
\providecommand{\PROSPINO} {\textsc{prospino}\xspace}
\providecommand{\PYTHIA} {{\textsc{pythia}}\xspace}
\providecommand{\SHERPA} {{\textsc{sherpa}}\xspace}
\providecommand{\TAUOLA} {\textsc{tauola}\xspace}
\providecommand{\TOPREX} {\textsc{TopReX}\xspace}
\providecommand{\XDAQ} {{\textsc{xdaq}}\xspace}

\newcommand {\DZERO}{D\O\xspace}     


\newcommand{\de}{\ensuremath{^\circ}}
\newcommand{\ten}[1]{\ensuremath{\times \text{10}^\text{#1}}}
\newcommand{\unit}[1]{\ensuremath{\text{\,#1}}\xspace}
\newcommand{\mum}{\ensuremath{\,\mu\text{m}}\xspace}
\newcommand{\micron}{\ensuremath{\,\mu\text{m}}\xspace}
\newcommand{\cm}{\ensuremath{\,\text{cm}}\xspace}
\newcommand{\mm}{\ensuremath{\,\text{mm}}\xspace}
\newcommand{\mus}{\ensuremath{\,\mu\text{s}}\xspace}
\newcommand{\keV}{\ensuremath{\,\text{ke\hspace{-.08em}V}}\xspace}
\newcommand{\MeV}{\ensuremath{\,\text{Me\hspace{-.08em}V}}\xspace}
\newcommand{\GeV}{\ensuremath{\,\text{Ge\hspace{-.08em}V}}\xspace}
\newcommand{\TeV}{\ensuremath{\,\text{Te\hspace{-.08em}V}}\xspace}
\newcommand{\PeV}{\ensuremath{\,\text{Pe\hspace{-.08em}V}}\xspace}
\newcommand{\keVc}{\ensuremath{{\,\text{ke\hspace{-.08em}V\hspace{-0.16em}/\hspace{-0.08em}c}}}\xspace}
\newcommand{\MeVc}{\ensuremath{{\,\text{Me\hspace{-.08em}V\hspace{-0.16em}/\hspace{-0.08em}c}}}\xspace}
\newcommand{\GeVc}{\ensuremath{{\,\text{Ge\hspace{-.08em}V\hspace{-0.16em}/\hspace{-0.08em}c}}}\xspace}
\newcommand{\TeVc}{\ensuremath{{\,\text{Te\hspace{-.08em}V\hspace{-0.16em}/\hspace{-0.08em}c}}}\xspace}
\newcommand{\keVcc}{\ensuremath{{\,\text{ke\hspace{-.08em}V\hspace{-0.16em}/\hspace{-0.08em}c}^\text{2}}}\xspace}
\newcommand{\MeVcc}{\ensuremath{{\,\text{Me\hspace{-.08em}V\hspace{-0.16em}/\hspace{-0.08em}c}^\text{2}}}\xspace}
\newcommand{\GeVcc}{\ensuremath{{\,\text{Ge\hspace{-.08em}V\hspace{-0.16em}/\hspace{-0.08em}c}^\text{2}}}\xspace}
\newcommand{\TeVcc}{\ensuremath{{\,\text{Te\hspace{-.08em}V\hspace{-0.16em}/\hspace{-0.08em}c}^\text{2}}}\xspace}

\newcommand{\pbinv} {\mbox{\ensuremath{\,\text{pb}^\text{$-$1}}}\xspace}
\newcommand{\fbinv} {\mbox{\ensuremath{\,\text{fb}^\text{$-$1}}}\xspace}
\newcommand{\nbinv} {\mbox{\ensuremath{\,\text{nb}^\text{$-$1}}}\xspace}
\newcommand{\percms}{\ensuremath{\,\text{cm}^\text{$-$2}\,\text{s}^\text{$-$1}}\xspace}
\newcommand{\lumi}{\ensuremath{\mathcal{L}}\xspace}
\newcommand{\Lumi}{\ensuremath{\mathcal{L}}\xspace}
%
\newcommand{\LvLow}  {\ensuremath{\mathcal{L}=\text{10}^\text{32}\,\text{cm}^\text{$-$2}\,\text{s}^\text{$-$1}}\xspace}
\newcommand{\LLow}   {\ensuremath{\mathcal{L}=\text{10}^\text{33}\,\text{cm}^\text{$-$2}\,\text{s}^\text{$-$1}}\xspace}
\newcommand{\lowlumi}{\ensuremath{\mathcal{L}=\text{2}\times \text{10}^\text{33}\,\text{cm}^\text{$-$2}\,\text{s}^\text{$-$1}}\xspace}
\newcommand{\LMed}   {\ensuremath{\mathcal{L}=\text{2}\times \text{10}^\text{33}\,\text{cm}^\text{$-$2}\,\text{s}^\text{$-$1}}\xspace}
\newcommand{\LHigh}  {\ensuremath{\mathcal{L}=\text{10}^\text{34}\,\text{cm}^\text{$-$2}\,\text{s}^\text{$-$1}}\xspace}
\newcommand{\hilumi} {\ensuremath{\mathcal{L}=\text{10}^\text{34}\,\text{cm}^\text{$-$2}\,\text{s}^\text{$-$1}}\xspace}


\newcommand{\zp}{\ensuremath{\mathrm{Z}^\prime}\xspace}


\newcommand{\kt}{\ensuremath{k_{\mathrm{T}}}\xspace}
\newcommand{\BC}{\ensuremath{{B_{\mathrm{c}}}}\xspace}
\newcommand{\bbarc}{\ensuremath{{\overline{b}c}}\xspace}
\newcommand{\bbbar}{\ensuremath{{b\overline{b}}}\xspace}
\newcommand{\ccbar}{\ensuremath{{c\overline{c}}}\xspace}
\newcommand{\JPsi}{\ensuremath{{J}/\psi}\xspace}
\newcommand{\bspsiphi}{\ensuremath{B_s \to \JPsi\, \phi}\xspace}
\newcommand{\AFB}{\ensuremath{A_\mathrm{FB}}\xspace}
\newcommand{\EE}{\ensuremath{e^+e^-}\xspace}
\newcommand{\MM}{\ensuremath{\mu^+\mu^-}\xspace}
\newcommand{\TT}{\ensuremath{\tau^+\tau^-}\xspace}
\newcommand{\wangle}{\ensuremath{\sin^{2}\theta_{\mathrm{eff}}^\mathrm{lept}(M^2_\mathrm{Z})}\xspace}
\newcommand{\ttbar}{\ensuremath{{t\overline{t}}}\xspace}
\newcommand{\stat}{\ensuremath{\,\text{(stat.)}}\xspace}
\newcommand{\syst}{\ensuremath{\,\text{(syst.)}}\xspace}

\newcommand{\HGG}{\ensuremath{\mathrm{H}\to\gamma\gamma}}
\newcommand{\gev}{\GeV}
\newcommand{\GAMJET}{\ensuremath{\gamma + \mathrm{jet}}}
\newcommand{\PPTOJETS}{\ensuremath{\mathrm{pp}\to\mathrm{jets}}}
\newcommand{\PPTOGG}{\ensuremath{\mathrm{pp}\to\gamma\gamma}}
\newcommand{\PPTOGAMJET}{\ensuremath{\mathrm{pp}\to\gamma +
\mathrm{jet}
}}
\newcommand{\MH}{\ensuremath{\mathrm{M_{\mathrm{H}}}}}
\newcommand{\RNINE}{\ensuremath{\mathrm{R}_\mathrm{9}}}
\newcommand{\DR}{\ensuremath{\Delta\mathrm{R}}}


\newcommand{\PT}{\ensuremath{p_{\mathrm{T}}}\xspace}
\newcommand{\pt}{\ensuremath{p_{\mathrm{T}}}\xspace}
\newcommand{\ET}{\ensuremath{E_{\mathrm{T}}}\xspace}
\newcommand{\HT}{\ensuremath{H_{\mathrm{T}}}\xspace}
\newcommand{\et}{\ensuremath{E_{\mathrm{T}}}\xspace}
\newcommand{\Em}{\ensuremath{E\!\!\!/}\xspace}
\newcommand{\Pm}{\ensuremath{p\!\!\!/}\xspace}
\newcommand{\PTm}{\ensuremath{{p\!\!\!/}_{\mathrm{T}}}\xspace}
\newcommand{\ETm}{\ensuremath{E_{\mathrm{T}}^{\mathrm{miss}}}\xspace}
\newcommand{\MET}{\ensuremath{E_{\mathrm{T}}^{\mathrm{miss}}}\xspace}
\newcommand{\ETmiss}{\ensuremath{E_{\mathrm{T}}^{\mathrm{miss}}}\xspace}
\newcommand{\VEtmiss}{\ensuremath{{\vec E}_{\mathrm{T}}^{\mathrm{miss}}}\xspace}

%

\newcommand{\ga}{\ensuremath{\gtrsim}}
\newcommand{\la}{\ensuremath{\lesssim}}
\newcommand{\swsq}{\ensuremath{\sin^2\theta_W}\xspace}
\newcommand{\cwsq}{\ensuremath{\cos^2\theta_W}\xspace}
\newcommand{\tanb}{\ensuremath{\tan\beta}\xspace}
\newcommand{\tanbsq}{\ensuremath{\tan^{2}\beta}\xspace}
\newcommand{\sidb}{\ensuremath{\sin 2\beta}\xspace}
\newcommand{\alpS}{\ensuremath{\alpha_S}\xspace}
\newcommand{\alpt}{\ensuremath{\tilde{\alpha}}\xspace}

\newcommand{\QL}{\ensuremath{Q_L}\xspace}
\newcommand{\sQ}{\ensuremath{\tilde{Q}}\xspace}
\newcommand{\sQL}{\ensuremath{\tilde{Q}_L}\xspace}
\newcommand{\ULC}{\ensuremath{U_L^C}\xspace}
\newcommand{\sUC}{\ensuremath{\tilde{U}^C}\xspace}
\newcommand{\sULC}{\ensuremath{\tilde{U}_L^C}\xspace}
\newcommand{\DLC}{\ensuremath{D_L^C}\xspace}
\newcommand{\sDC}{\ensuremath{\tilde{D}^C}\xspace}
\newcommand{\sDLC}{\ensuremath{\tilde{D}_L^C}\xspace}
\newcommand{\LL}{\ensuremath{L_L}\xspace}
\newcommand{\sL}{\ensuremath{\tilde{L}}\xspace}
\newcommand{\sLL}{\ensuremath{\tilde{L}_L}\xspace}
\newcommand{\ELC}{\ensuremath{E_L^C}\xspace}
\newcommand{\sEC}{\ensuremath{\tilde{E}^C}\xspace}
\newcommand{\sELC}{\ensuremath{\tilde{E}_L^C}\xspace}
\newcommand{\sEL}{\ensuremath{\tilde{E}_L}\xspace}
\newcommand{\sER}{\ensuremath{\tilde{E}_R}\xspace}
\newcommand{\sFer}{\ensuremath{\tilde{f}}\xspace}
\newcommand{\sQua}{\ensuremath{\tilde{q}}\xspace}
\newcommand{\sUp}{\ensuremath{\tilde{u}}\xspace}
\newcommand{\suL}{\ensuremath{\tilde{u}_L}\xspace}
\newcommand{\suR}{\ensuremath{\tilde{u}_R}\xspace}
\newcommand{\sDw}{\ensuremath{\tilde{d}}\xspace}
\newcommand{\sdL}{\ensuremath{\tilde{d}_L}\xspace}
\newcommand{\sdR}{\ensuremath{\tilde{d}_R}\xspace}
\newcommand{\sTop}{\ensuremath{\tilde{t}}\xspace}
\newcommand{\stL}{\ensuremath{\tilde{t}_L}\xspace}
\newcommand{\stR}{\ensuremath{\tilde{t}_R}\xspace}
\newcommand{\stone}{\ensuremath{\tilde{t}_1}\xspace}
\newcommand{\sttwo}{\ensuremath{\tilde{t}_2}\xspace}
\newcommand{\sBot}{\ensuremath{\tilde{b}}\xspace}
\newcommand{\sbL}{\ensuremath{\tilde{b}_L}\xspace}
\newcommand{\sbR}{\ensuremath{\tilde{b}_R}\xspace}
\newcommand{\sbone}{\ensuremath{\tilde{b}_1}\xspace}
\newcommand{\sbtwo}{\ensuremath{\tilde{b}_2}\xspace}
\newcommand{\sLep}{\ensuremath{\tilde{l}}\xspace}
\newcommand{\sLepC}{\ensuremath{\tilde{l}^C}\xspace}
\newcommand{\sEl}{\ensuremath{\tilde{e}}\xspace}
\newcommand{\sElC}{\ensuremath{\tilde{e}^C}\xspace}
\newcommand{\seL}{\ensuremath{\tilde{e}_L}\xspace}
\newcommand{\seR}{\ensuremath{\tilde{e}_R}\xspace}
\newcommand{\snL}{\ensuremath{\tilde{\nu}_L}\xspace}
\newcommand{\sMu}{\ensuremath{\tilde{\mu}}\xspace}
\newcommand{\sNu}{\ensuremath{\tilde{\nu}}\xspace}
\newcommand{\sTau}{\ensuremath{\tilde{\tau}}\xspace}
\newcommand{\Glu}{\ensuremath{g}\xspace}
\newcommand{\sGlu}{\ensuremath{\tilde{g}}\xspace}
\newcommand{\Wpm}{\ensuremath{W^{\pm}}\xspace}
\newcommand{\sWpm}{\ensuremath{\tilde{W}^{\pm}}\xspace}
\newcommand{\Wz}{\ensuremath{W^{0}}\xspace}
\newcommand{\sWz}{\ensuremath{\tilde{W}^{0}}\xspace}
\newcommand{\sWino}{\ensuremath{\tilde{W}}\xspace}
\newcommand{\Bz}{\ensuremath{B^{0}}\xspace}
\newcommand{\sBz}{\ensuremath{\tilde{B}^{0}}\xspace}
\newcommand{\sBino}{\ensuremath{\tilde{B}}\xspace}
\newcommand{\Zz}{\ensuremath{Z^{0}}\xspace}
\newcommand{\sZino}{\ensuremath{\tilde{Z}^{0}}\xspace}
\newcommand{\sGam}{\ensuremath{\tilde{\gamma}}\xspace}
\newcommand{\chiz}{\ensuremath{\tilde{\chi}^{0}}\xspace}
\newcommand{\chip}{\ensuremath{\tilde{\chi}^{+}}\xspace}
\newcommand{\chim}{\ensuremath{\tilde{\chi}^{-}}\xspace}
\newcommand{\chipm}{\ensuremath{\tilde{\chi}^{\pm}}\xspace}
\newcommand{\Hone}{\ensuremath{H_{d}}\xspace}
\newcommand{\sHone}{\ensuremath{\tilde{H}_{d}}\xspace}
\newcommand{\Htwo}{\ensuremath{H_{u}}\xspace}
\newcommand{\sHtwo}{\ensuremath{\tilde{H}_{u}}\xspace}
\newcommand{\sHig}{\ensuremath{\tilde{H}}\xspace}
\newcommand{\sHa}{\ensuremath{\tilde{H}_{a}}\xspace}
\newcommand{\sHb}{\ensuremath{\tilde{H}_{b}}\xspace}
\newcommand{\sHpm}{\ensuremath{\tilde{H}^{\pm}}\xspace}
\newcommand{\hz}{\ensuremath{h^{0}}\xspace}
\newcommand{\Hz}{\ensuremath{H^{0}}\xspace}
\newcommand{\Az}{\ensuremath{A^{0}}\xspace}
\newcommand{\Hpm}{\ensuremath{H^{\pm}}\xspace}
\newcommand{\sGra}{\ensuremath{\tilde{G}}\xspace}
\newcommand{\mtil}{\ensuremath{\tilde{m}}\xspace}
\newcommand{\rpv}{\ensuremath{\rlap{\kern.2em/}R}\xspace}
\newcommand{\LLE}{\ensuremath{LL\bar{E}}\xspace}
\newcommand{\LQD}{\ensuremath{LQ\bar{D}}\xspace}
\newcommand{\UDD}{\ensuremath{\overline{UDD}}\xspace}
\newcommand{\Lam}{\ensuremath{\lambda}\xspace}
\newcommand{\Lamp}{\ensuremath{\lambda'}\xspace}
\newcommand{\Lampp}{\ensuremath{\lambda''}\xspace}
\newcommand{\spinbd}[2]{\ensuremath{\bar{#1}_{\dot{#2}}}\xspace}

\newcommand{\MD}{\ensuremath{{M_\mathrm{D}}}\xspace}
\newcommand{\Mpl}{\ensuremath{{M_\mathrm{Pl}}}\xspace}
\newcommand{\Rinv} {\ensuremath{{R}^{-1}}\xspace}

%
%
\hyphenation{en-viron-men-tal}

\cmsNoteHeader{09-011}
\title{Performance of the CMS Cathode Strip Chambers\\with Cosmic Rays}%
%
%
\address[cern]{CERN}
\author[cern]{The CMS Collaboration}
\date{\today}
\abstract{
The Cathode Strip Chambers (CSCs) constitute the primary muon tracking
device in the CMS endcaps.  Their performance has been evaluated
using data taken during a cosmic ray run in fall 2008.
Measured noise levels are low, with the number of noisy
channels well below 1\%.  Coordinate resolution was measured
for all types of chambers, and fall in the range $47~\mu$m
to $243~\mu$m.  The efficiencies for local charged track
triggers, for hit and for segments reconstruction were 
measured, and are above 99\%.   The timing resolution
per layer is approximately $5$~ns.
}
\hypersetup{%
pdfauthor={Michael Schmitt},%
pdftitle={Performance of the CMS Cathode Strip Chambers with Cosmic Rays},%
pdfsubject={CRAFT},%
pdfkeywords={CMS, CSC, CRAFT}}
\maketitle
\section{Introduction}
\par
The primary goal of the Compact Muon Solenoid (CMS) experiment~\cite{CMS} 
is to explore particle physics at the TeV energy scale, exploiting the 
proton-proton collisions delivered by the Large Hadron Collider 
(LHC) at CERN~\cite{LHC}.  The central feature of the CMS
apparatus is a superconducting solenoid, of 6~m internal diameter,
providing a field of 3.8~T.  Within the field volume are the silicon
pixel and strip tracking detectors, the crystal electromagnetic
calorimeter and the brass/scintillator hadron calorimeter.  Muons
are measured in gas-ionization detectors embedded in the steel return
yolk.  In addition to the barrel and endcap detectors, CMS has
extensive forward calorimetry.
\par
The Cathode Strip Chambers (CSCs) constitute an essential component
of the CMS muon detector, providing precise tracking and triggering
of muons in the endcaps.  Their performance is critical to many
physics analyses based on muons.  An early assessment of
their performance is possible using data recorded during the
fall of 2008 as part of the {Cosmic Run At Four Tesla} (CRAFT)
exercise.  This paper summarizes the results obtained from
the analysis of those data.
\par
The CRAFT campaign involved all installed subdetector systems, most
of which were nearly fully operational, as described in Ref.~\cite{MAINCRAFT}.
Approximately 270~million cosmic ray muon triggers were recorded while the 
magnet was operating  at a field of~3.8~T.
Of these, roughly a fifth were triggered by the CSCs.
\par
In the sections that follow, a selection of distributions characteristic
of the flux of cosmic ray muons through the CSCs is shown, followed
by an assessment of the electronics noise, measurements of the
efficiency and resolution of the chambers, and finally some basic
information about the timing capabilities of the CSCs.  This paper
begins with a brief description of the CSC muon system
and of the basics of offline moun reconstruction.

\section{The CSC System}
\par
\begin{figure}
\begin{center}
\includegraphics[width=0.85\textwidth]{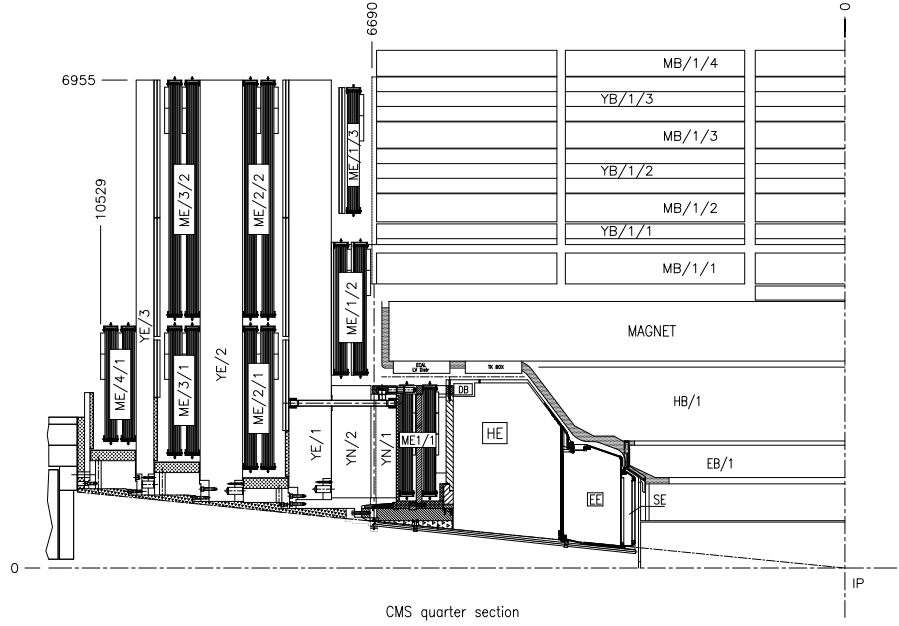}
\end{center}
\caption{\label{fig:CSCDET}
A cross-sectional view of a quarter of the CMS detector, 
highlighting the CSCs.}
\end{figure}

The CSC subdetector is composed of rings of trapezoidal chambers
mounted on eight disks - four in each endcap~\cite{MTDR}.  
There are 468~chambers in total.  The rings of 
chambers are designated by ME$\pm S/R$, where ``ME'' stands for ``Muon 
Endcap,'' the $\pm$ sign indicates which endcap, $S$ indicates the disk 
(or ``station'')  and $R$ is the ring number.  The chambers in the outer 
rings, such as ME$\pm 2/2$ and ME$\pm 3/2$, are considerably larger than 
the chambers closer to the beam pipe, such as ME$\pm 1/1$ and ME$\pm 1/2$.  
A drawing of CMS highlighting the CSC subdetector is shown in 
Fig.~\ref{fig:CSCDET}.

\par
Every chamber contains six detecting layers each composed of
an anode wire plane stretched between two planar copper cathodes,
one continuous, the other segmented in strips to provide
position measurement.
The distance between anode planes is 2.54~cm, except for the
ME$\pm 1/1$ chambers, for  which it is 2.2~cm.  
The wires are read out in groups, of which the width
varies between 1.5 and 5~cm for different chambers. 
The high voltage is supplied to ranges of wire groups,
depending on the size of the chamber; the largest chambers
have five such high-voltage segments.
The strips are read out individually, and their average widths
vary between 5 and 12~mm.  They are trapezoidal in shape,
like the chambers themselves.  
The strips in alternating layers
are staggered, except in ME$\pm 1/1$.
The strips in the ME$\pm 1/1$ chambers are cut along a line
parallel to the short sides of the trapezoid in order to reduce
the rate on any one strip.  The strips closer to the beam line
constitute ME$\pm 1/1a$, and the others, ME$\pm 1/1b$.
The studies presented in this paper concern ME$\pm 1/1b$ only.
The smaller chambers tend to have a lower electronics noise 
due to smaller capactive coupling between the wire and strip planes,
better resolution due to smaller strip widths, and, in the case 
of ME$\pm 1/1$, higher gas gain.
A synopsis of relevant cathode strip parameters
is given in Table~\ref{tab:specs}.

\begin{table}[b]
\begin{center}
\caption{\label{tab:specs}
Selected physical specifications of the cathode strip chambers.
The range of strip width is given, and the average width in square brackets.
For more information, see Ref.~\cite{MTDR}.
}
\begin{tabular}{|ccccc|}
\hline
Ring & Chambers per ring & Strips per chamber & Strip width (mm) & Pitch (mrad)  \\
\hline
ME$\pm 1/1a$ & 36 & 48 & $4.11$ -- $5.82$ ~[4.96]  & $3.88$ \\
ME$\pm 1/1b$ & 36 & 64 & $4.44$ -- $7.6$ ~[6.0]  & $2.96$ \\
ME$\pm 1/2$ & 36 & 80 &  $6.6$ -- $10.4$ ~[8.5]  & $2.33$ \\
ME$\pm 1/3$ & 36 & 64 & $11.1$ -- $14.9$ ~[13.0] & $2.16$ \\
ME$\pm 2/1$ & 18 & 80 & $ 6.8$ -- $15.6$ ~[11.2] & $4.65$ \\
ME$\pm 2/2$ & 36 & 80 & $ 8.5$ -- $16.0$ ~[12.2] & $2.33$ \\
ME$\pm 3/1$ & 18 & 80 & $ 7.8$ -- $15.6$ ~[11.7] & $4.65$ \\
ME$\pm 3/2$ & 36 & 80 & $ 8.5$ -- $16.0$ ~[12.2] & $2.33$ \\
ME$\pm 4/1$ & 18 & 80 & $ 8.6$ -- $15.6$ ~[12.1] & $4.65$ \\
\hline
\end{tabular}
\end{center}
\end{table}

The CSCs are designed to measure the azimuthal coordinates ($\phi$) of muon tracks well,
as the bending of the muon trajectories in the magnetic
flux returned through the steel disks is mainly about 
the direction of a unit vector pointing away from the beam line. 
The strips describe constant $\phi$ values.
High precision is achieved by exploiting the shape of 
the charge distribution on three consecutive strips; this 
allows an adequate measurement of the muon momentum as needed 
for triggering purposes. The anode wires run perpendicular to
the central strip, and hence parallel to the two
parallel sides of the chamber; they provide an
approximate measure of the radial coordinate.  They are tilted
by $29^\circ$ in ME$\pm 1/1$ to compensate for the average effect of
the magnetic field on the drift.
In terms of the local coordinate system, defined at the level
of a single chamber, the six layers are parallel to the $xy$
plane, with the $y$ axis perpendicular to the wires, and
the $x$ axis nearly perpendicular to the centermost strip.
Thus, the wires measure the local $y$ coordinate, and the
strips dominate the measurement of the local $x$ coordinate.
\par
The readout of a CSC is triggered by the presence
of anode and cathode local charged track patterns, referred to as
ALCT and CLCT, respectively, which are defined in the trigger 
logic~\cite{TDR,LCT}.    
A set of regional processors called the CSC Track Finder~\cite{CSCTF}
builds the CSC muon trigger from the trigger primitives
generated by individual chambers and sends it to the global
muon trigger processor. For CRAFT, 
events were recorded with a very loose CSC trigger based on the 
logical ``OR'' of the trigger signals of all individual chambers.
The rate of this loose trigger was about 60~Hz.  
\par
The ALCT wire patterns and the CLCT strip patterns were designed to be 
efficient only for muons originating from the interaction point.  
The range of track inclination ($dy/dz$ in local
coordinates) which should give efficient ALCT response is 
$-0.69 < dy/dz < 0$ for smaller chambers, and
$-1.97 < dy/dz < 0$ for larger chambers. 
(The minus sign is a matter of convention.)
Similarly, for the 
CLCT response the range is $|dx/dz| < 0.24$ for smaller, and 
$0.63$ for larger chambers.   For collision data, the muons will 
naturally have inclination angles within these ranges.  
Muons from cosmic rays, however, arrive with a much wider
angular distribution.

\par
The wire group signal is relatively fast and serves 
to establish the beam crossing number (BX) for
a signal.  Usually the anode signal extends over only
one or two 25~ns beam crossings.
The cathode strip signal is integrated
and extends over several hundred nanoseconds.
The shape of the cathode pulse can be used to
infer the time of the signal to a fraction of
a beam crossing number.  To this end, the pulse is sampled
every 50~ns (2~BX) with the results from eight time slices 
stored in a switched capacitor array (SCA).  The arrival of 
the pulse is arranged so that the first two time bins are free
from signal, allowing a dynamical estimate of the
signal base line.  A good description of the pulse
shape recorded in the SCA is given by a 5-pole
semi-Gaussian function:
\begin{displaymath}
 S(t) \propto \left( \frac{t-T_S}{T_0} \right)^4
  \exp\left[ -\frac{(t-T_S)}{T_0}\right]
\end{displaymath}
valid for $t > T_S$, the start time.  Given the
fixed exponent of the first factor, the shape of the
pulse is determined by the decay constant $T_0$, and
the maximum occurs at $t = T_S + 4T_0$.
Cross-talk is approximately 12\% of the signal
and is taken into account when calculating strip 
coordinates~\cite{MTDR}.

\par
The assembly of the CSCs included a comprehensive commissioning
regimen to verify chamber performance during production. 
This set of tests was performed again on each chamber upon
arrival at CERN, and multiple times following installation
on the endcap disks on the surface during 2005-7.  In 2007,
the disks were lowered into the CMS cavern at Point~5, and the
full set of services and infrastructure became available early
in~2008.  At this time, the scope of the commissioning program 
was expanded from checking one chamber at a time to covering 
the entire set of 468 chambers as a subdetector system.
\par
\begin{figure}
\begin{center}
\includegraphics[width=0.35\textwidth]{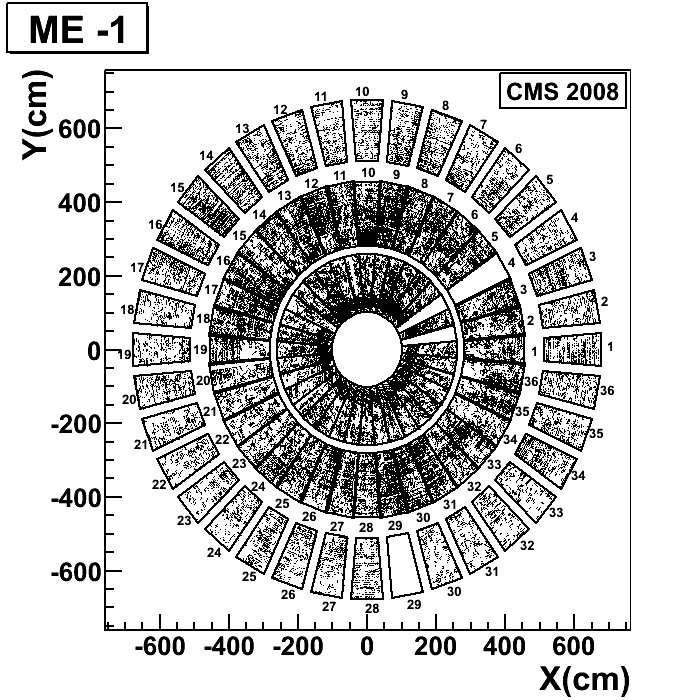}
\includegraphics[width=0.35\textwidth]{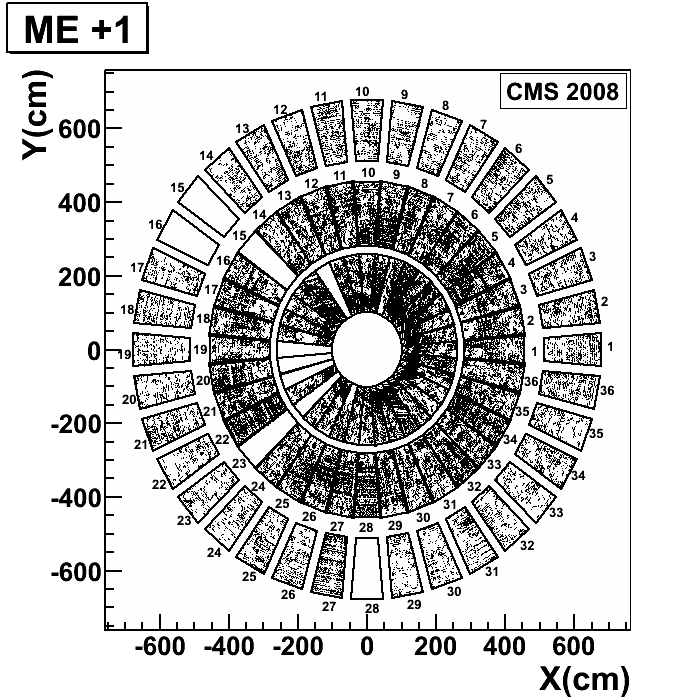}
\includegraphics[width=0.35\textwidth]{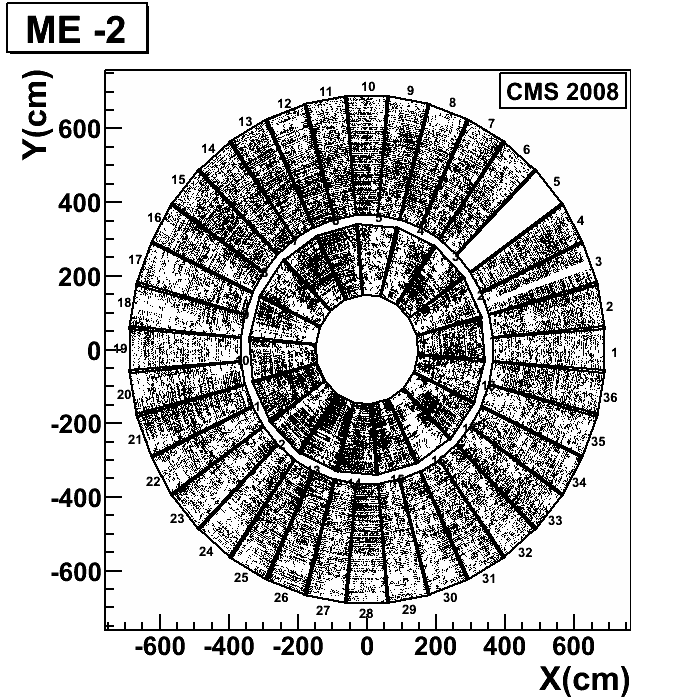}
\includegraphics[width=0.35\textwidth]{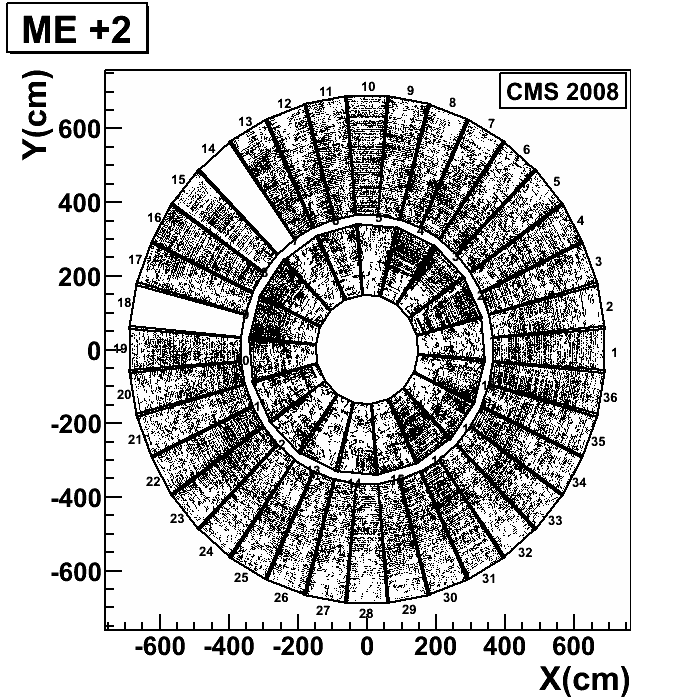}
\includegraphics[width=0.35\textwidth]{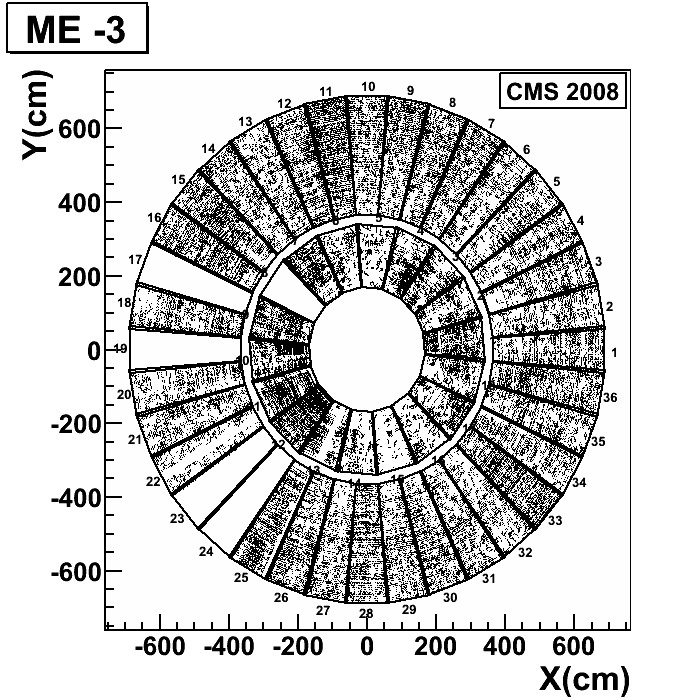}
\includegraphics[width=0.35\textwidth]{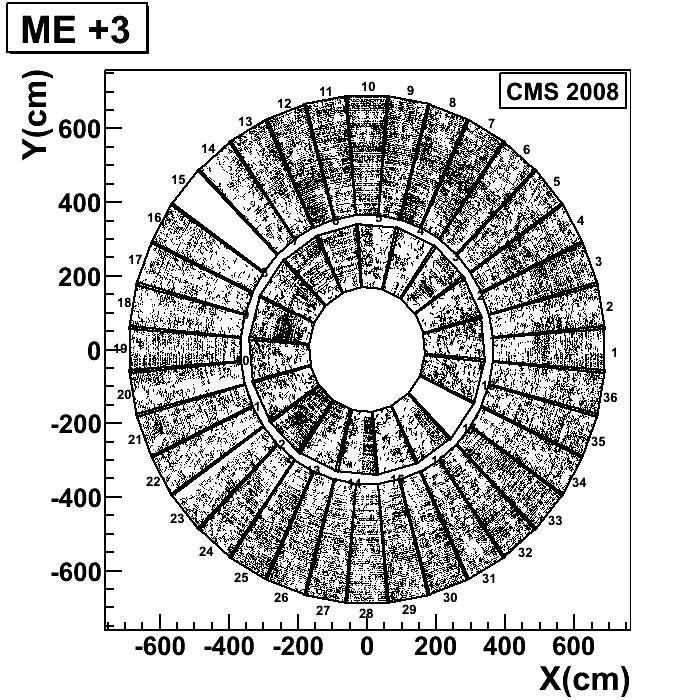}
\includegraphics[width=0.35\textwidth]{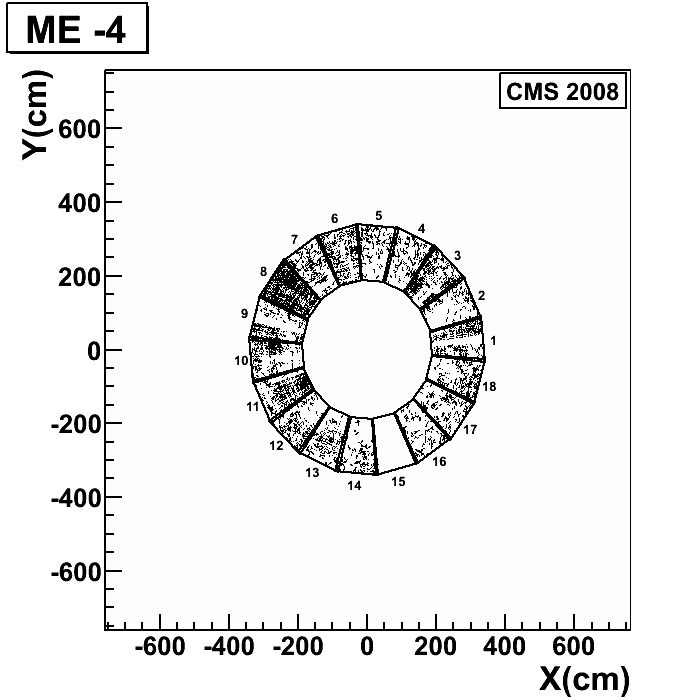}
\includegraphics[width=0.35\textwidth]{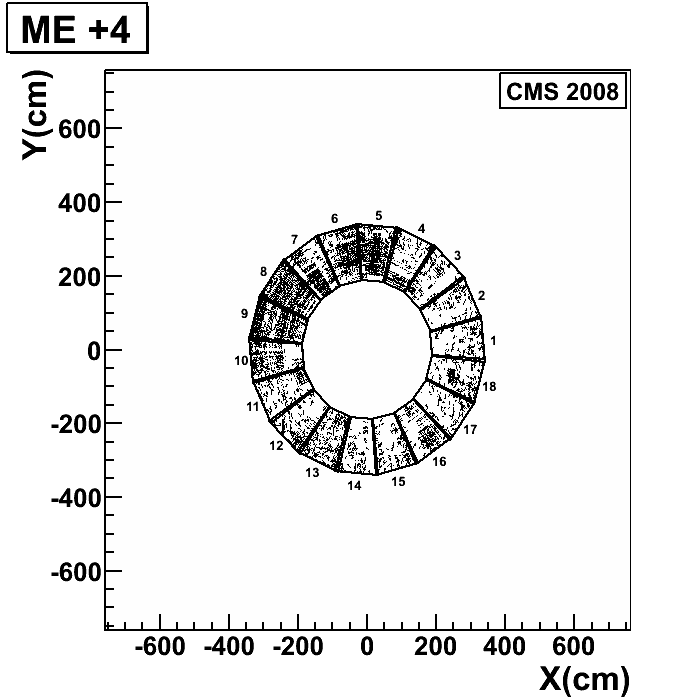}
\end{center}
\caption{\label{fig:rechits_global}
Distributions of hits reconstructed from a portion of 
the CRAFT data.  Nearly all of the chambers were fully operational.  
A few inoperative chambers can be seen as white trapezoids; very
thin white trapezoids indicate missing signals from a group
of 16~strips.
}
\end{figure}

The commissioning effort included the following tasks:
establishing inter-component communication, loading
new versions of firmware on the electronics boards,
turning on and configuring all components in a robust
way, and measuring the parameters necessary to ensure
synchronization of the system.  
The development of a suite of software tools was 
essential to bring the CSC system online.
During CRAFT, the CSCs were included
in the global readout about 80\% of the time, and 
more than 96\% of the readout channels were live.  
Figure~\ref{fig:rechits_global} shows that hits could be 
reconstructed successfully in nearly all of the chambers.
The chambers that did not provide data during CRAFT have 
been repaired since then.

\section{Reconstruction of Muon Track Segments}
\par
Raw data from the detector are unpacked offline into integer-based
objects called ``digis.''  There are digi collections for the strip signals,
the wire signals, and the local charged tracks (LCTs).
The information stored in the digis
is processed to produce a collection of objects called ``rechits'' 
with measured $x$ and $y$ coordinates at a known $z$ coordinate.
These represent the measurement of the intersection point between
the track and a CSC layer. 
The rechits reconstructed in a given chamber are used to form a
straight-line segment, which is fit to provide a measure of the
muon trajectory in the chamber.  
Only one rechit is used from any given layer, and at least three
rechits are required.  The majority of segments have six rechits,
while a modest fraction have fewer due to the impact of
$\delta$-ray electrons and the boundaries of the chamber.
These segments are used to seed the reconstruction of 
muon tracks based on muon chamber data only -- these are
called ``stand-alone muons''~\cite{MuonReco}.
Due to the very broad range of cosmic ray incident angles,
only a small fraction of the stand-alone muons 
can be matched to reconstructed tracks in the silicon tracker,
especially in the endcaps.
\par
Simulated data sets were produced using a
Monte Carlo event generator~\cite{CRMC} which is configured 
to reproduce the CRAFT data as closely as possible.  The
CSC detector simulation reproduced approximately the number
and distribution of inoperative chambers.
The simulated data, the reconstructed CRAFT data, and the
results presented in this paper are based on CMS
reconstruction code releases dating from the spring of 2009.

\section{Basic Information from Cosmic Rays}
\par
Most cosmic ray muons above ground have an energy of at 
most a few~GeV~\cite{PDG}.
In the underground cavern at Point~5, the energy spectrum 
is shifted to somewhat
higher values.  Muons must have energies of at least a few~GeV
in order to pass through three consecutive CSC stations, since
the steel disks between them are approximately $34~X_0$ thick.  
Most reconstructed muons have only a few~GeV, so 
multiple scattering in the steel yoke can displace the muon's 
trajectory by several centimeters with respect to the ideal trajectory.
\par
Most of the muons triggered in the 
endcaps are not useful because their trajectories are steeply
inclined or pass through only an edge of one of the endcaps.
Only a minute fraction of the recorded cosmic ray muons
follow a useful path through the endcaps, and satisfy the 
nominal geometric requirements for the efficient triggering
and readout of the CSCs, as explained in detail below.
\par
In order to secure a sample of useful events, a filter was applied
to the primary data set to select events in which at least three
chambers had hits, and in which at least two segments
had been reconstructed.  Events with very many rechits
or segments were excluded, since they were likely to contain
muon-induced showers. These criteria reduced the
data sample with CSC triggers by a factor of twenty, and enabled direct
comparisons of the simulated data to the CRAFT data.
\par
\begin{figure}
\begin{center}
\includegraphics[width=0.45\textwidth]{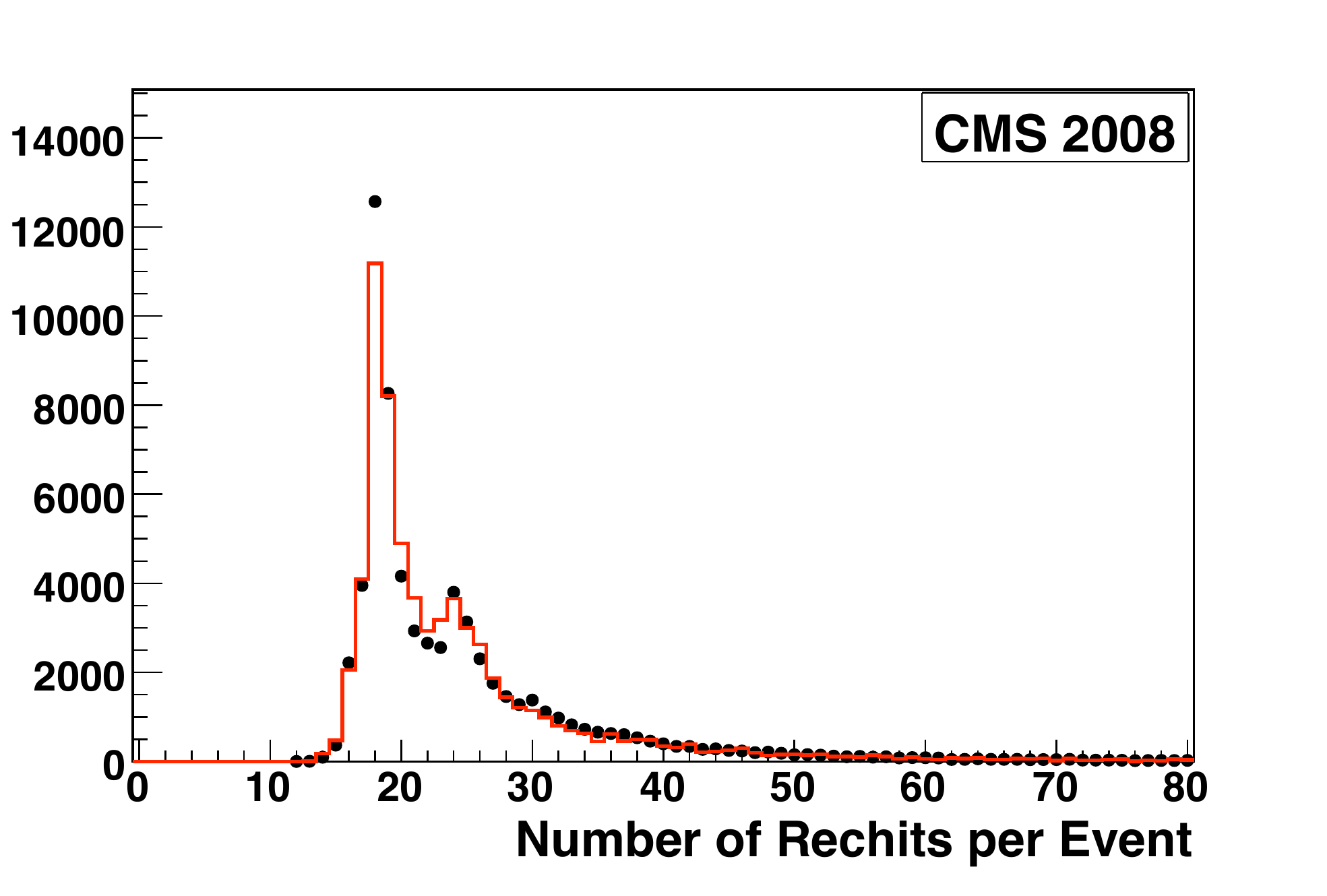}
\includegraphics[width=0.45\textwidth]{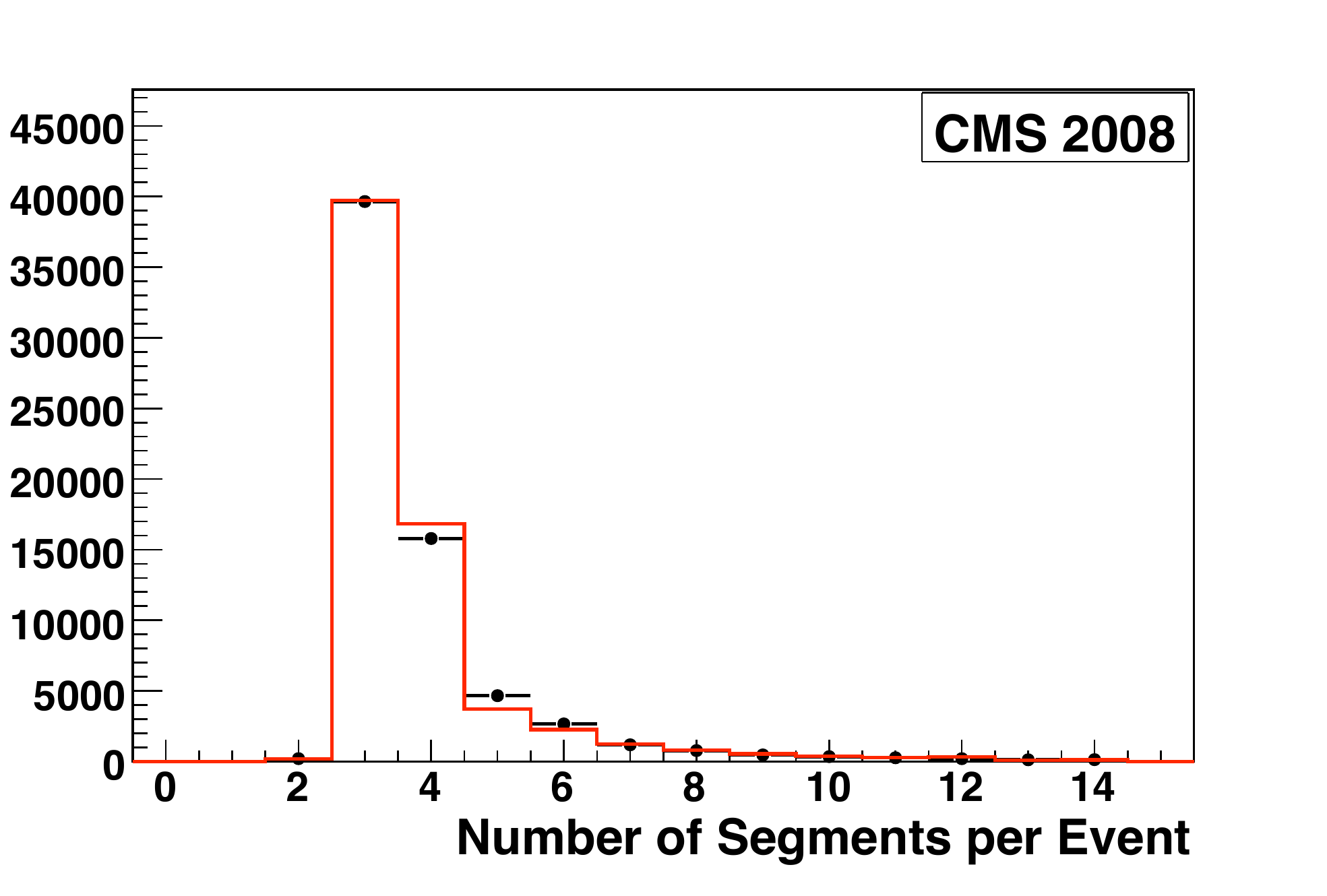}
\end{center}
\caption{\label{fig:AK_global} 
A comparison of the simulated events (solid line histogram)
to the CRAFT events (points) for simple global quantities.
Left: total number of rechits per event.
Right: total number of segments per event.}
\end{figure}

Distributions of the total number of rechits
per event and the number of segments per event are shown
in Fig.~\ref{fig:AK_global}.  The requirement of three
chambers with hits suppresses entries at the low end of
these distributions.  In the left-hand plot, the spikes
at 18 and 24~rechits correspond to muons which have passed
through three and four chambers.
\par
\begin{figure}
\begin{center}
\includegraphics[width=0.325\textwidth]{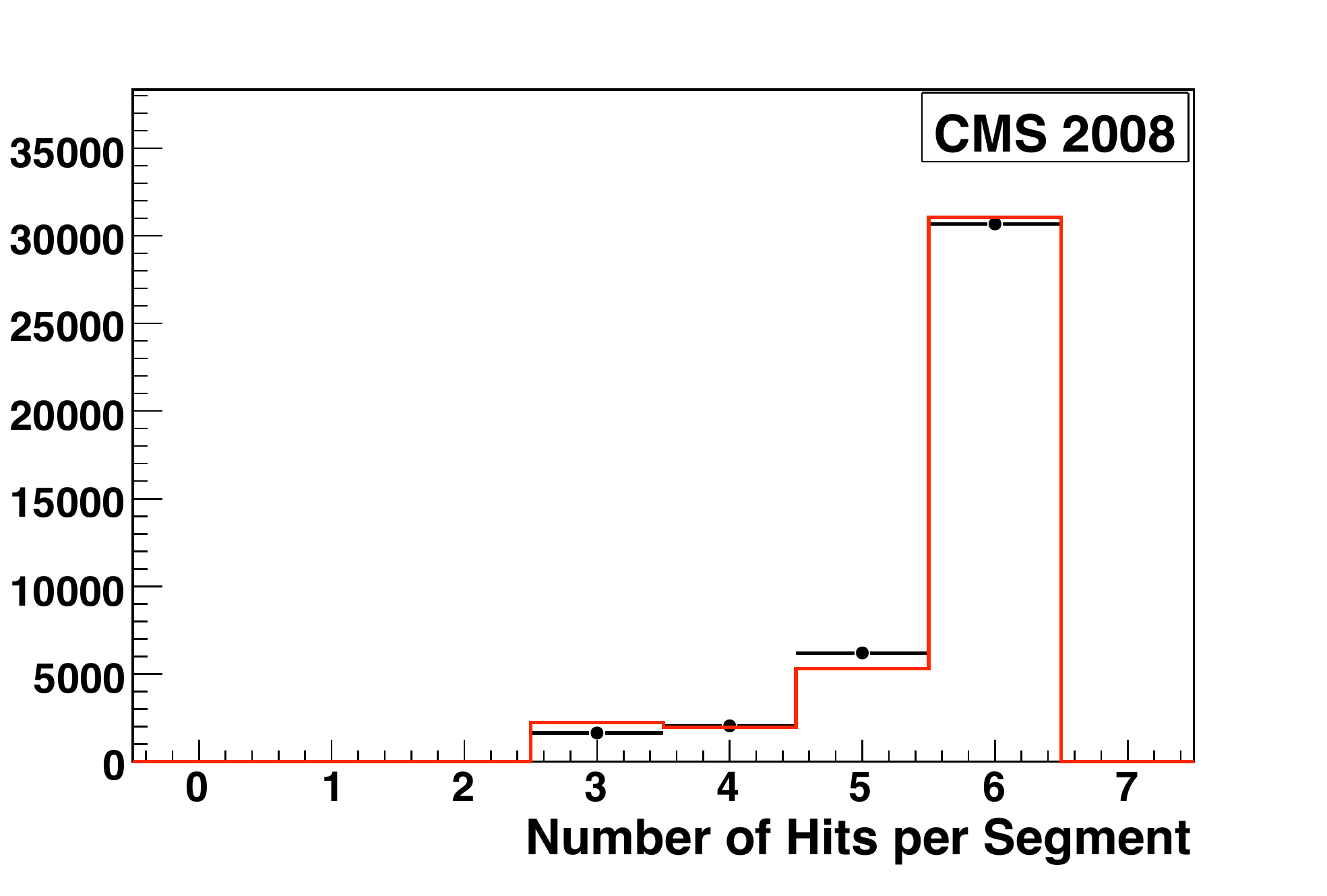}
\includegraphics[width=0.325\textwidth]{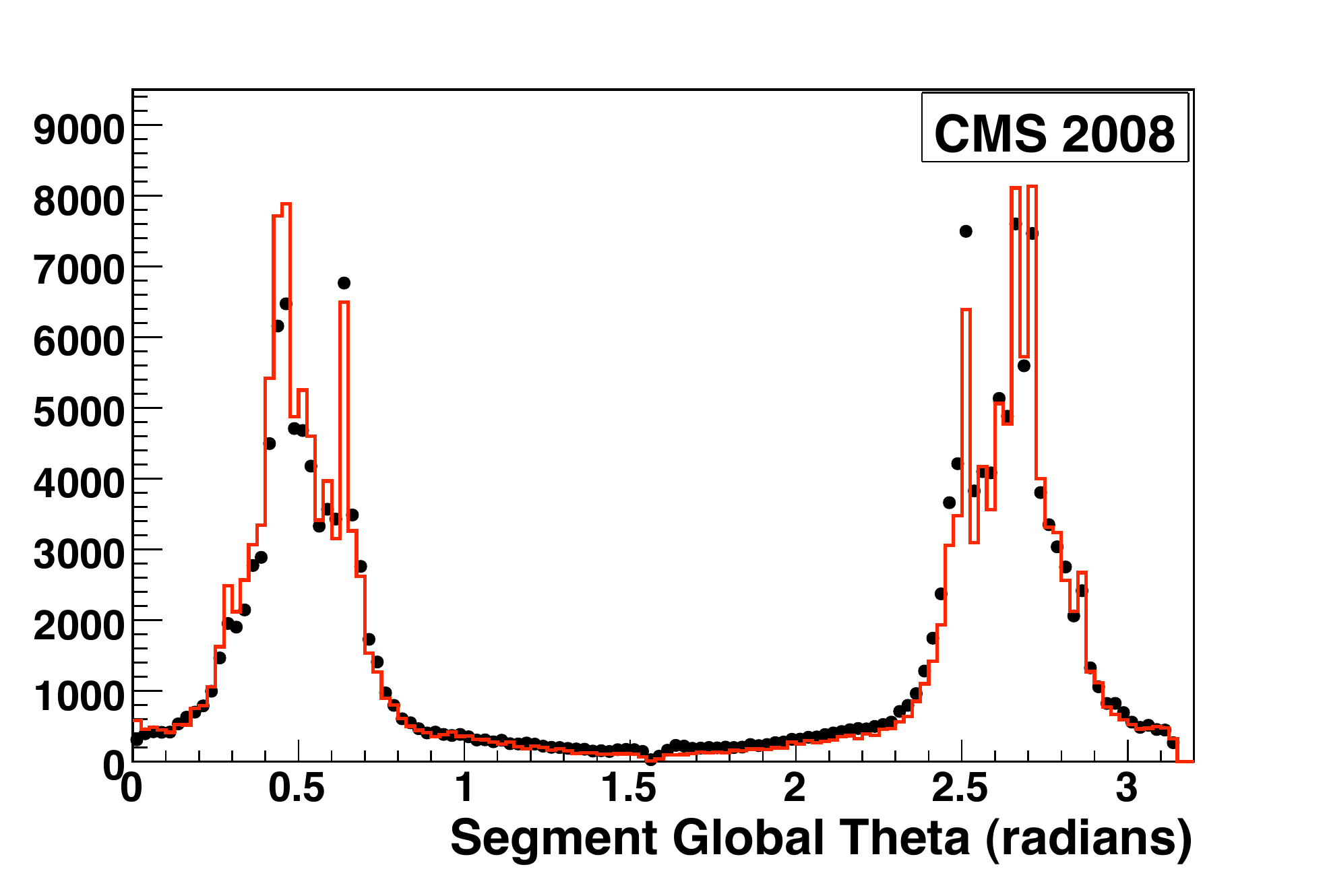}
\includegraphics[width=0.325\textwidth]{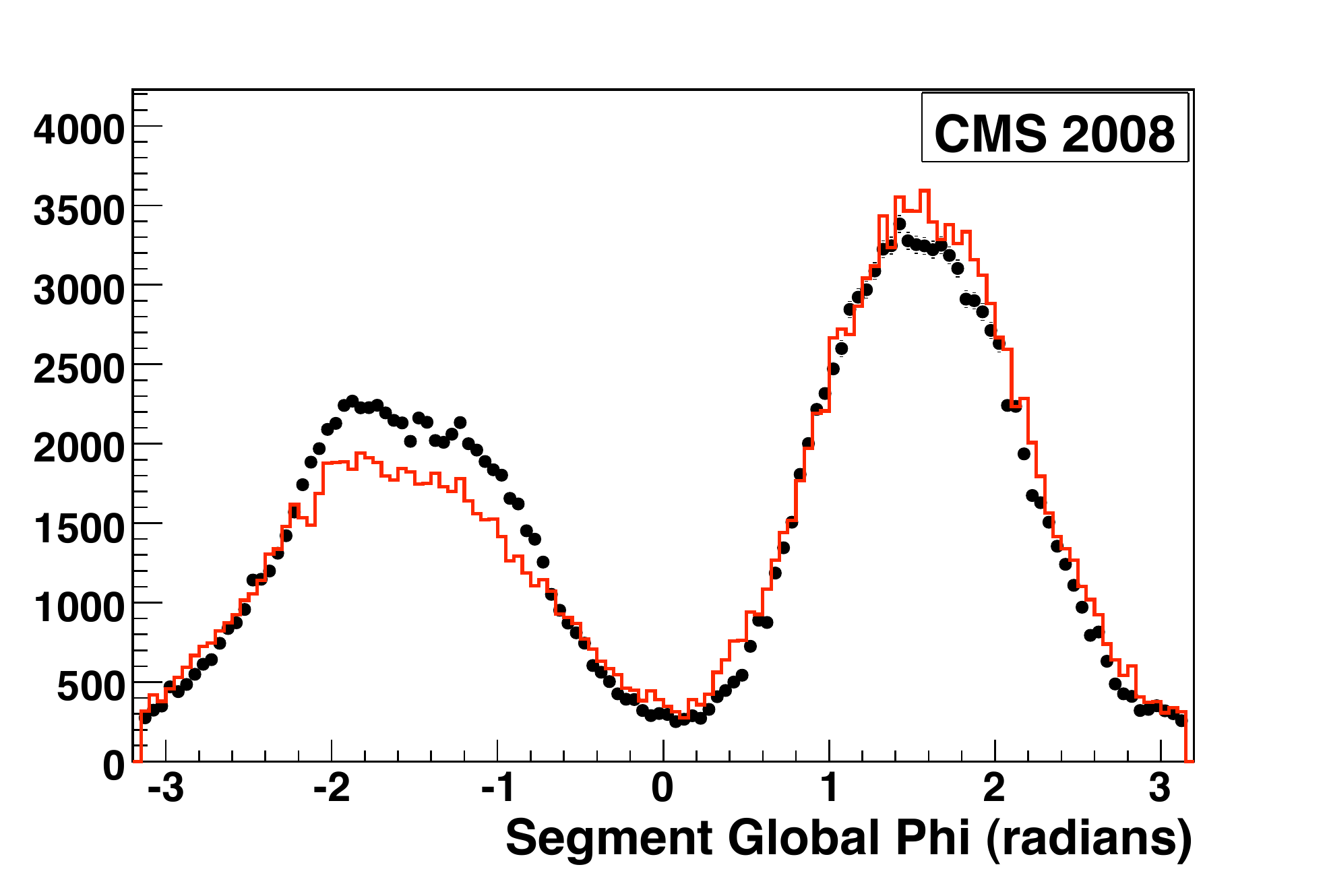}
\end{center}
\caption{\label{fig:AK_segment} 
A comparison of the simulated events to the CRAFT events
for reconstructed segment quantities.
Left: number of hits per segment.
Middle: global polar angle.  The two endcaps are clearly visible
(ME+ at $\theta \approx 0.5$ and ME- at $\theta \approx 2.7$).
The narrow spikes are defined by the boundaries of the CSC rings
and the event selection requirements.
Right: global azimuthal angle.  The bump at $\phi \approx 1.8$
corresponds to the upward vertical direction, and $\phi\approx -1.8$, 
to the downward.}
\end{figure}

Further information about the reconstructed segments is
shown in Fig.~\ref{fig:AK_segment}.  The first plot shows
the number of hits on a segment, which must be at least
three and cannot be more than six.  Most segments have
one rechit in every layer, and this is well reproduced
by the simulation.  The second and third plots show
the inclinations of the segments, namely, the polar
angle (``global theta'') and the azimuthal angle
(``global phi'').  These distributions reflect the
vertical nature of the cosmic ray flux as well as the
geometry of the muon endcap detector, and
are fairly well reproduced by the simulation.
\par
\begin{figure}
\begin{center}
\includegraphics[width=0.45\textwidth]{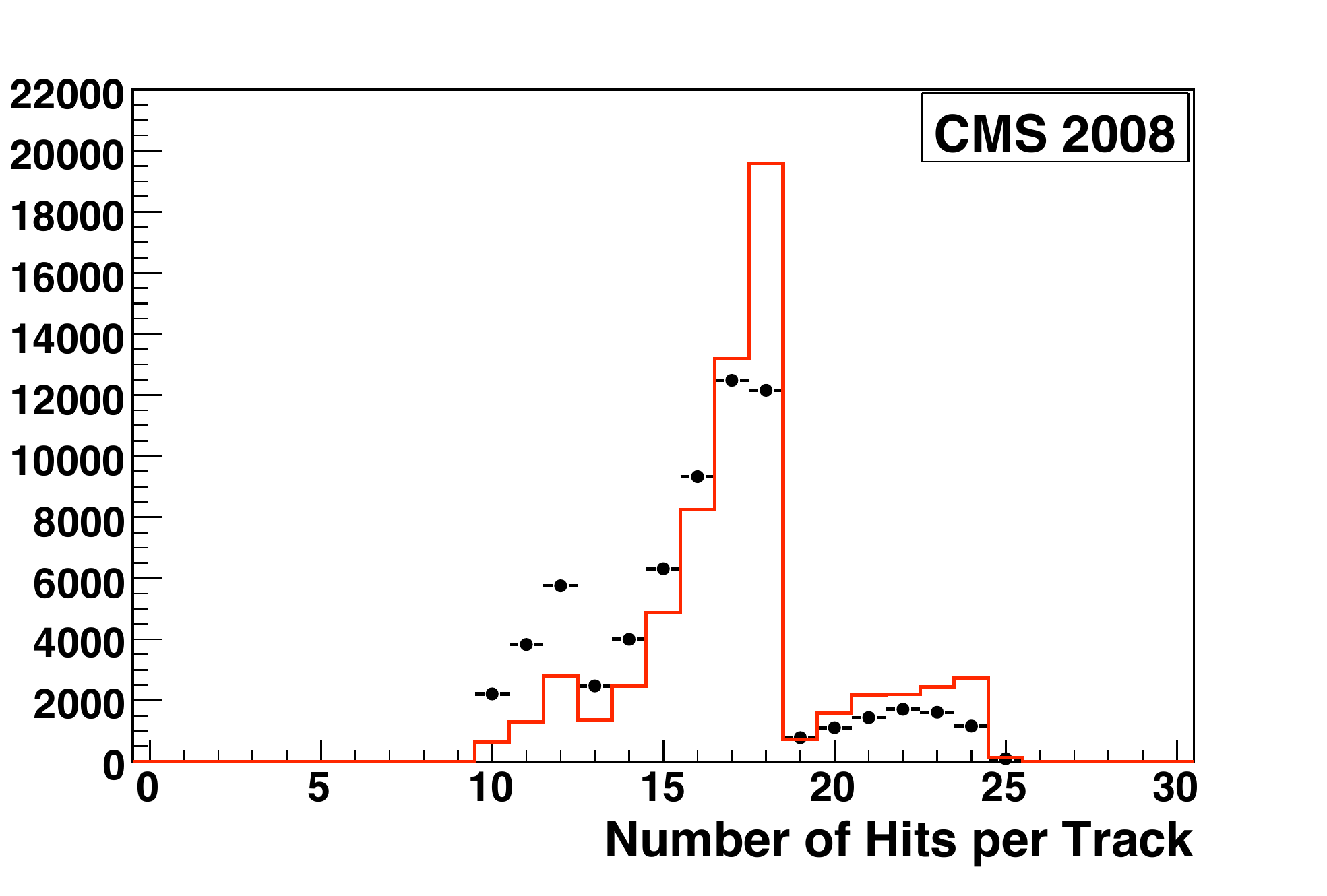}
\includegraphics[width=0.45\textwidth]{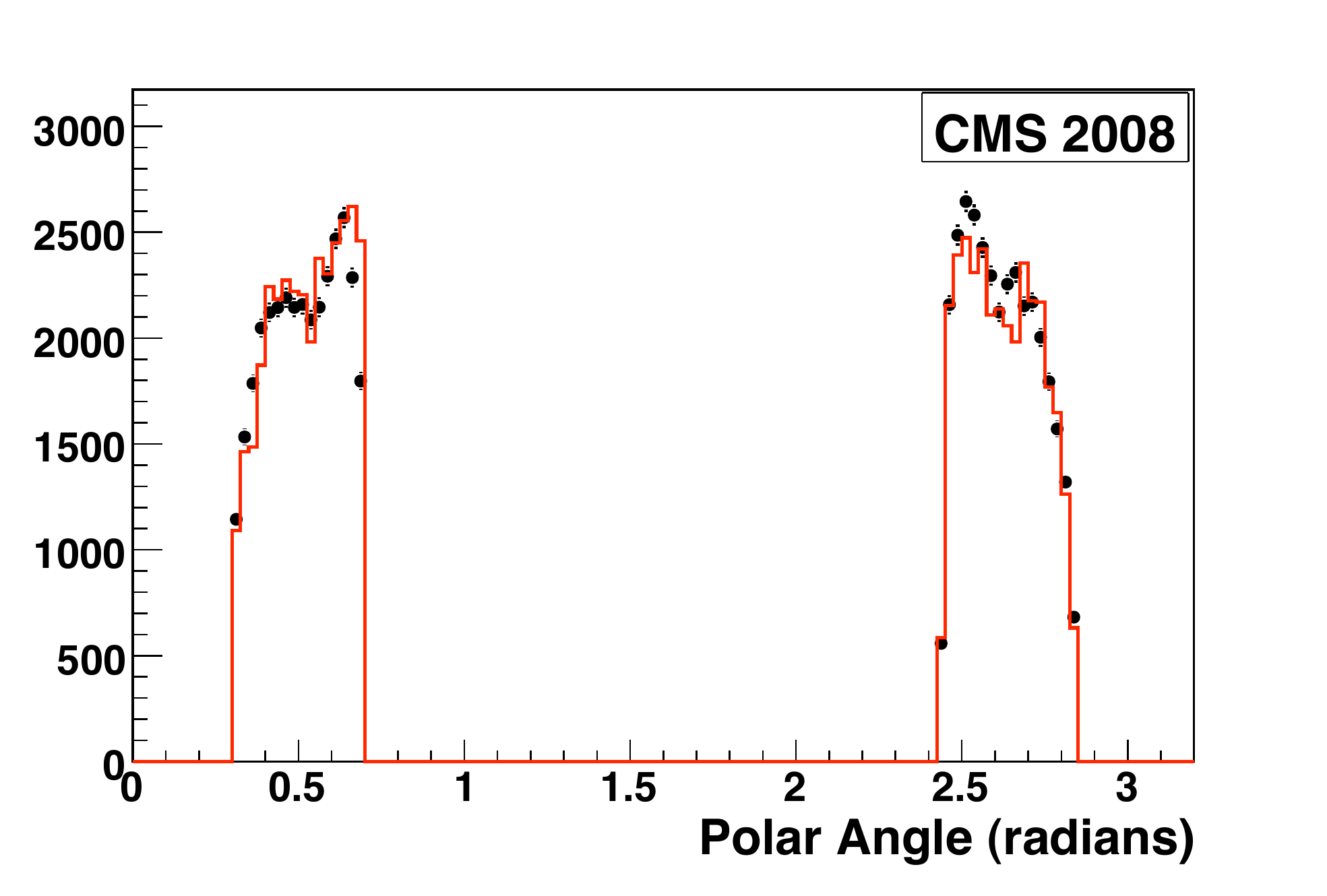}
\end{center}
\caption{\label{fig:AK_sa} 
A comparison of the simulated events to the CRAFT events
for stand-alone muon tracks.
Left: number of hits per track.
Right: global polar angle.  }
\end{figure}

Finally, basic distributions for stand-alone muons in
the endcaps are presented in Fig.~\ref{fig:AK_sa}.
The first plot shows the distribution of the number of
CSC rechits on the track.  The distribution of simulated
events differs from the CRAFT distribution in part because
the residual misalignments were not fully expressed in
the simulation.
The second plot shows the distribution of polar angles computed at the
point on the stand-alone muon track closest to the
center of the detector.  The agreement is very good.



\section{Noise}
\par
An assessment of the fraction of non-functional and noisy channels
must be made before any discussion of efficiencies or resolution.
Setting aside the few chambers that were turned off due to problems 
with high voltage, low voltage, or a very small number of malfunctioning
electronics boards, the number of anode wire and cathode strip
channels that failed to give data were below 1\% of the total.
Given the six-layer redundancy of each chamber, and the redundancy
of the four disks in each endcap, the impact of these very few
dead channels is negligible.
\par
Noise can have two different deleterious effects, in principle: 
it can generate extra hits which interfere with the reconstruction
of muon tracks, and it can smear or distort the measurement of
the charge registered on the strips, thereby smearing or distorting
the coordinates calculated from the strip information.  We have used 
the CRAFT data to make a basic assessment of the noise on both the 
anode wire and cathode strip channels.
\par
The first two out of eight 50~ns time slices of a strip signal 
are free of signal, by design, so that an average of these two 
ADC values can be used as an estimate of the base line.
Consequently, the difference in the ADC values
recorded for the first two time bins, $Q_1-Q_0$,
should be zero, aside from any random fluctuations
due to electronics noise.
In order to ensure that no signal
contributes to $Q_1$ and $Q_0$, strip channels were omitted
which have a sum of charges 13~ADC counts or more above base line.
\par
\begin{figure}
\begin{center}
\includegraphics[width=0.48\textwidth]{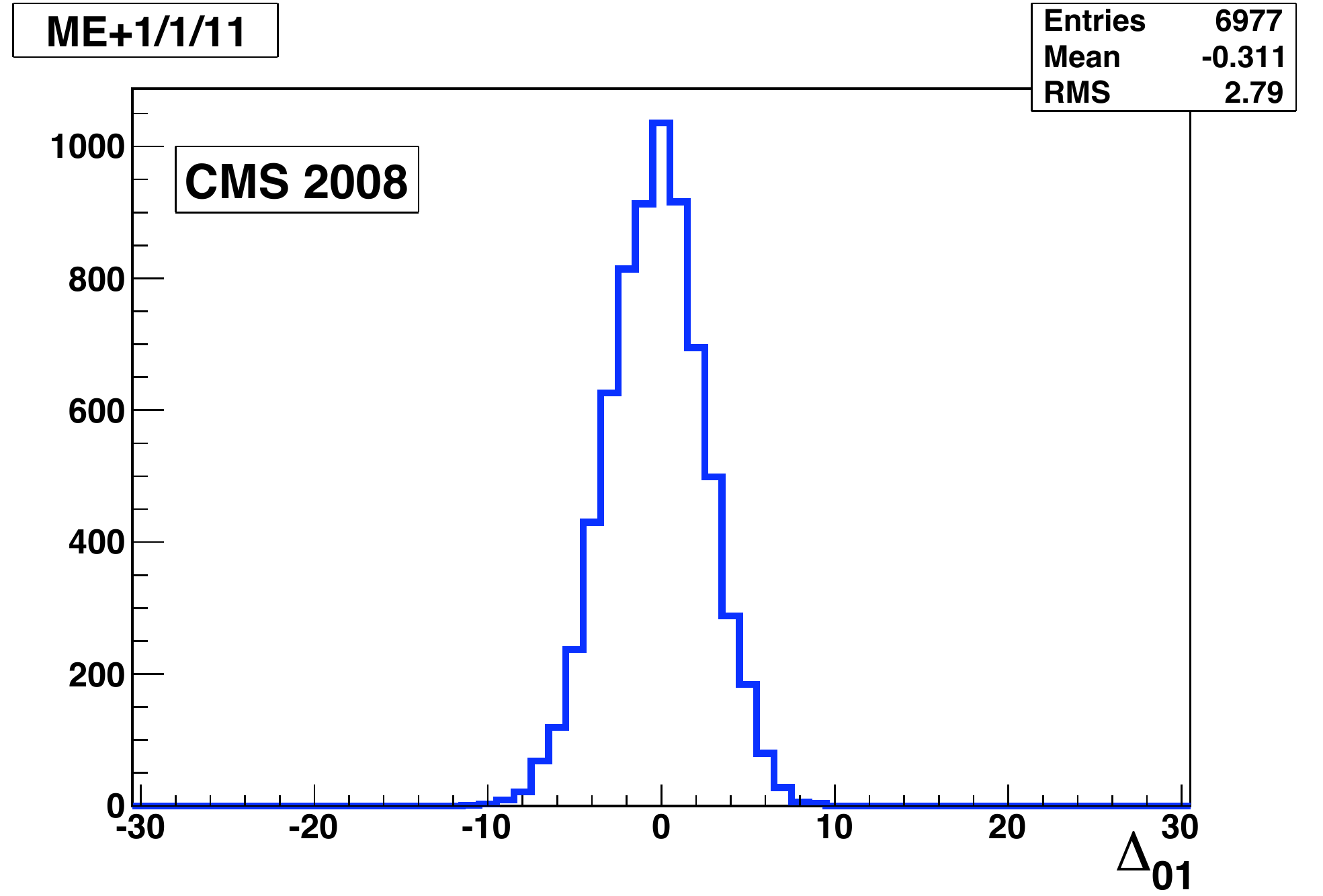}
\includegraphics[width=0.48\textwidth]{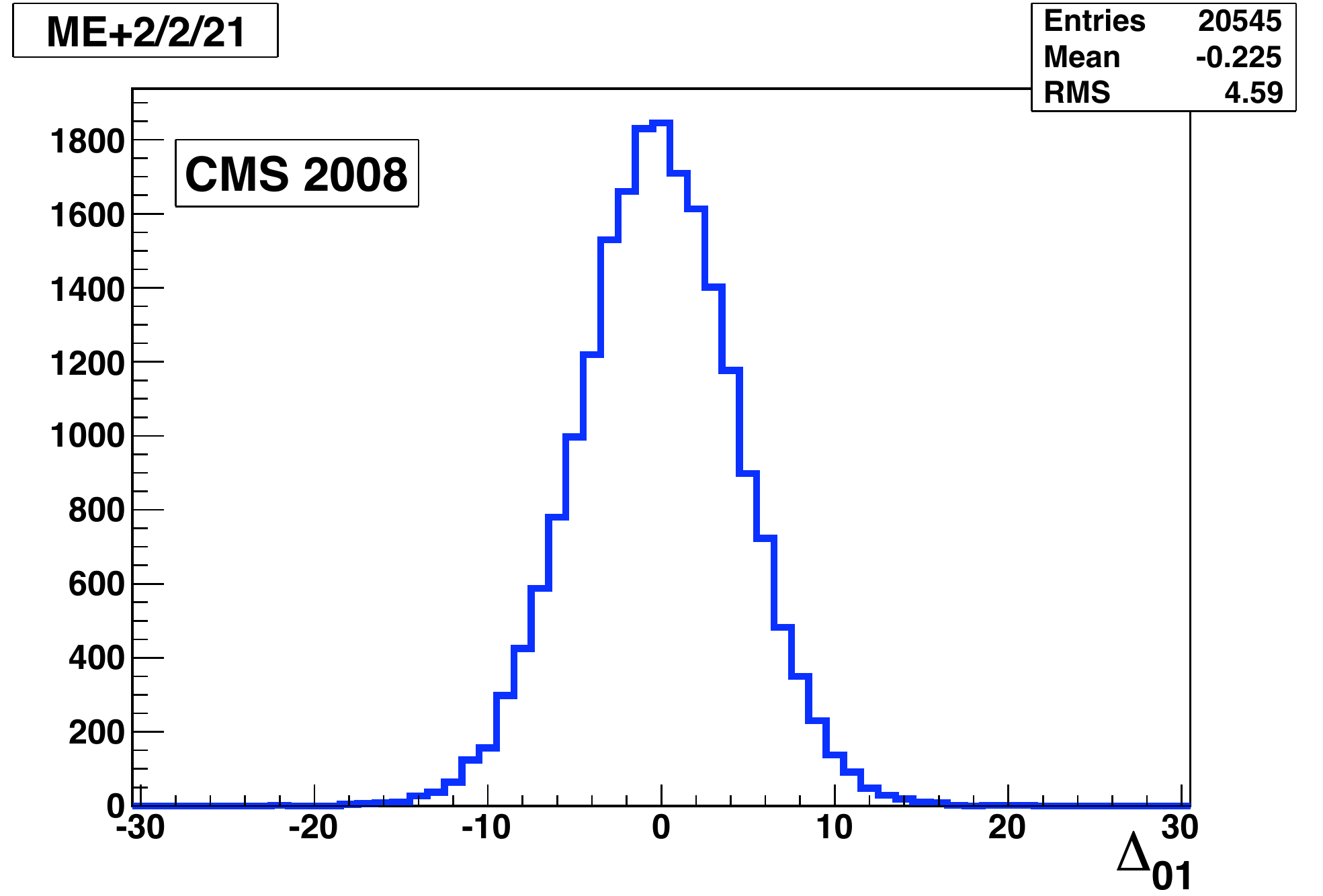}
\end{center}
\caption{\label{fig:rms_examples}
Two examples of $\Delta_{01}$ distributions, where
$\Delta_{01}$ is the difference in the first two ADC
readings for a strip.
On the left, a small chamber (ME$+1/1/11$), and on the
right, a large chamber (ME$+2/2/21$).
}  
\end{figure}

The rms of the distribution of $\Delta_{01} \equiv Q_1-Q_0$,
$\sigma_{01}$, is taken to be a measure of noise, and was
obtained for all sets of 16 strip channels handled by the
cathode front-end boards, for all chambers.
Figure~\ref{fig:rms_examples} displays two example
distributions for $\Delta_{01}$ showing that the distributions 
have no tails or asymmetry.
One ADC count corresponds to approximately~$0.54$~fC.
\par
\begin{figure}
\begin{center}
\includegraphics[width=0.49\textwidth]{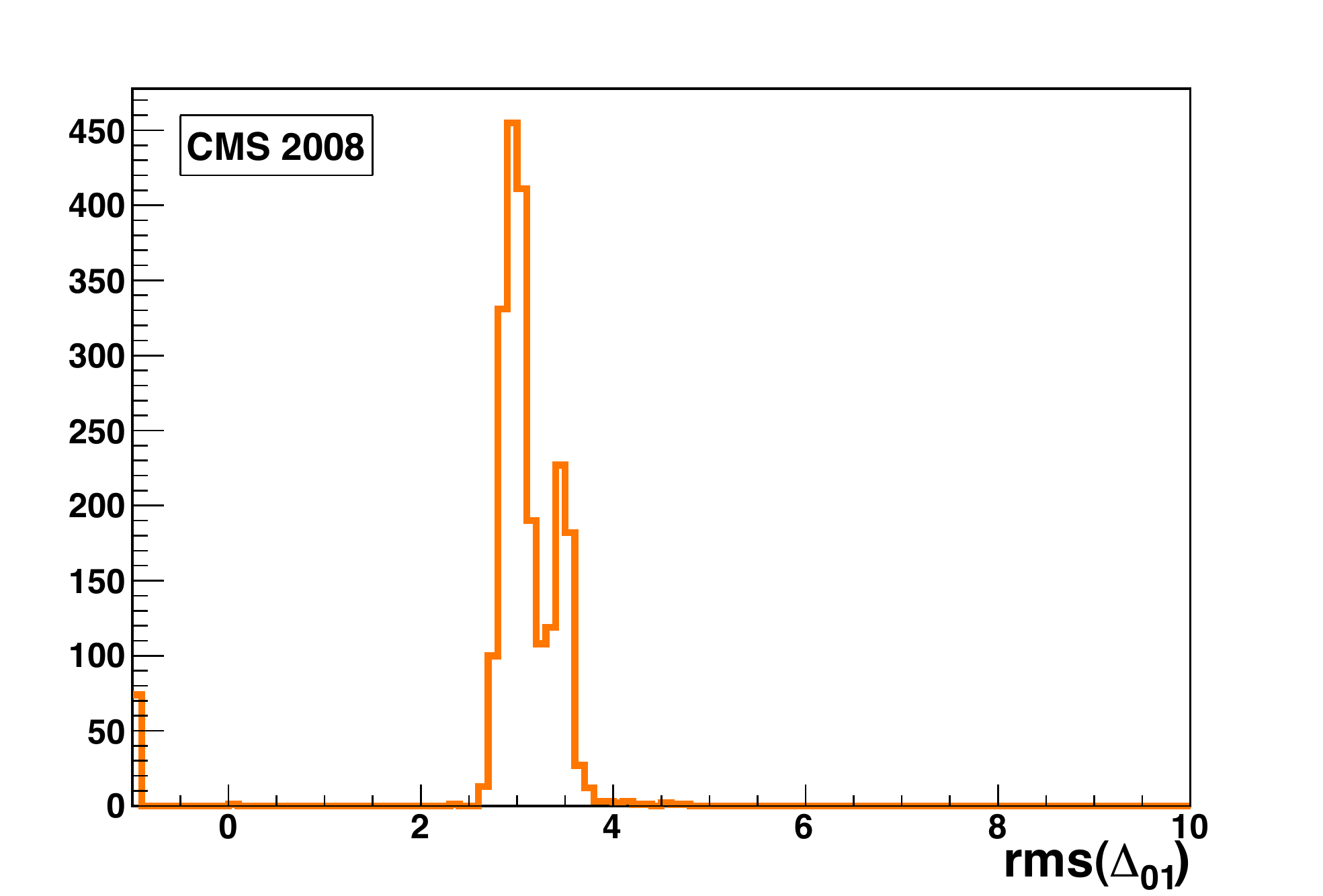}
\includegraphics[width=0.49\textwidth]{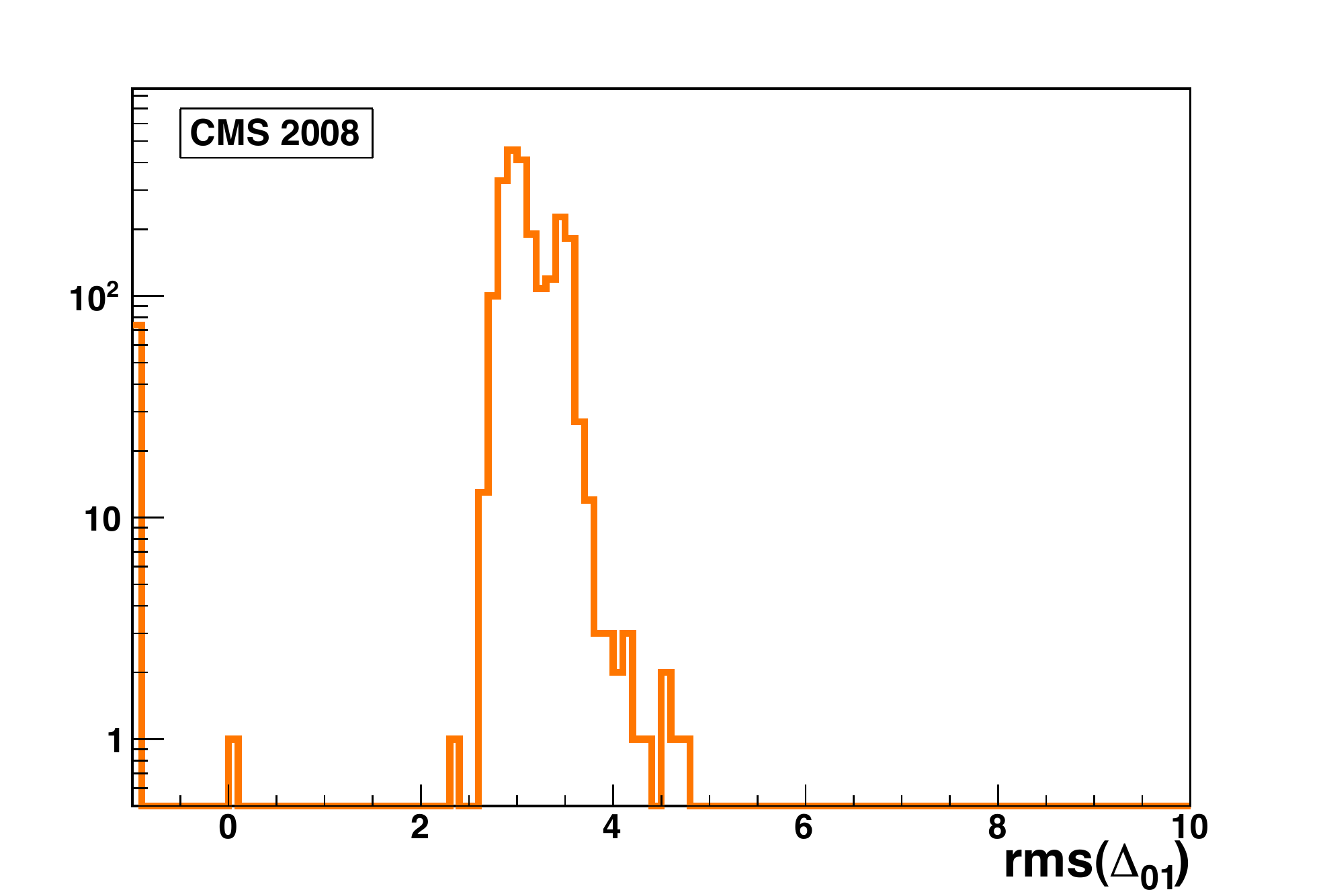}
\end{center}
\caption{\label{fig:sigma01}
Distribution of all $\sigma_{01}$ values, 
\ie, the rms of the difference in the first two ADC readings,
on a linear scale (left) and a log scale (right).
There is one entry per chamber, and the
entries at $\Delta_{01} = -1$
correspond to channels that were turned off.
The single entry at $\Delta_{01} = 0$ comes from
a single nonfunctional channel.}
\end{figure}

Figure~\ref{fig:sigma01} shows the distribution of all $\sigma_{01}$ values
which are typically about 3 ADC counts or slightly larger;
the spread of the distribution is small indicating excellent uniformity.
There are no large values, indicating
no oscillating or otherwise noisy channels.
The two populations in Fig.~\ref{fig:sigma01}
correspond to the smaller and larger chambers.
\par
The time integration of the amplifier leads to an
auto-correlation manifested as a correlation coefficient
of $0.26$ between consecutive time slices which reduces
slightly $\sigma_{01}$ with respect to the uncorrelated
case.  We repeated this noise analysis using the first
and the last time bins, and found that the rms values
increased by about 10\%, due partly to the lack of 
correlation between the first and last time slices.
We also observed some sensitivity to signal in the last
time slice, due to cross-talk, which explains the rest
of the 10\% increase with respect to $\sigma_{01}$.


\par
\begin{figure}
\begin{center}
\includegraphics[width=0.6\textwidth]{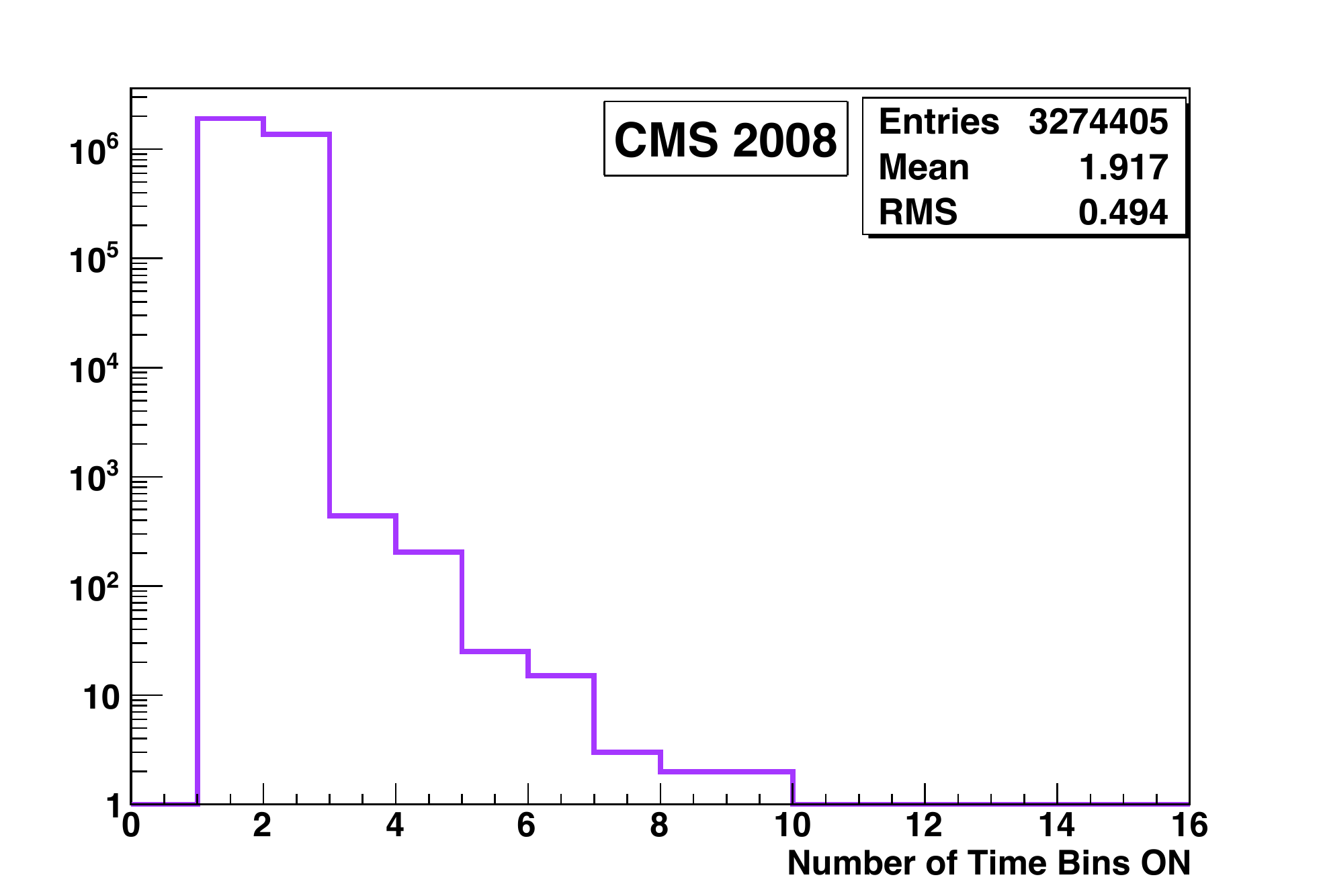}
\end{center}
\caption{   \label{fig:nmb_time_occup_ME-2/1/9} 
A semi-log plot of $N_{\mathrm{on}}$ 
(the number of time bins for which there is signal)
for all anode wire channels in ME-2/1/9.}
\end{figure}

The anode wire signals normally extend over one or two 25~ns time bins.
A noisy channel, however, will rise above threshold in more time bins,
so a useful quantity to identify noisy channels
is the number of time bins for which a given anode
hit is {\sl on}, denoted here by $N_{\mathrm{on}}$.
The distribution of $N_{\mathrm{on}}$ for all anode channels in
a particular chamber is shown in Fig.~\ref{fig:nmb_time_occup_ME-2/1/9}, on a
semi-log plot.  A very small tail for $N_{\mathrm{on}} > 2$ can be seen.
The number of noisy anode wire channels is estimated to be less than~0.1\%.

\typeout{=========== START OF CSC EFFICIENCY ===========}
\section{Efficiency}
\par
The goal of this study is to measure the absolute efficiency of 
each step in the reconstruction of muons in the CSCs,
from the generation of ALCTs and CLCTs to
segment reconstruction.  By design, for good muons
coming from the interaction point, all steps should
be highly efficient.  The method described here uses
two chambers to ``tag'' a muon that passes through a
designated ``probe'' chamber.  When computing the 
efficiency of each step, the same tagged sample
(\ie, the denominator in the efficiency calculation)
is used for all steps.
\par
For efficiency measurements, we need a well-defined muon 
track which is independent of the measurements in the 
chamber under investigation. We use muon tracks
reconstructed in several CSCs without any information 
from the silicon tracker.
The number of useful stand-alone muons is adequate for the
present purposes, thanks to the redundancy of the muon endcap system.
To minimize the impact of 
multiple scattering, energy loss, and tracking in a strong
magnetic field, a chamber is probed only if it lies between 
the endpoints of the track.  Consequently, at least two independent
measurements of the muon track are needed, and only interpolation 
and not extrapolation to the probe chamber is used.  
Some rings, namely ME$\pm 1/1$, ME$\pm 4/1$ and ME$-3/2$ cannot be 
covered by this study, although hits in the CMS Resistive Plate
Chambers allow coverage of ME$+3/2$.
\par
A typical event selected for these efficiency measurements 
contains three or four CSCs contributing to a good stand-alone
muon track.  Since the trigger efficiency is generally high
(see below), and a trigger from any one of these chambers
sufficed to produce a trigger for read out of CMS, we assume
that any trigger bias in these results is negligible.
\par
\begin{figure}
{\centering
\includegraphics[width=0.48\textwidth]{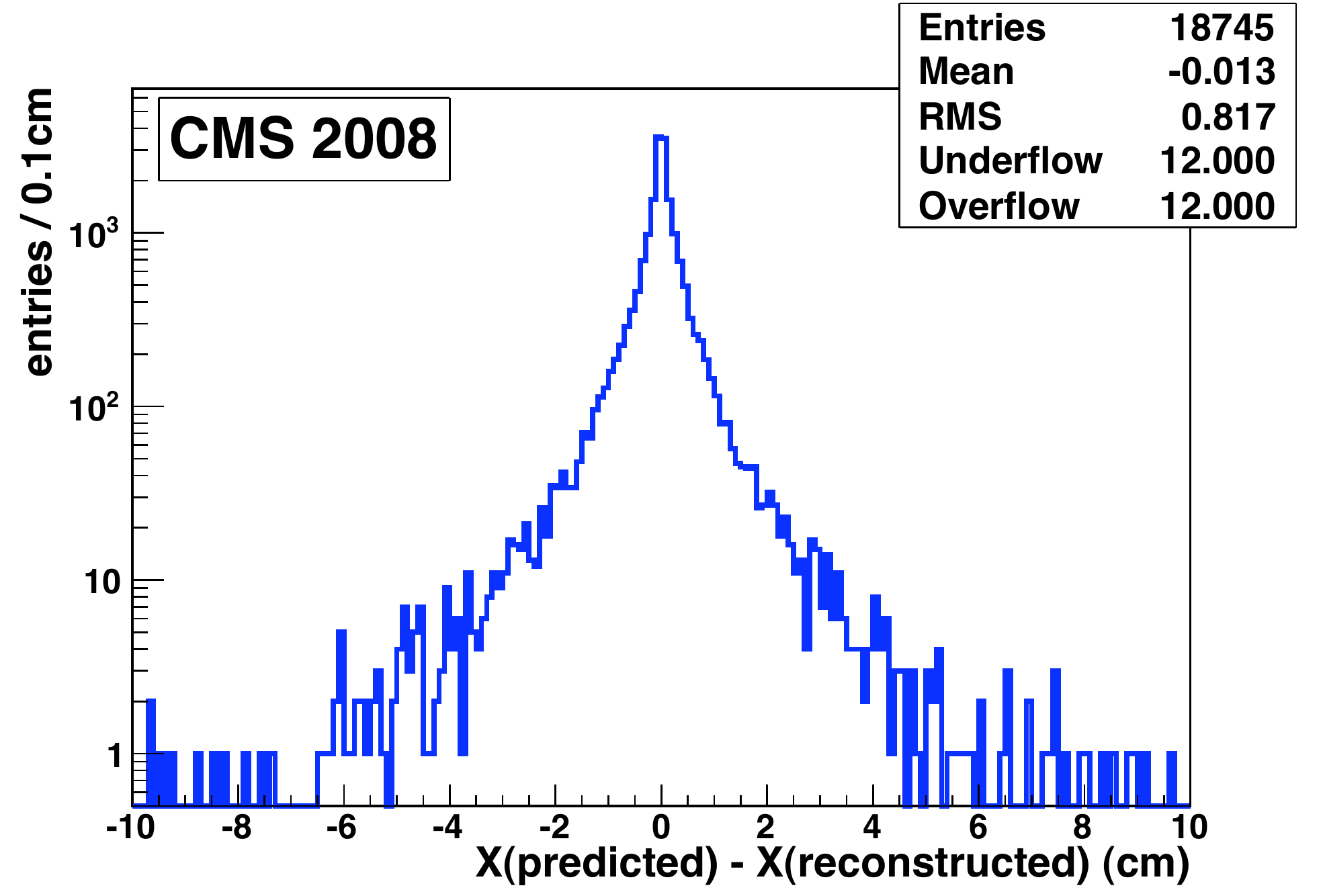}
\includegraphics[width=0.48\textwidth]{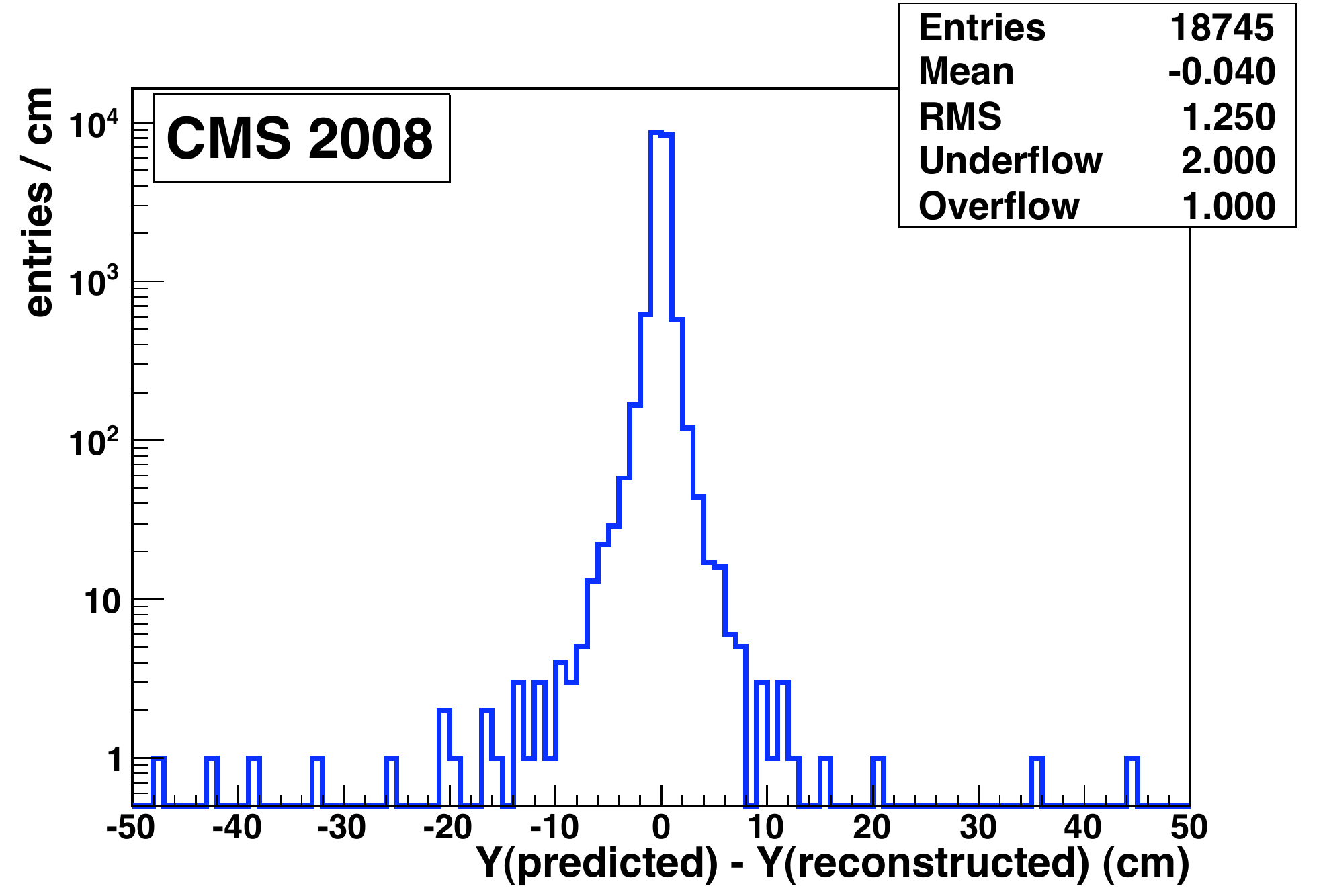}
}
\caption{\label{diffXY}
Differences between the predicted positions of a segment and 
the position of the reconstructed segment in the probe chamber.
$\Delta x$ is on the left, and $\Delta y$ is on the right, where
$x$ and $y$ are local coordinates.  $x$ is measured primarily by
the strips, and $y$ by the wires.}
\end{figure}

We place cuts on the predicted position of the muon in the
probe chamber to avoid losses due to insensitive regions
at the periphery of the chamber and at the boundaries
of the high voltage segments.
Figure~\ref{diffXY} shows distributions of the difference 
between the measured position of a segment in the probe chamber
and the predicted position, obtained by propagating the
muon track from another station to the probe chamber,
taking the magnetic field, multiple scattering and energy
loss into account.
In this figure, the local coordinate $x$ runs parallel to
the wires, and is measured primarily by the strips, while
$y$ runs perpendicular to the wires, and is measured by
the wire signals.  According to these distributions, 
nearly all of the tracks fall within 10~cm of the predicted position.

\par
A set of stringent criteria is used to select ``good''
tracks for the denominator of all efficiency calculations. 
Only one stand-alone muon track is allowed in an endcap.
This track has to have a minimum number of hits,
and to be reconstructed well, as indicated by the $\chi^2$
and the relative error on the momentum.  
The momentum is required to be in the range 
$25 < p < 100$~GeV/$c$. A track 
satisfying these requirements is propagated
to a designated ring of CSC chambers to 
ascertain which chamber is the probe chamber.  If the interpolated
point lies within 10~cm of the edges of the chamber or dead regions 
defined by high voltage segment boundaries, then the chamber is skipped.
The tracks which pass all of these criteria are the ``probe'' tracks.

\par
The following sections report the details of the measurements
and the values of the efficiency for each step in the
CSC local reconstruction.
\subsection{LCT Efficiencies  \label{sec:LCTEff} }
\par
The ALCT and CLCT efficiencies are measured independently.  
For a given chamber,
the ALCT and CLCT digis are unpacked to test for the presence of a
valid ALCT or CLCT.  If they are present anywhere in the chamber,
then the trial is a ``success'' and the chamber is ``efficient'' for
that event.  
\par
To suppress the muons which are not likely to fire the ALCT and/or 
CLCT triggers, we apply cuts on the slopes of the muon tracks 
interpolated through the chamber:
\begin{displaymath}
  -0.8 < \frac{dy}{dz} <-0.1
 \qquad {\mathrm{and}} \qquad
  \left| \frac{dx}{dz} \right| < 0.2 .
\end{displaymath}
One could adjust these ranges for the various rings of chambers,
but the impact on the efficiency measurements is negligible.
All the efficiencies measured with CRAFT data include these 
requirements in the  event selection.
\par
\begin{figure}
{\centering
\includegraphics[width=0.48\textwidth]{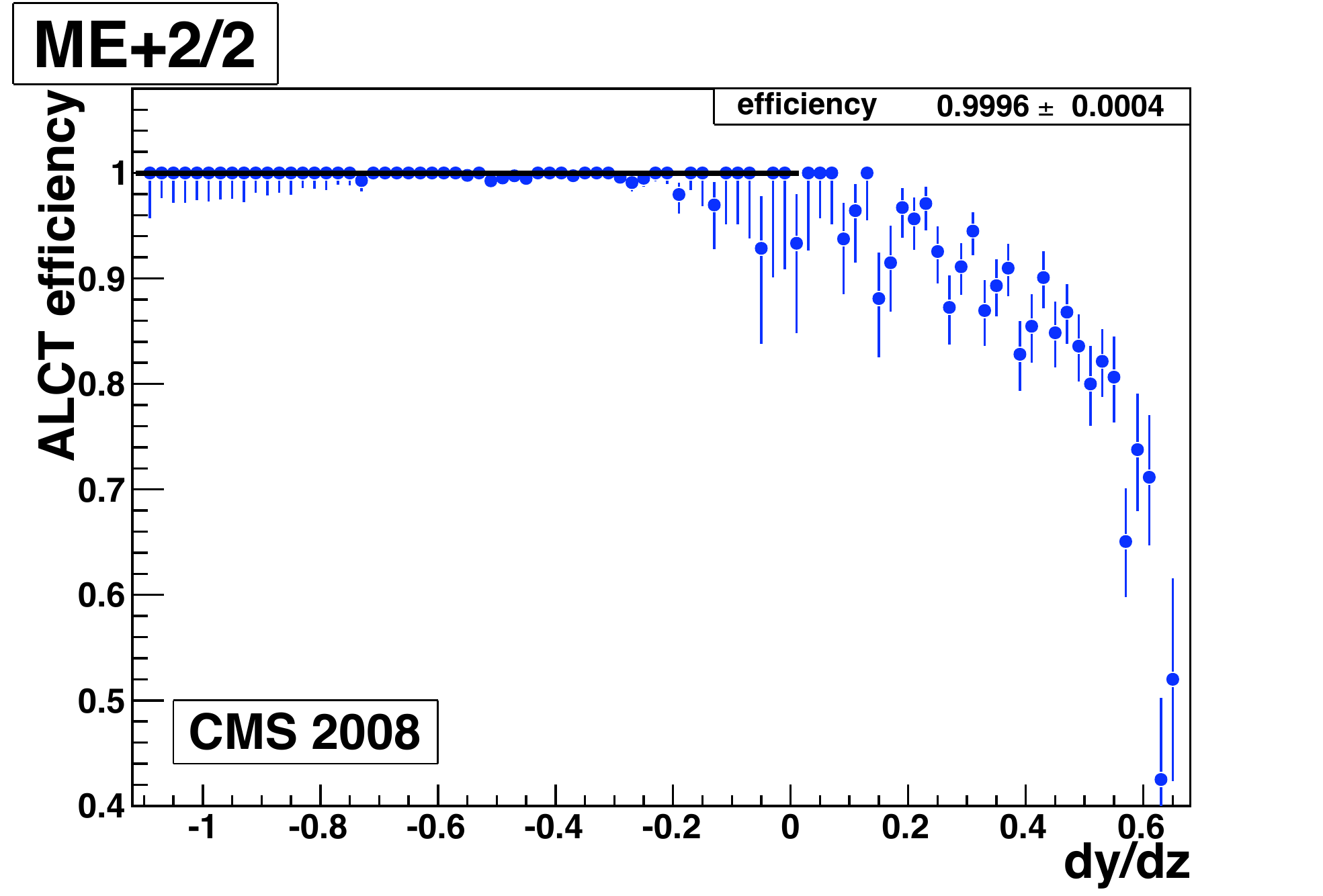}
\includegraphics[width=0.48\textwidth]{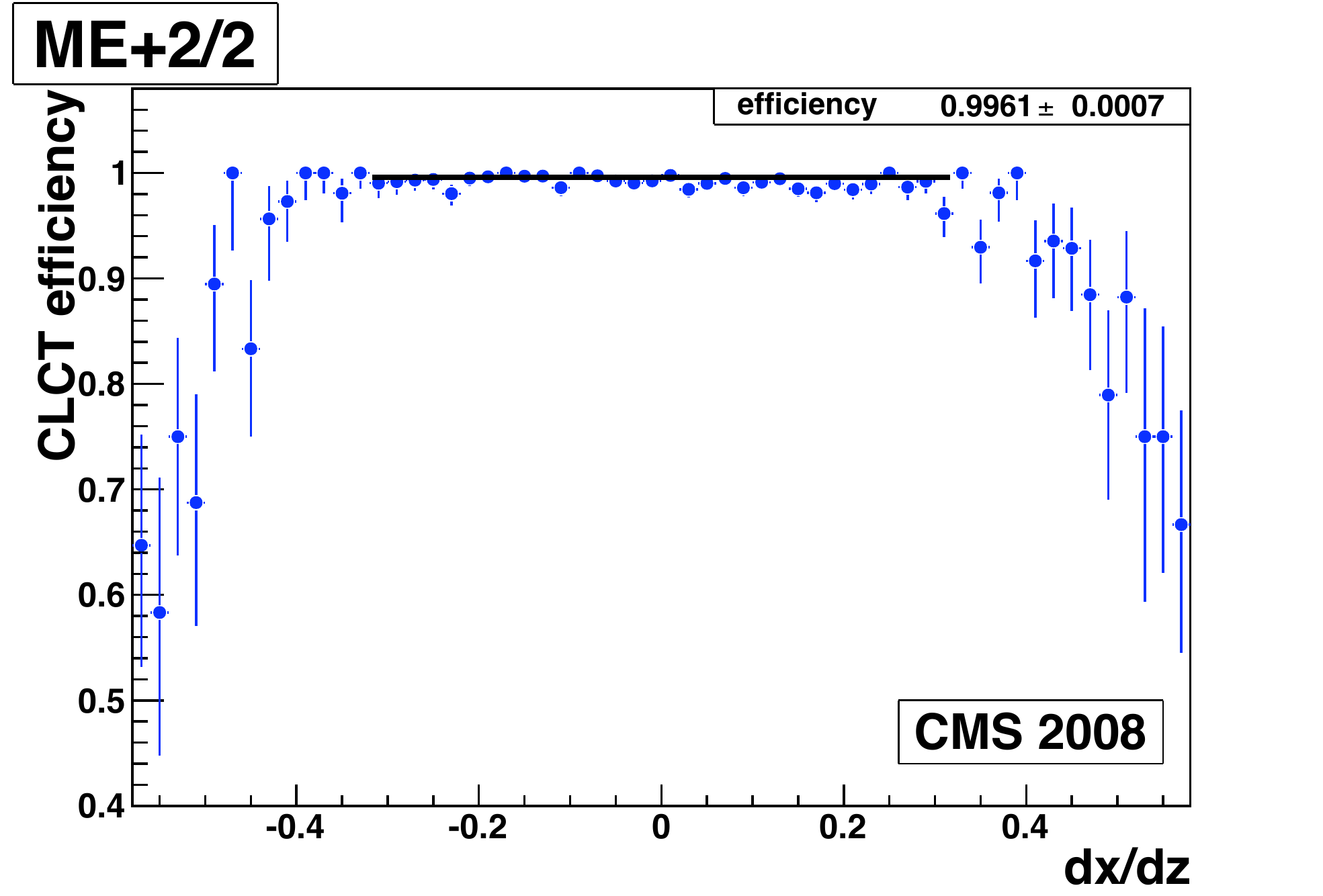}
\caption{Left: ALCT efficiency as a function of the track inclination, 
$dy/dz$ in local coordinates.
Right: CLCT efficiency as a function of the track inclination,
$dx/dz$ in local coordinates.}
\label{ALCT_dydz}
}
\end{figure}

The variation of the ALCT efficiency as a function of $dy/dz$
is shown in Fig.~\ref{ALCT_dydz} (left).  For this figure, the cut
on $dy/dz$ was not applied, although the cut on $dx/dz$ was applied.
Similarly, the variation of the CLCT efficiency as a function
of $dx/dz$ is shown in Fig.~\ref{ALCT_dydz} (right), with the cut on
$dx/dz$ relaxed, and the cut on $dy/dz$ applied.
The results shown in these plots are based
on data from chambers 5--13 in ring ME$+2/2$ which are known to
have been operating well during CRAFT.
In both figures, clear plateaus can be seen which were fit 
with level functions to ascertain the efficiency.
Very high values in excess of $0.99$ are observed,
confirming earlier results obtained with cosmic rays~\cite{MTCC_data}.

\subsection{Strip and Wire Group Efficiencies}
\par
The presence of an ALCT and CLCT should trigger the readout of the chamber,
and hence, signals on the wires and strips should be present in the raw data,
or equivalently, in the strip and wire digis.  The efficiency for
strip and wire digis are measured independently.  The probe is given
by a good track passing through the given chamber.  
\par
The efficiencies of strips, wire groups and rechits are defined naturally per layer.
If the layer measurements are independent, then the average efficiency per chamber 
would be 
\begin{equation}
\label{eff1}
\bar{\epsilon} = \frac{\sum_i \epsilon_i}{L} =  \frac{\sum_i {n_i}}{N \times L}
\end{equation}
with an estimated uncertainty of
\begin{equation}
\label{err_end}
\Delta \bar {\epsilon} =
        \sqrt{\frac{\bar{\epsilon} \times (1-\bar{\epsilon}) }{L \times N}} ,
\end{equation}
where $L = 6$ is the number of layers,
$\epsilon_i$ is the efficiency in layer $i$ ($i=1,..,6$),
$n_i$ is the number of efficient cases (``successes'') for layer $i$, and
$N$ is the number of probe tracks.
In principle, there might be events with a simultaneous loss of 
information from all six layers, in which case Eq.~(\ref{err_end}) 
is incorrect.  
There is no evidence for any such correlated losses.
\par
\begin{figure}
{\centering
\includegraphics[width=0.48\textwidth]{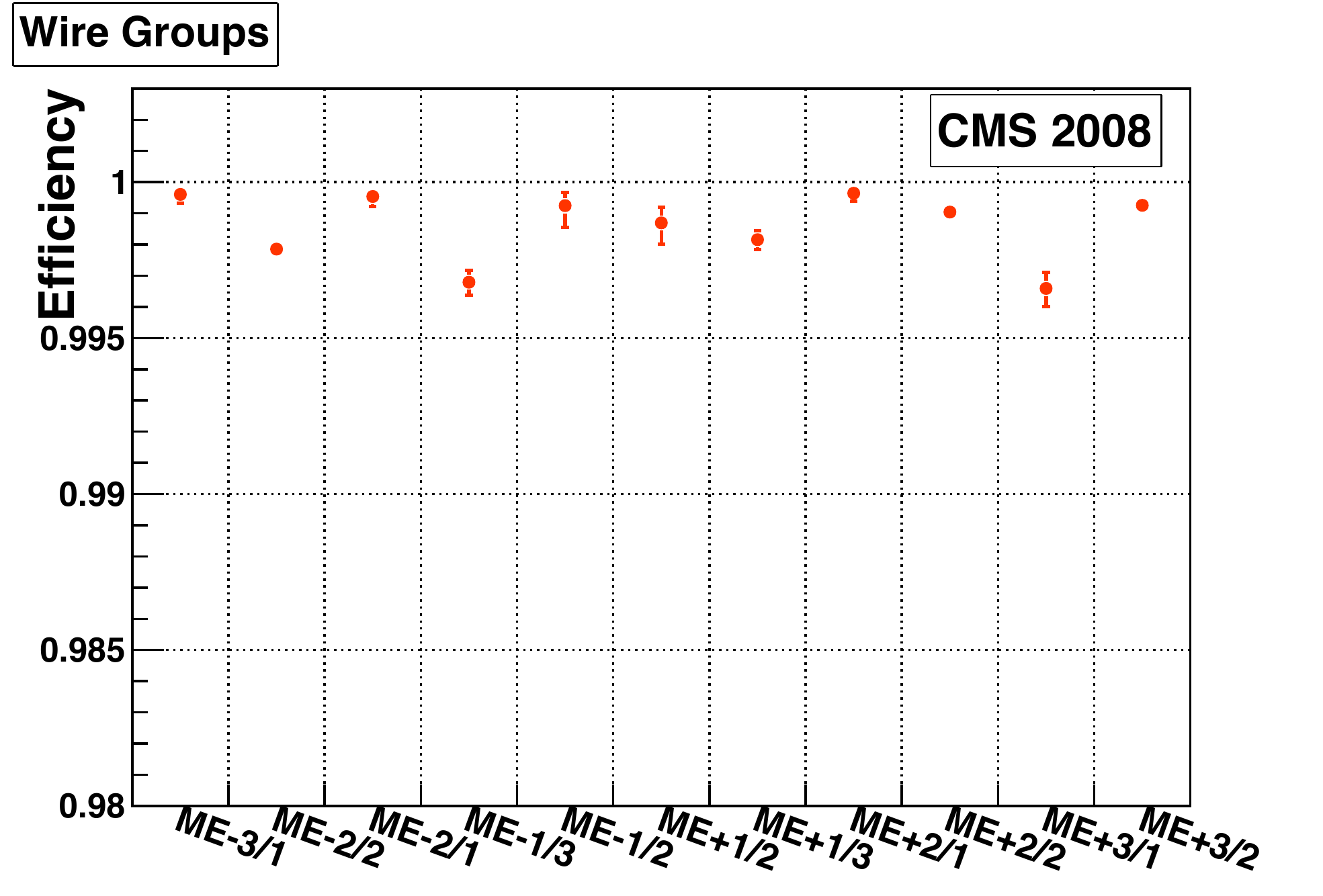}
\includegraphics[width=0.48\textwidth]{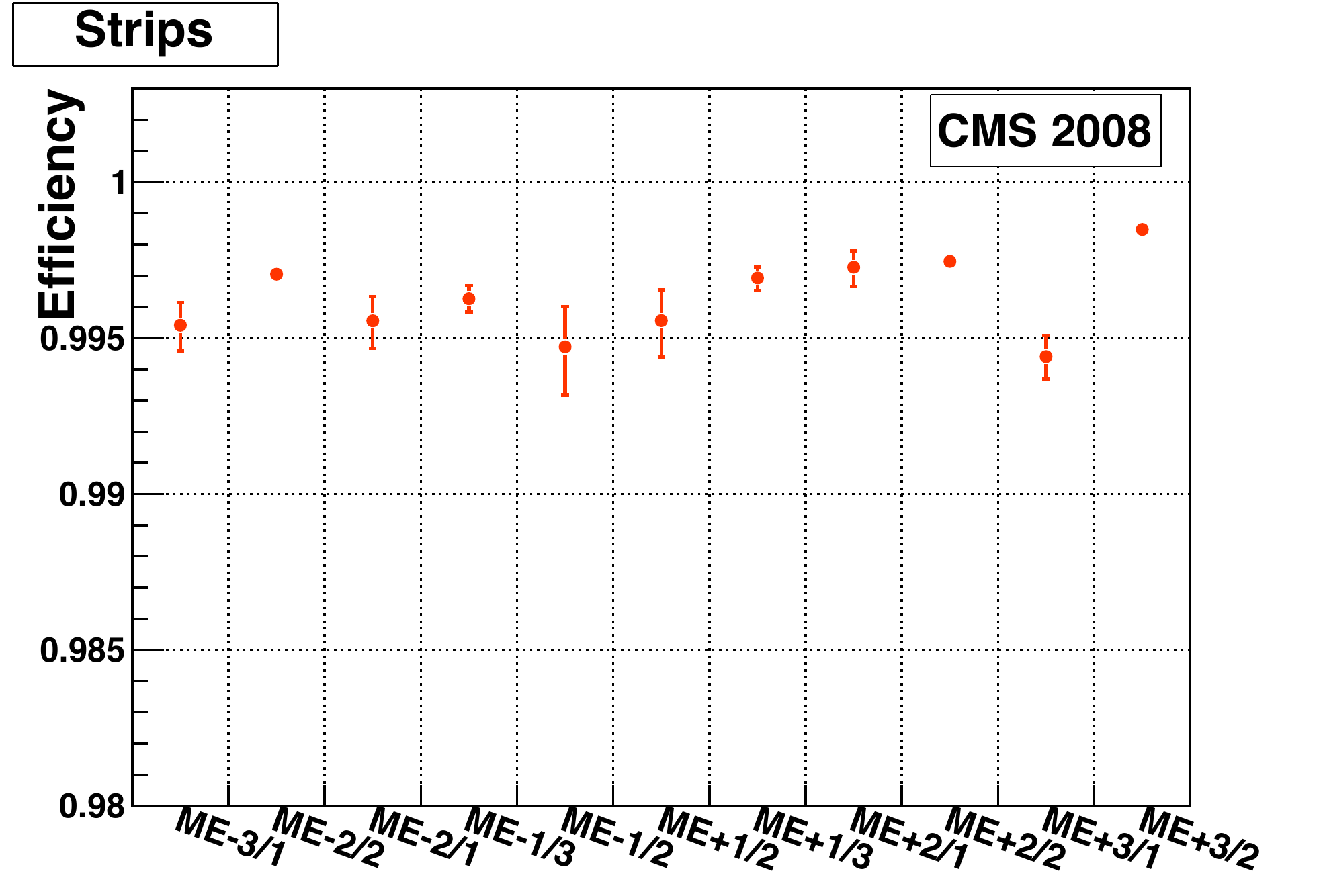}
\caption{\label{Eff_WG}
A summary of wire group (left) and strip (right) digi efficiencies,
over all functioning chambers in a ring.  Some rings are inaccessible
in this study with CRAFT data.}
}
\end{figure}

The average wire group and strip digi efficiencies are shown in Fig.~\ref{Eff_WG}.
Typically, all six layers have high efficiency, greater than 99.4\%.

\subsection{Rechit Efficiency}
\par
The efficiency for reconstructing a rechit is measured for each layer
in a chamber.  The chamber is efficient if the rechits are found in a 
given layer - there is no requirement on the distance between
the rechit and the interpolated point.  Also, no
quality requirements are placed on the individual rechits as part
of the measurement of rechit efficiency.
\par
The rechit efficiency will be a convolution of the strip and wire group
digi efficiencies.  It might also depend on some of the details of
the rechit reconstruction algorithm, especially as regards quality 
or other criteria applied to the strip and wire signals.
The rechit efficiency for all the accessible CSC rings is above 99.3\%,
as shown in Fig.~\ref{Eff_rechit}~(left).

\subsection{Segment Efficiency}
\par
It should be possible to build a segment if at least three good
rechits are recorded along the muon trajectory.  The chamber is 
efficient if a segment has been reconstructed.  No matching
criteria have been applied because the reconstructed segments
are found close to the extrapolated positions, as shown
in Fig.~\ref{diffXY}.
\par
\begin{figure}
{\centering
\includegraphics[width=0.48\textwidth]{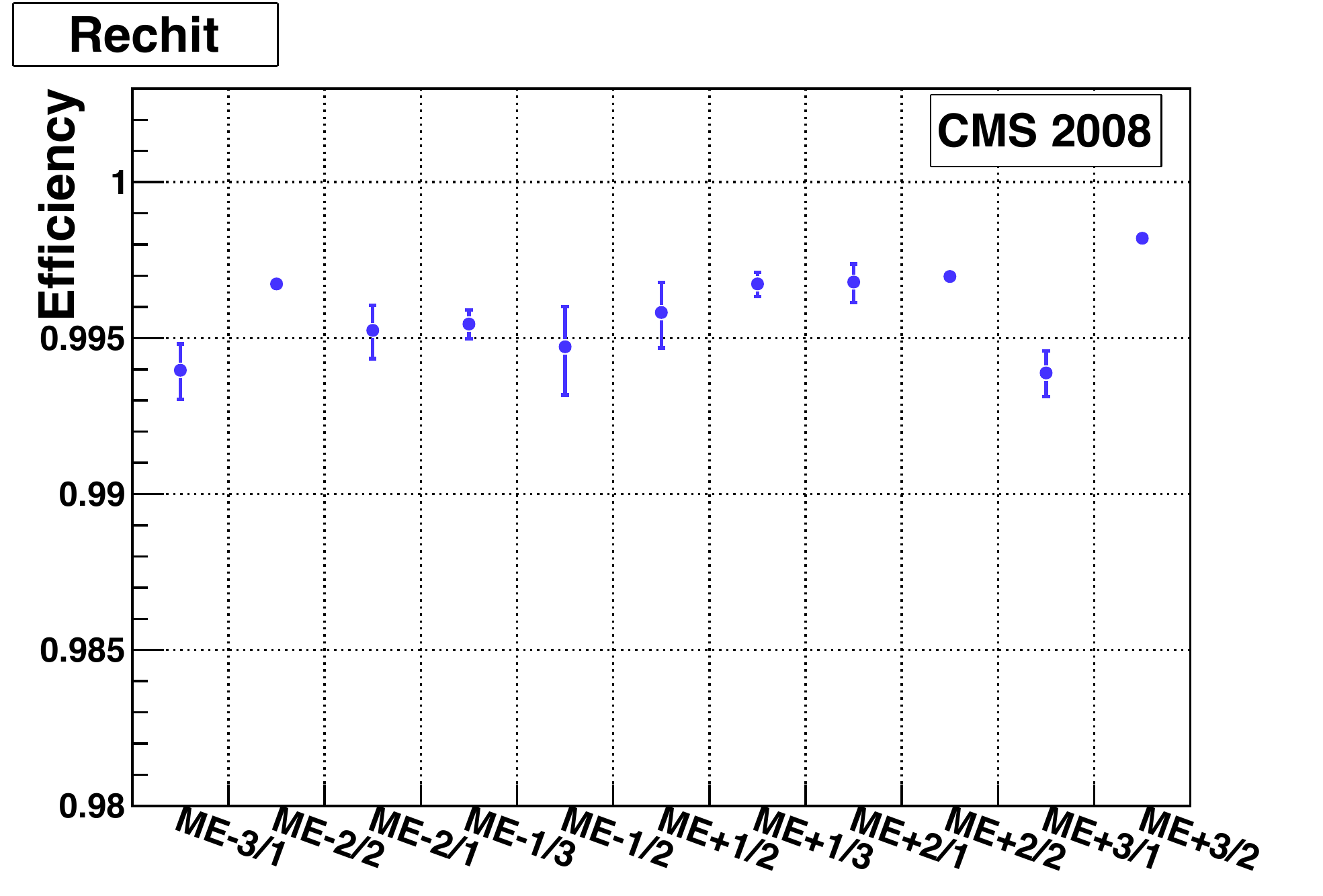}
\includegraphics[width=0.48\textwidth]{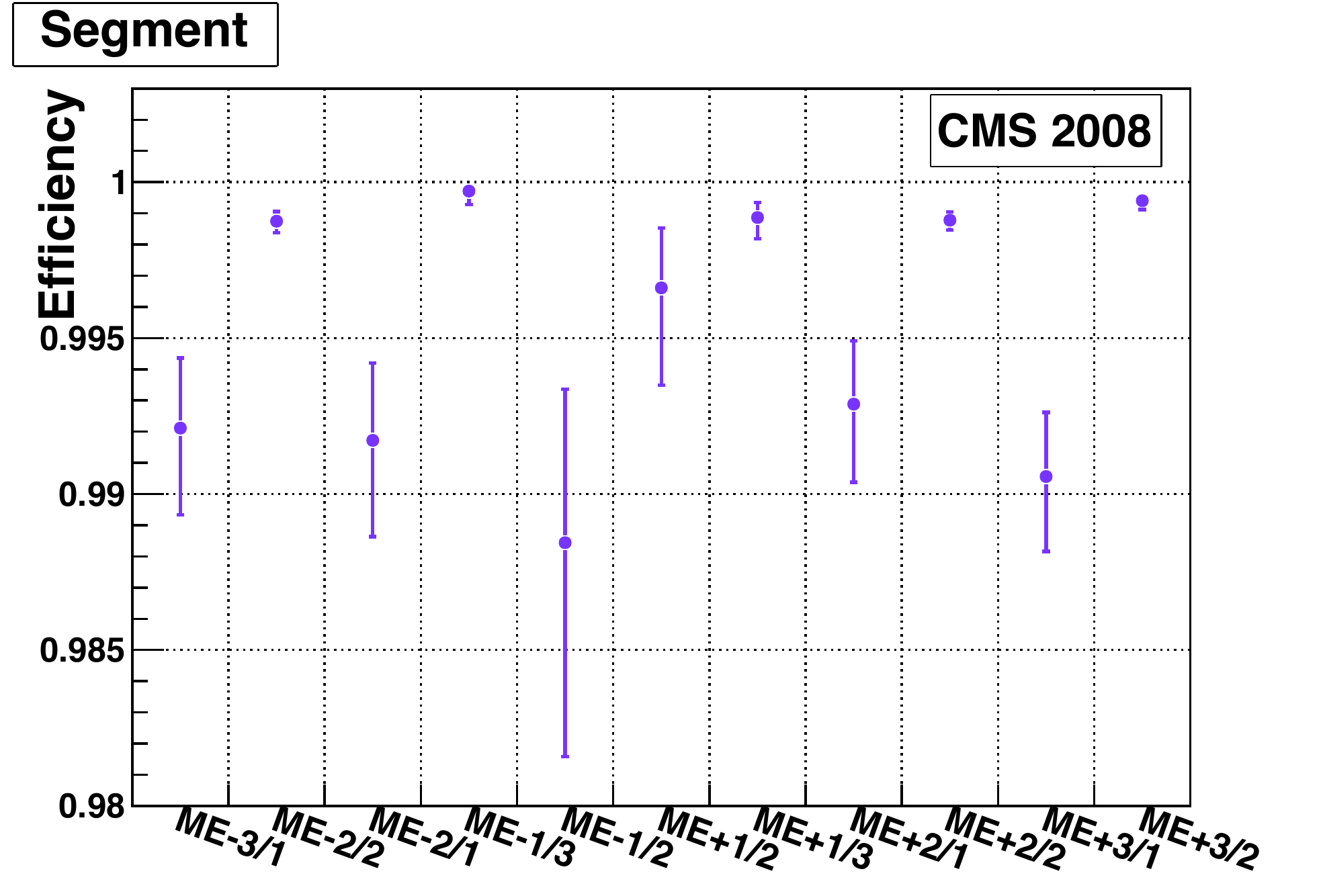}
\caption{\label{Eff_rechit}
Summaries of rechit and segment efficiencies, analogous to Fig.~\ref{Eff_WG}.}
}
\end{figure}

Ideally, the segment efficiency would be related in a simple and
direct way to the rechit efficiency.  The segment reconstruction 
algorithm, however, also places requirements on the rechits used
to build segments.  It does not find segments
in chambers with very many hits, due to prohibitive combinatorial
problems -- this will register as an inefficiency in the present
study.  
The segment efficiency for all the rings in the CSC system is shown 
in Fig.~\ref{Eff_rechit}~(right).  For cosmic rays, the segment
efficiency is above 98.5\%.

\subsection{Attachment Efficiency}      
\par
\begin{figure}
{\centering
\includegraphics[width=0.48\textwidth]{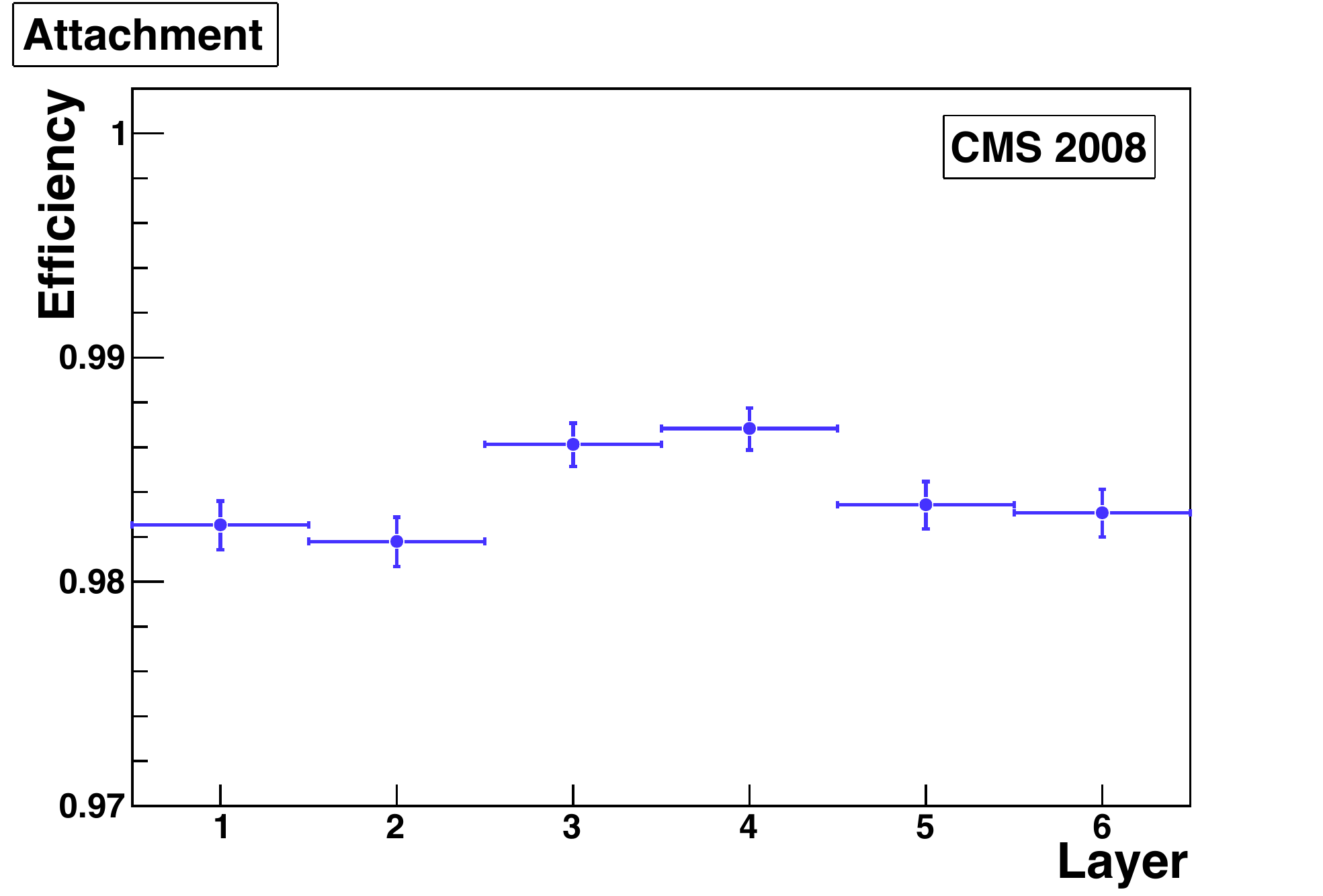}
\caption{The attachment efficiency for each layer.}
\label{Eff_att}
}
\end{figure}

The attachment efficiency is a characteristic of the segment builder. It is defined
as the probability of the segment to use a rechit from a given layer if there are 
rechits in that layer.  The segment finder could reject some rechits if their
quality were poor, or if they were producing a bad fit, so one can anticipate
a small inefficiency with respect to the efficiency for producing rechits.
What is important is that this inefficiency should be the same for all layers.
Any significant variation with layer number would be a hint of a problem -- 
for example, an unacceptable dependence on the track angle.
Figure~\ref{Eff_att} shows that there is no bias in the CRAFT data.

\begin{table}
\begin{center}
\caption{\label{tab:eff_summary}
Summary of efficiencies for chambers in good operating condition.}
\begin{tabular}{|cc|}
\hline
quantity & typical efficiency (\%) \\
\hline
ALCT   & $>99.9$ \\
CLCT   & $>99.5$ \\
wire digis & $>99.5$ \\
strip digi & $>99.4$ \\
rechit & $>99.3$ \\
segment & $>98.5$ \\
\hline
\end{tabular}
\end{center}
\end{table}

\par
In summary, all the basic efficiencies have been shown to be high,
for chambers in good operating condition during CRAFT, as listed
in Table~\ref{tab:eff_summary}.

\typeout{=========== END OF CSC EFFICIENCY ===========}

\typeout{=========== START OF RESOLUTION  ===========}
\section{Resolution}
\par
The CRAFT data were used to study and measure the
spatial resolution of the CSCs as they are meant to be
operated for early physics. (The current high voltage
settings are intentionally lower than what was used for the
test beam studies, in order to avoid aging the chambers unnecessarily
during commissioning periods.  This has a significant impact
on the spatial resolution, as described below.)
The purpose of this study is to verify that all working
chambers perform as they should, before colliding beams commence.
Earlier studies of CSC spatial resolution can
be found in Ref.~\cite{Florida2}. 

\subsection{Methodology}
\par
The {\sl resolution} is the typical
measurement error.  It is determined by the design parameters
of the chamber (width of the cathode strip, distance to the
anode wire plane, high voltage, anode wire radius and pitch,
gas mixture, electronics noise and cross talk) 
as well as certain characteristics of each muon track (angle, position
with respect to the center of the struck strip, and amount
of charge collected), the physics of multi-wire
proportional chambers (electron diffusion, magnetic field
influence) and the reconstruction (reduction of data and
knowledge of misalignments). The distribution of hit residuals
with respect to the muon trajectory can give a good measure
of the resolution.  A {\sl residual} is the difference
between the measured coordinate and the predicted coordinate.  
\par
For the purposes of the study, the coordinate
of interest is the coordinate measured by the strips.
In global coordinates, this would be $R\phi$, 
but most of the results presented here are expressed in
{\sl strip coordinates}.  The strip coordinate, $s$,
is the $R\phi$ coordinate relative to the center
of the strip, divided by the strip width
at the position of the hit.  Apart from resolution effects,
one has $-0.5 \le s \le 0.5$.  
In order to obtain a resolution in physical units,
we multiply by the mean width of a strip in the given
chamber, reported in Table~\ref{tab:specs}.  
\par
\begin{figure}
\begin{center}
\includegraphics[width=0.45\textwidth]{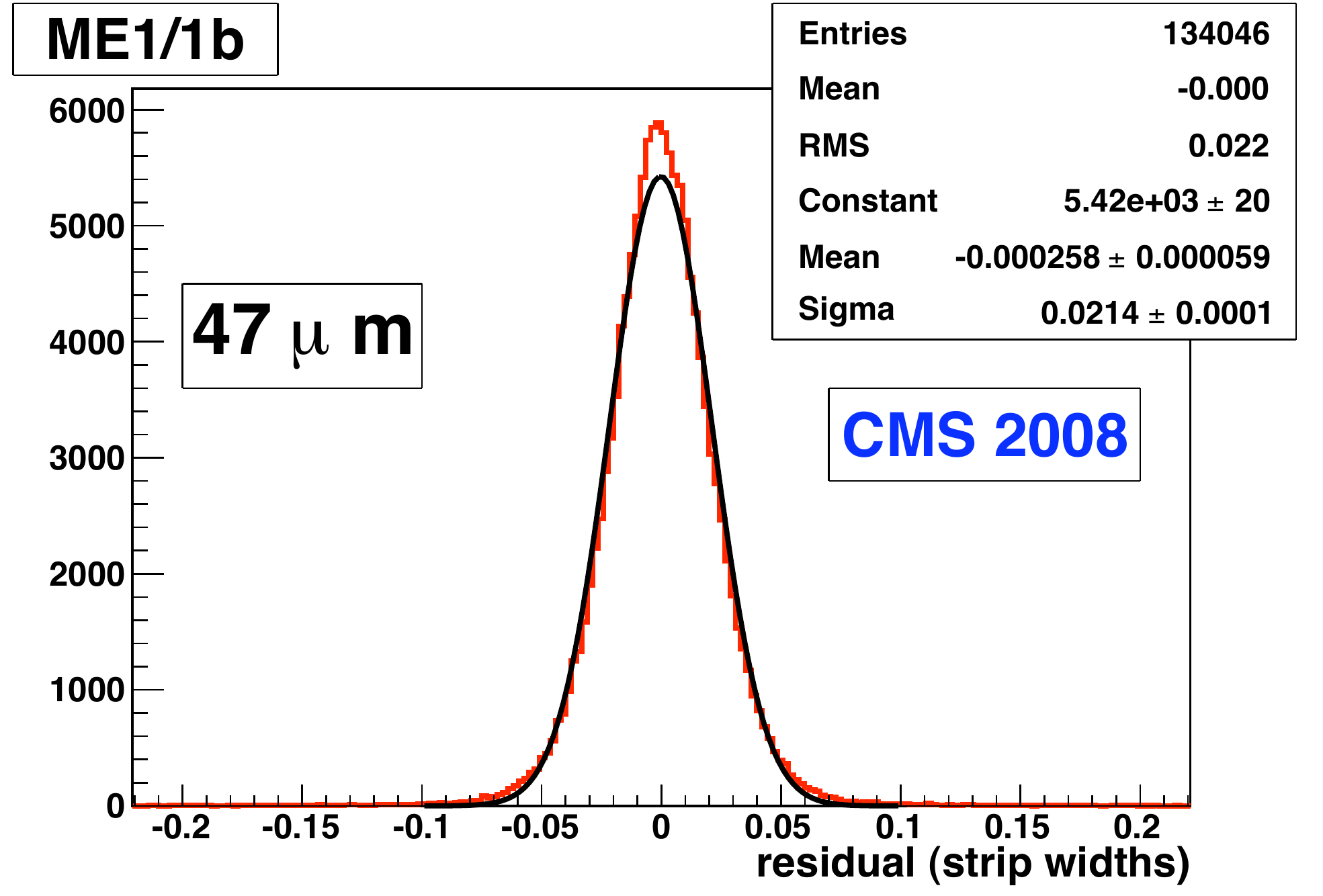}
\includegraphics[width=0.45\textwidth]{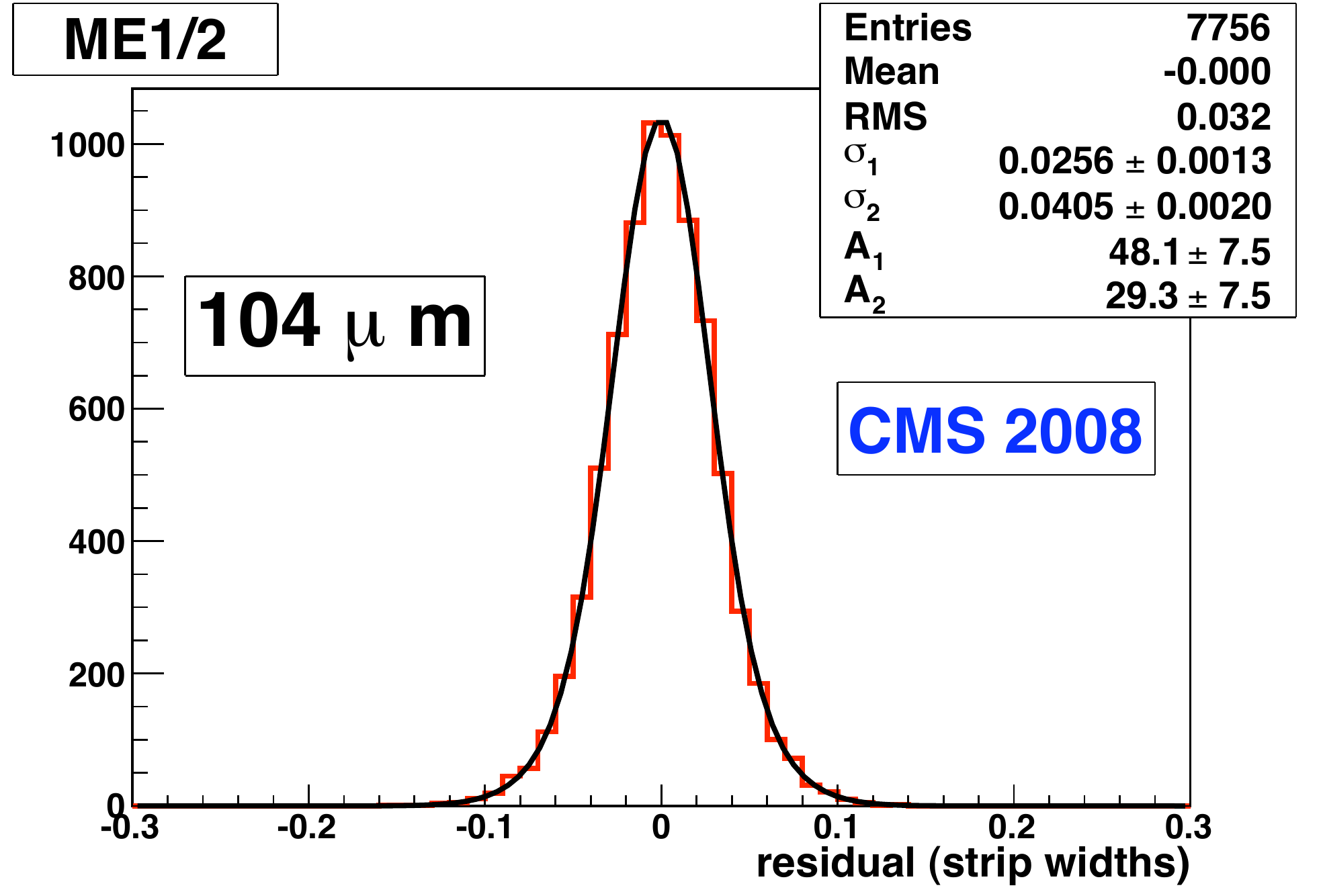}
\includegraphics[width=0.45\textwidth]{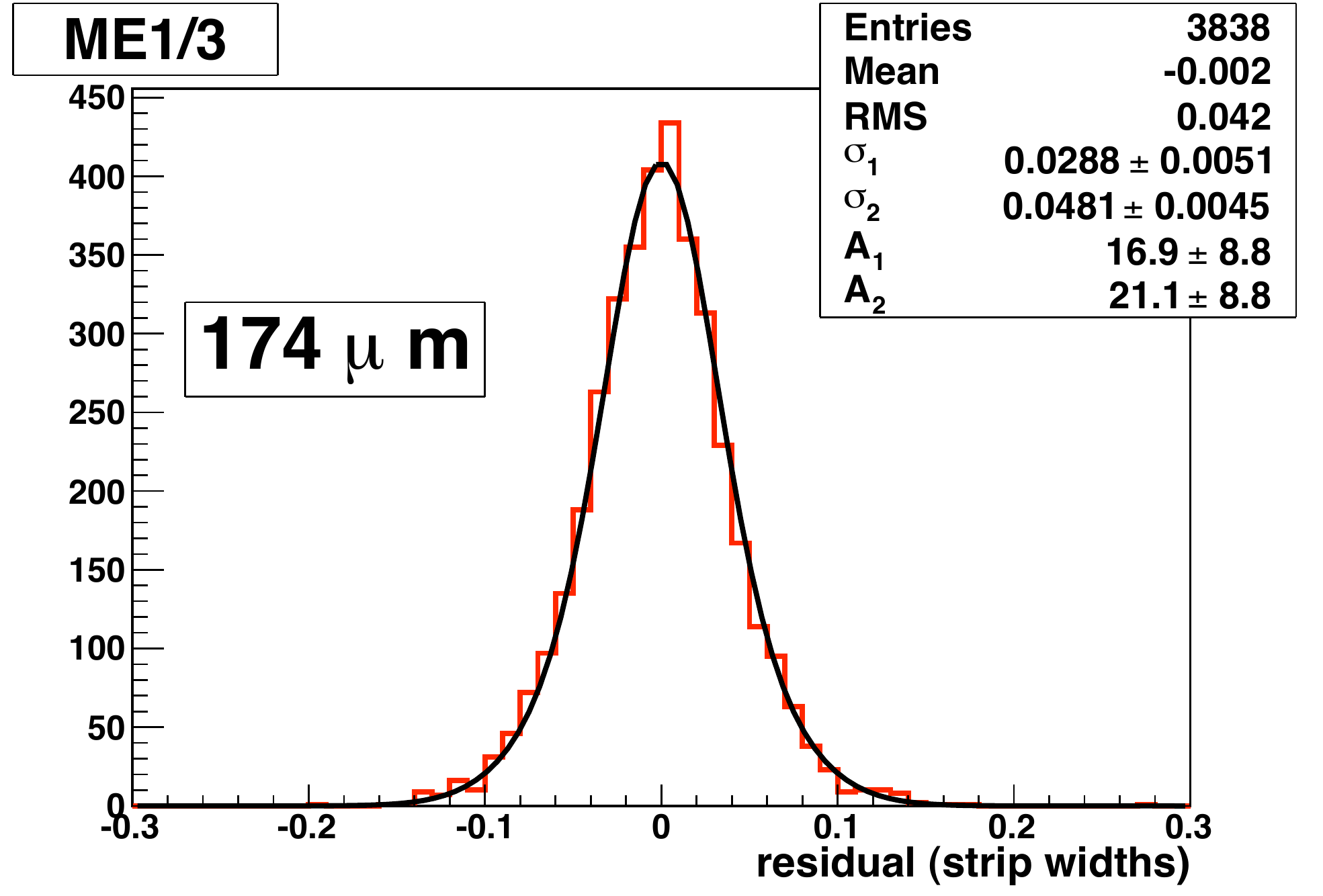}
\includegraphics[width=0.45\textwidth]{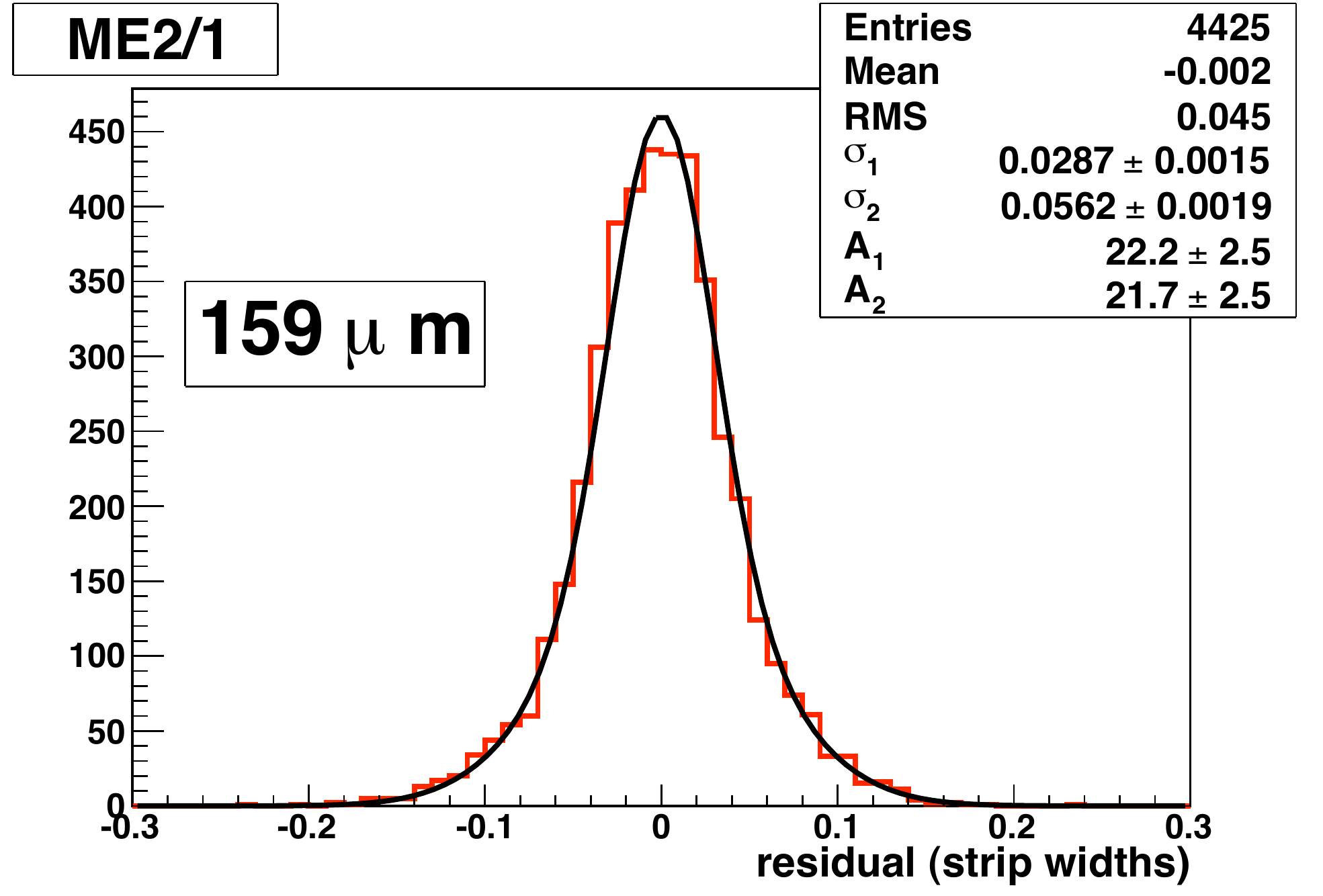}
\includegraphics[width=0.45\textwidth]{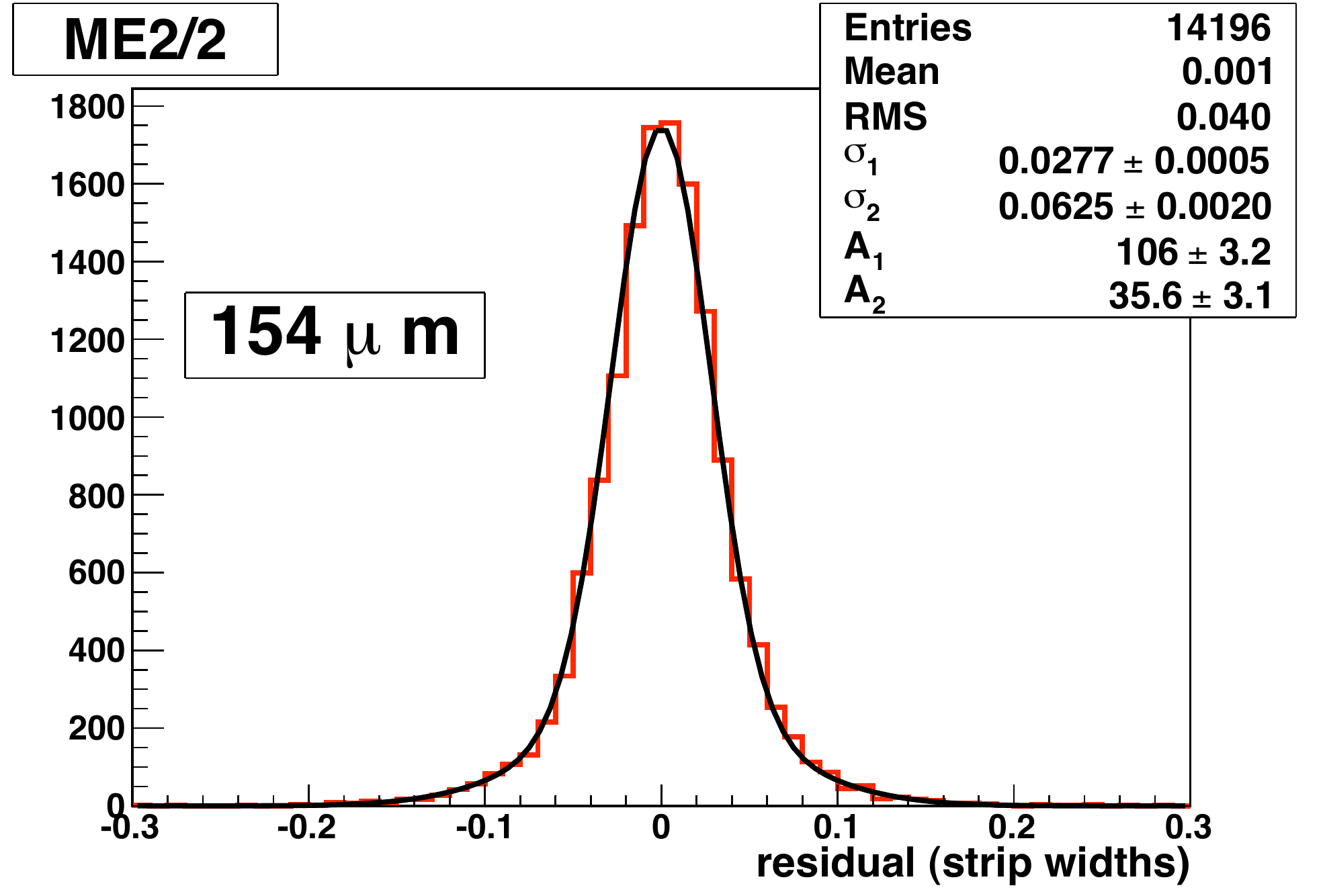}
\includegraphics[width=0.45\textwidth]{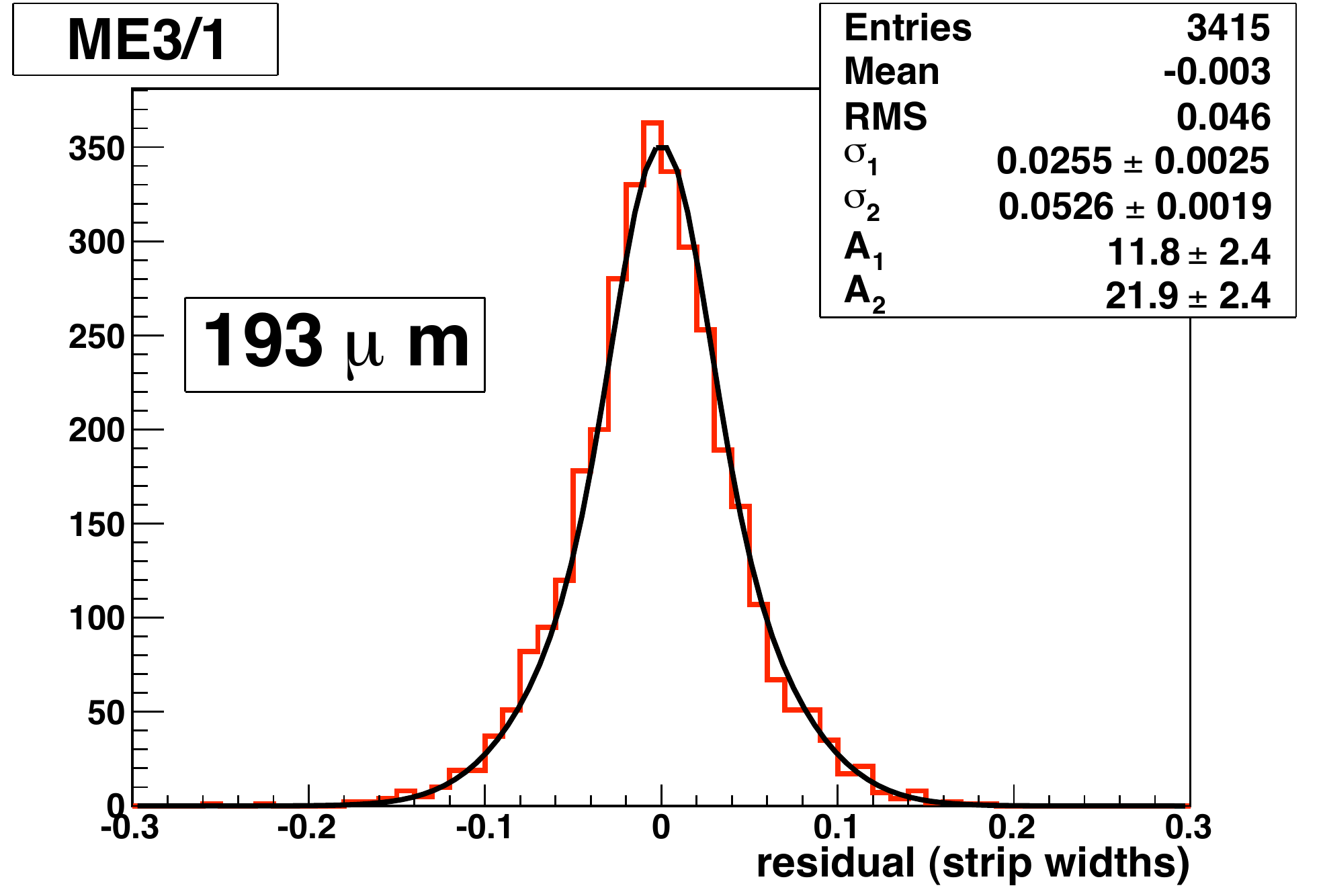}
\includegraphics[width=0.45\textwidth]{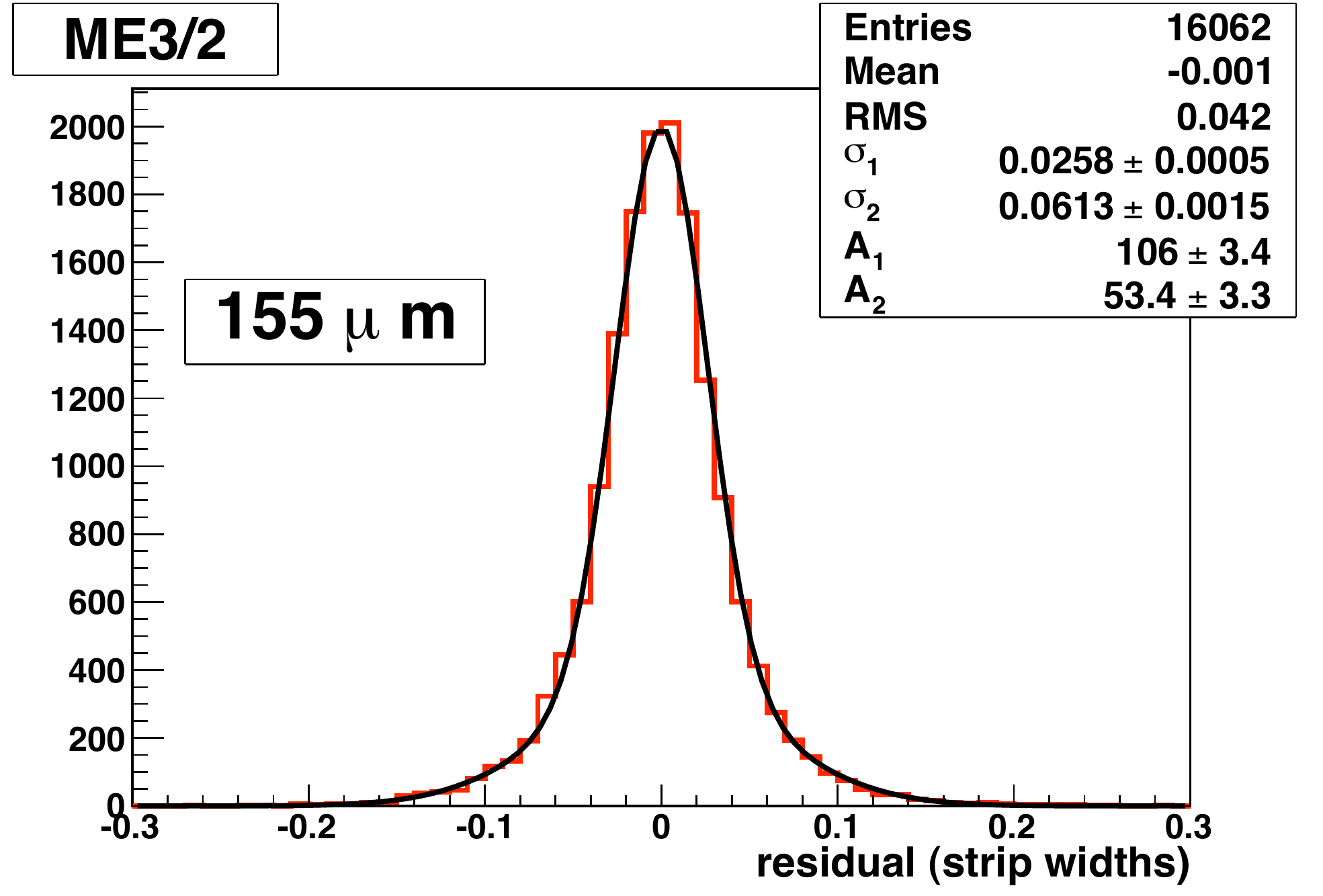}
\includegraphics[width=0.45\textwidth]{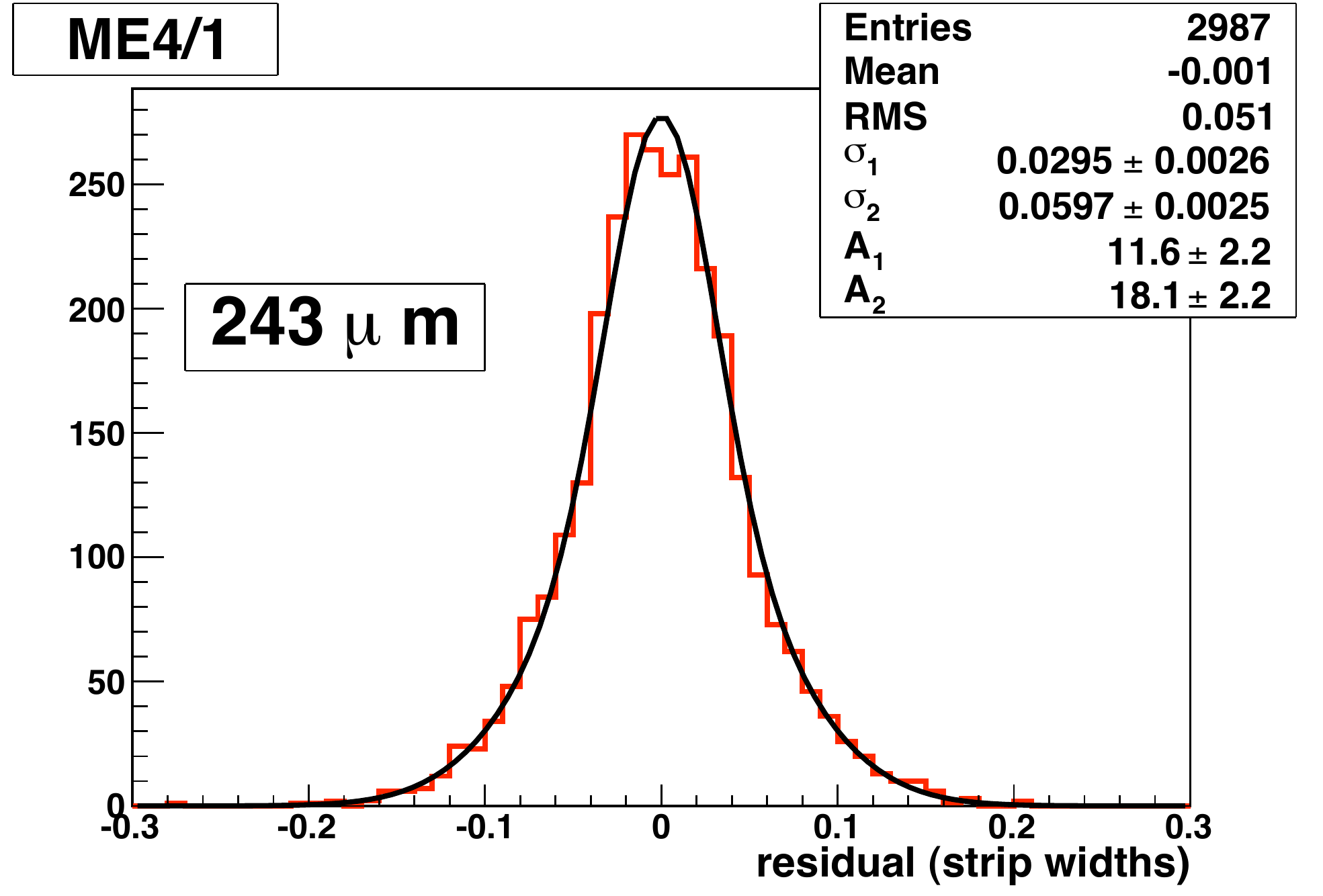}
\end{center}
\caption{\label{fig:dRes1}
Distributions of residuals fit to the double-Gaussian function given 
in Eq.~(\ref{eq:fitGG}), except for the ME$\pm 1/1$ chambers,
which are fit to a single Gaussian.
The numbers in boxes correspond to the
{\sl chamber resolution}, obtained from Eq.~(\ref{eq:res_segment_simple})
and the average strip widths given in Table~\ref{tab:specs}.}
\end{figure}

The residuals distribution is not Gaussian, in general,
so one must settle on a measure of the residuals distribution
to be identified with the ``resolution'' of the given chamber.
We fit the distribution with a sum of two Gaussian functions,
with zero mean, using the functional form:
\begin{equation}
\label{eq:fitGG}
 f(x) \equiv 
 \frac{A_1}{\sqrt{2\pi}\sigma_1} \, \exp\left(\frac{-x^2}{2\sigma_1^2}\right)
   +
 \frac{A_2}{\sqrt{2\pi}\sigma_2} \, \exp\left(\frac{-x^2}{2\sigma_2^2}\right)
\end{equation}
where values for the parameters $\sigma_1$, $\sigma_2$, $A_1$ and $A_2$
are obtained from the fit. We take the resolution to be:
\begin{equation}
\label{eq:resolution}
 \bar{\sigma} = \sqrt{\frac{A_1\sigma_1^2+A_2\sigma_2^2}{A_1+A_2}} .
\end{equation}
If one Gaussian suffices, then we take simply the 
$\sigma$ parameter of the single Gaussian.
We do not take the rms as the residual distributions
often have long non-Gaussian tails which inflate
the rms - these tails are caused by $\delta$-ray
electrons and fall outside a discussion of the core
resolution. 
The residuals distributions of eight chamber types
with fits to Eq.~(\ref{eq:fitGG}) are given in Fig.~\ref{fig:dRes1}.

\par
As defined, the resolution $\bar{\sigma}$ pertains to
a hit in a {\sl single layer}.  The resolution of a
chamber is more complicated, since it depends on the
number of hits in the segment, the direction of the segment,
the generally non-normal angle between wire groups and
strips, and the fact that the strips are staggered
layer-by-layer for all chambers except ME$\pm1/1$.
We can take the special case of
segments with six hits that are normal to the chamber
and pass through the center.  If the residuals
distribution for hits near the edge of a strip ($|s| > 0.25$) has Gaussian
width $\sigma_e$, and for hits near the center of a strip ($|s| < 0.25$), 
$\sigma_c$, then to a good approximation, the 
resolution for the segment is 
\begin{equation}
\label{eq:res_segment_simple}
   \sigma_{\mathrm{seg}} = \left(
      \frac{3}{\sigma_e^2} + \frac{3}{\sigma_c^2}
                    \right)^{-1/2} .
\end{equation}
We will use this expression to characterize the chamber
resolution.
\par
Another method for measuring the resolution does not rely
on the residuals of a single layer, but rather on the value
of $\chi^2$ for the linear fit to all six hits.  We
define the {\sl unweighted} $\chi^2$ as follows:
\begin{equation}
\label{eq:unwchisq}
 \chi^2_{0} \equiv 
 \sum_{i=1}^{6}
 \left[ s_i - (a + bi) \right]^2
\end{equation}
where $a$ and $b$ are free parameters, and the layer number~$i$
plays the role of the $z$ coordinate.  Since there are two
free parameters and six data points,
$\langle  \chi^2_{0} \rangle = 4\sigma_0^2$,
where $\sigma_0$ is the effective uncertainty on~$s_i$.
\par
We do not have a good exterior measure of the position of the
muon, so we have to use the segment itself.  
We fit the hits in layers 1, 2, 4, 5 and 6 to a straight line 
to predict the ``correct'' position in layer~3, and then compare
to the measured position in layer~3.  The estimated error for
those five hits are used in the fit.  Monte Carlo studies
show that the width of the residuals distribution is inflated
by about 10\% due to the measurement error from the five-hit
fit; this uncertainty is larger for layers 1, 2, 5 or 6.
We do not remove this 10\% inflation for the results reported
in this paper.
Also, no attempt was made to remove layer-by-layer misalignments,
as these are known to be small compared to the resolution.

\subsection{Results from CRAFT \label{sec:craft} }
\par
The resolution is known to vary with several quantities, 
including the charge recorded for that hit,
the position within the strip,
the physical width of the strip,
the inclination of the track and the magnetic field, 
among others~\cite{Gatti,Matheison,Moissenz}.
The charge usually extends across three strips, which we label
$Q_L$, $Q_C$ and $Q_R$, where by definition the charge on the central 
strip is larger than that on the left and right side strips.  
We take the charge on these three strips, measured for three
consecutive time slices centered on the peak of the signal, 
and form the sum, $Q_{3\times 3}$~\cite{Florida2}.
\par
Events were selected which contained a good segment from which
residuals distributions for layer~3 could be formed.  A good segment
was one which contained six rechits and $\chi^2 < 200$ (unreduced).
An event was selected if it contained at least one good segment.
In order to retain only clean events, any event with more than
eight segments of any quality were rejected, as well as events
with more than fifty rechits.  The event was also rejected if
any chamber contained more than four segments of any quality.
\par
Further criteria were applied when filling residuals distributions:
\begin{enumerate}
\item
The estimated errors on the six rechits have to be smaller than
0.2 strip widths.  This eliminates rechits based on a single
strip or anomalous charge distributions.
\item
The sum of charges for three strips and three time slices
for layer~3 could not be too small or too large:
$250 < Q_{3\times 3} < 1000$ ADC counts ($4000$~ADC counts
for the ME$\pm1/1$ chambers).
\item
The segment inclination should correspond to tracks
originating roughly from the interaction point:
\begin{equation}
\label{eq:slopes}
 -1 < \frac{dy}{dz} < -0.15
 \qquad {\mathrm{and}} \qquad
 \left| \frac{dx}{dz} \right| < 0.15 .
\end{equation}
\item
The strip coordinates were fit to a straight line.
The resulting $\chi^2$ values were required to be
less than $9$ for the 5-hit fit, and less than $50$ 
for the 6-hit fit.
\end{enumerate}
These cuts were relaxed singly when checking the impact
of these criteria.

%

\par
\begin{figure}
\begin{center}
\includegraphics[width=0.48\textwidth]{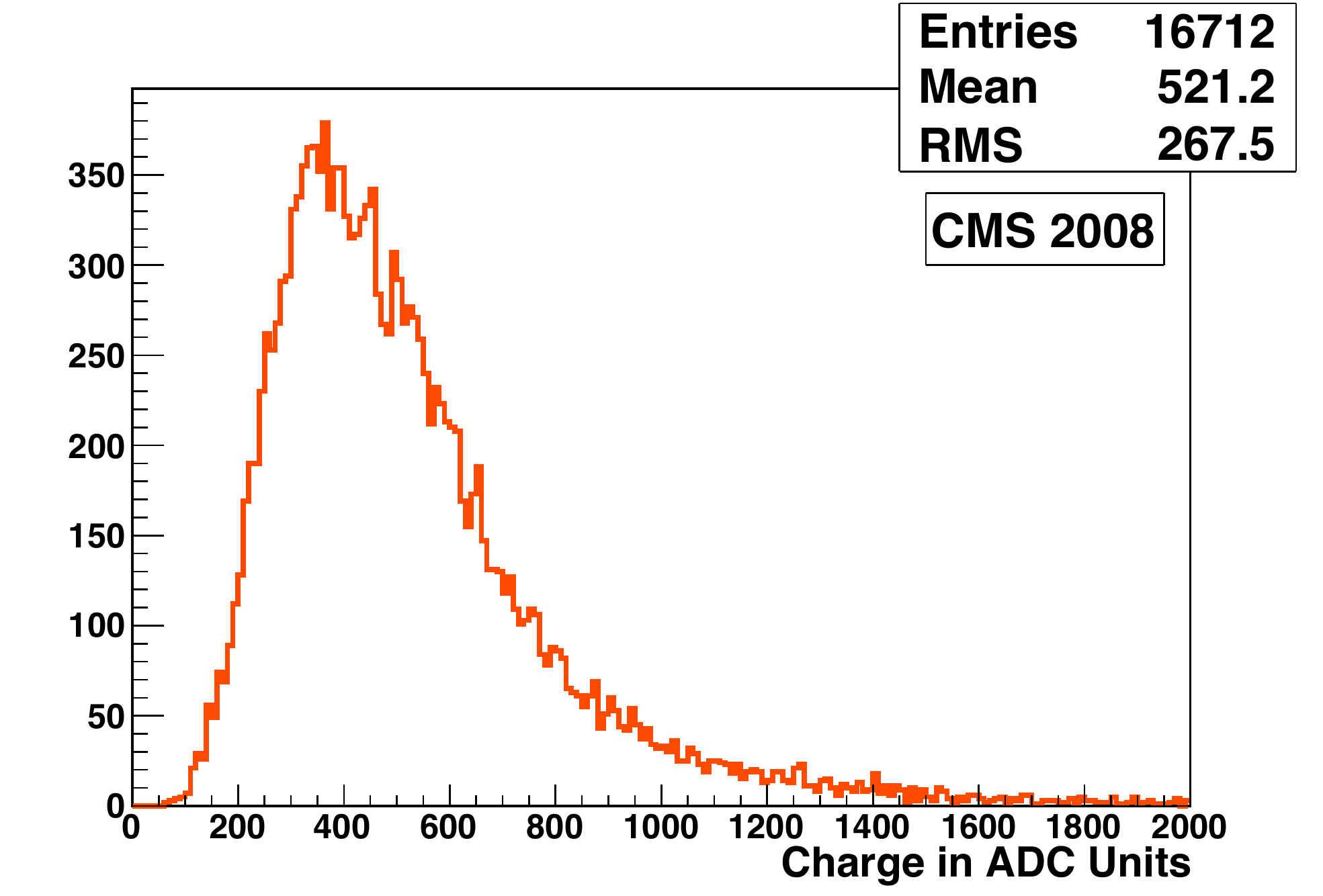}
\includegraphics[width=0.48\textwidth]{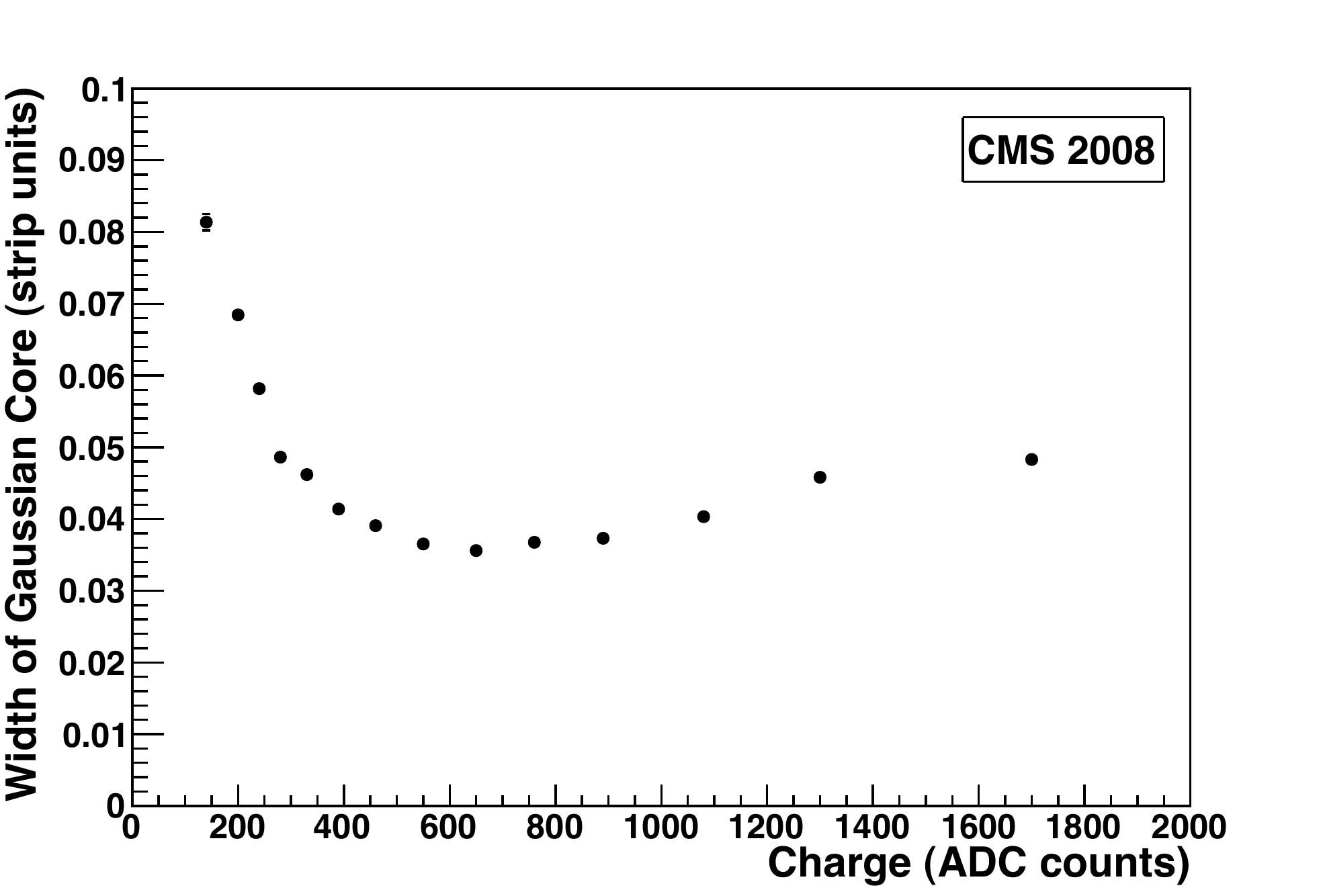}
\end{center}
\caption{\label{fig:q3x3}
Left: Observed charge distribution, $Q_{3\times 3}$, in ADC counts.
Right: Variation of the {\em per layer} resolution as a function of $Q_{3\times 3}$.
This measurement was made using chambers in ME$\pm2/2$ and ME$\pm3/2$;
other chambers give very similar results.}
\end{figure}

The registered charge depends on several factors, including the gas
composition, pressure, high voltage, amplifier gain, and 
the ionization of the gas by the muon.  
A distribution of $Q_{3\times 3}$ for the CRAFT data is shown 
in Fig.~\ref{fig:q3x3}~(left).
The distribution has a long tail, similar to that expected
from the Landau distribution.  
\par
The variation of the resolution as a function of charge
is illustrated in Fig.~\ref{fig:q3x3}~(right).  Chambers in rings ME$\pm2/2$
and ME$\pm3/2$ were selected for this plot, since they have the 
largest number of events in CRAFT.  The cuts on the $\chi^2$
of the fits to strip coordinates were relaxed for this study,
so that the impact of $\delta$-ray electrons is evident at 
large ionization charge.
If the cuts are imposed, then the rise for $Q_{3\times 3} > 800$~ADC 
counts is eliminated.
\par
\begin{figure}
\begin{center}
\includegraphics[width=0.48\textwidth]{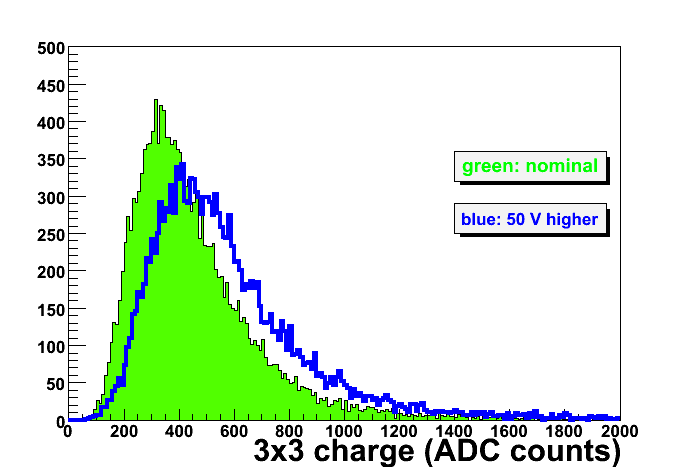}
\includegraphics[width=0.48\textwidth]{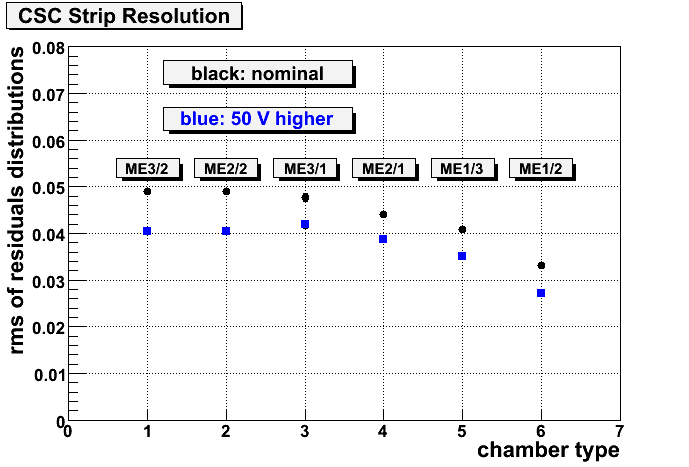}
\end{center}
\caption{\label{fig:dHV}
Left: Charge distributions for two consecutive runs.  The solid
histogram corresponds to the nominal setting, and the open histogram
corresponds to an increase of 50~V.
Right: Comparison of the {\sl per layer} resolution for the same 
two runs, in strip units.}
\end{figure}

Another demonstration of the sensitivity of the resolution to charge
is provided by two runs taken outside of the CRAFT exercise, in
which the high voltage was raised by 50~V from 3600~V.  Since the number of
events was modest, the event and segment selection was somewhat
looser than described above.   The increase in the observed charge 
is about 20\% and the improvement in resolution is 
about 20\%, consistent with expectations - see Fig.~\ref{fig:dHV}.


\par
\begin{figure}
\begin{center}
\includegraphics[width=0.48\textwidth]{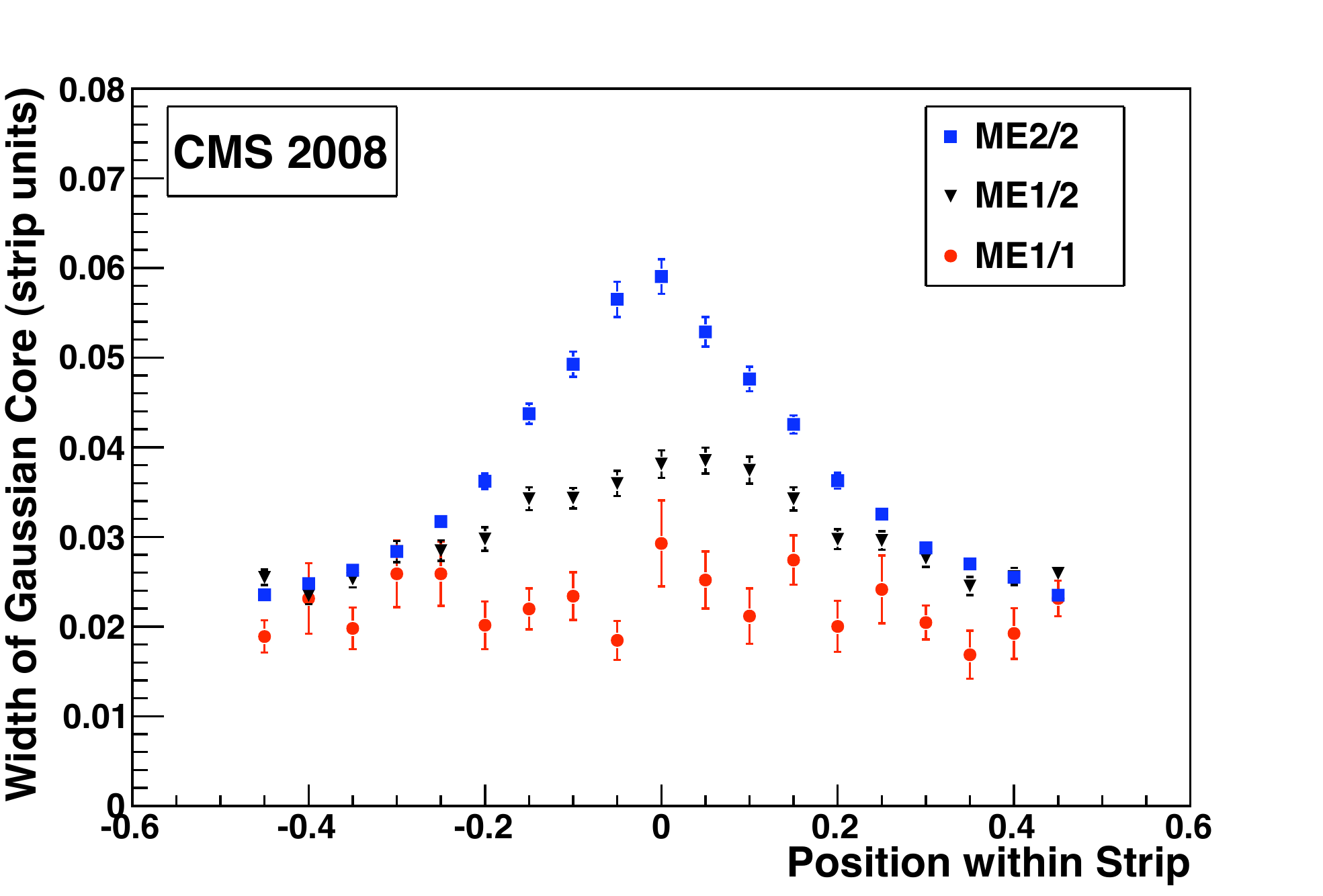}
\includegraphics[width=0.48\textwidth]{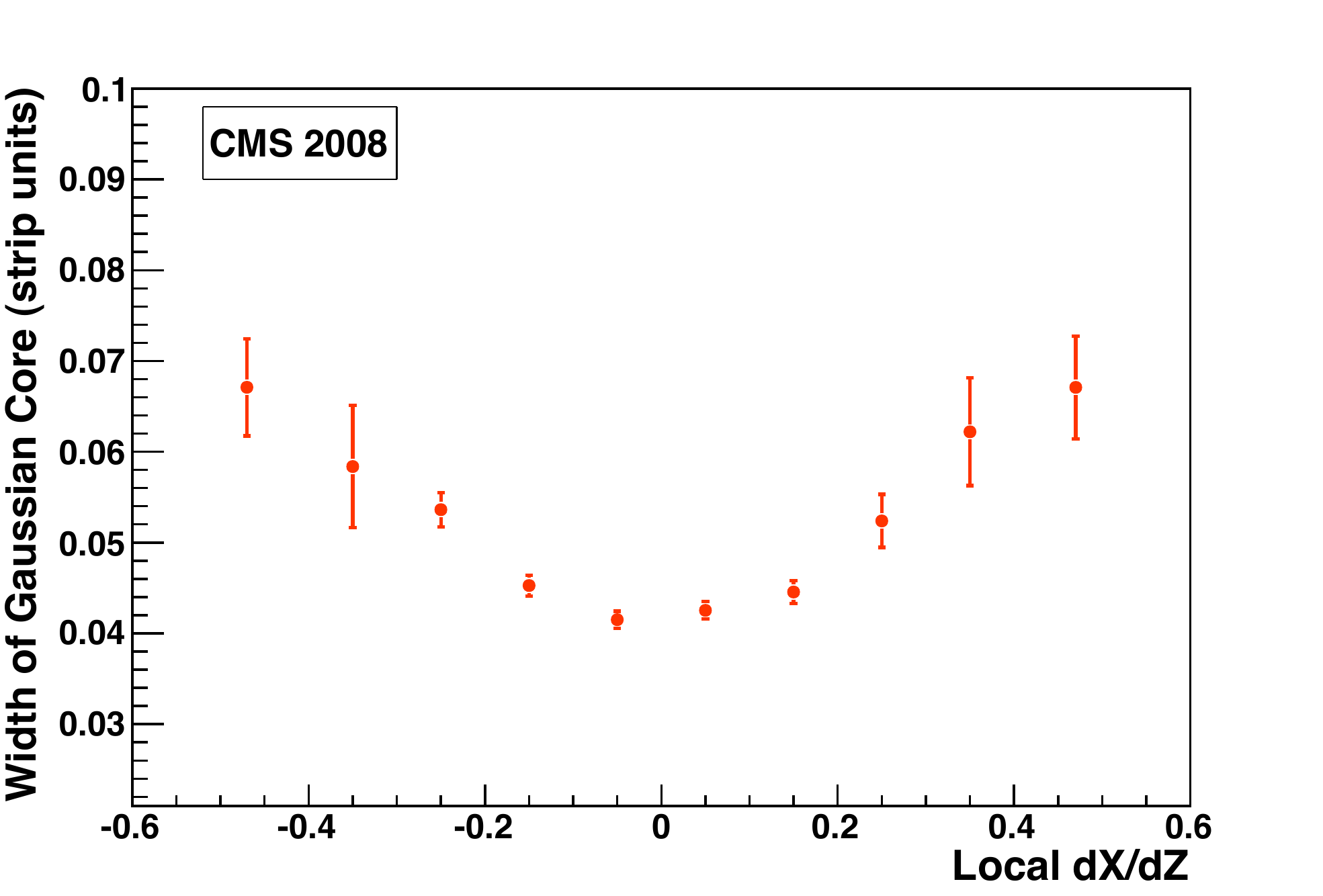}
\end{center}
\caption{\label{fig:res_s}
Left: Variation of the {\em per layer} resolution as a function of $s$,
the position within the strip, for three different types of chambers.
Right: Variation as a function of local $dx/dz$.
These measurements were done with the ME$\pm2/2$ chambers.}
\end{figure}

The variation of the resolution with the position within a strip, $s$, is shown
in Fig.~\ref{fig:res_s}~(left).  For the ME$\pm2/2$ chambers, the resolution
in the center of the strip is worse by about a factor of two
than at the edge.  This variation is weaker for chambers with
thinner strips, such as ME$\pm1/2$ and ME$\pm1/1$.
\par
Most of the analysis presented here is done in terms of the normalized
strip width,~$s$.  The physical width of the strip matters, too.
For broad strips, most of the charge is collected on the central
strip, leaving a small amount for $Q_L$ and $Q_R$, leading to
a poorer resolution.  For this reason, the smaller chambers in ME$\pm 1/1$
have a much better resolution than the larger chambers.
Within a chamber, there is a mild variation of the resolution
along the strip, since the strip is narrower at the narrow
end of the chamber and wider at the broad end.
\par
The results described above were derived for
muon trajectories that were nearly perpendicular to the
strips.  For low-momentum muons coming from the interaction
point, however, more oblique trajectories are possible.
We have observed a clear variation of the resolution
as a function of $dx/dz$ in chambers from ring ME$\pm 2/2$,
see Fig.~\ref{fig:res_s}~(right).
For all other results reported in this note, a tight
cut on $|dx/dz|$ has been applied, as listed in Eq.~(\ref{eq:slopes}).

\subsection{Measurements of the Resolution}
\par
The results in the previous section demonstrate the expected
behavior of the resolution.  In this section, we
quantify the resolution of the CSCs, as measured with CRAFT data,
in order to verify that they are performing as designed.
\par
Residuals distributions for chambers in each ring 
were fit to the sum of two Gaussian functions as 
in Eq.~(\ref{eq:fitGG}),
and the resolution computed according to Eq.~(\ref{eq:resolution}).
These distributions are shown in Fig.~\ref{fig:dRes1}, and
Table~\ref{tab:layer_res} lists the {\em per layer}
resolution obtained in this manner.  The values given
in~$\mu m$ are obtained by multiplying the resolution
in strip widths by the average width of the strip
(see Table~\ref{tab:specs}).
\par
The estimated uncertainty is computed
taking into account variations as a function of charge, position
within a strip, and strip width.
Distributions of normalized residuals (``pull distributions'')
allow us to check those calculations.
A summary of the pulls for all chamber types is
given in Table~\ref{tab:layer_res}.  Overall, the pulls
are somewhat too wide, especially for the ME$\pm1/1$ chambers,
indicating that the uncertainties are slightly underestimated.
It will be possible to adjust the error estimates on
the basis of the CRAFT data.
\par
We formed distributions of $\chi^2_{0}$ defined in Eq.~(\ref{eq:unwchisq}) 
for each chamber type, computing $\sigma_0$ 
and converting to an uncertainty in~$\mu$m
using the average physical strip width.  The results are listed
in Table~\ref{tab:layer_res}.  These values agree well
with the values obtained from the fit to Gaussian functions.
\par
The resolution of a chamber, given six good rechits,
can be estimated on the basis of the {\em per layer}
resolution.  One can simply take the numbers listed
in Table~\ref{tab:layer_res} and divide by $\sqrt{6}$,
or one can perform a slightly more refined analysis
indicated by Eq.~(\ref{eq:res_segment_simple}).
The latter gives systematically lower values for the
resolution than the former. Table~\ref{tab:chamber_res}
lists both sets of values, which can be compared to the 
design values~\cite{MTDR}.  Most observed
values are somewhat higher, except for the ME$\pm 1/1$
chambers, which are significantly better than design.
The fact that the high voltage is set to a somewhat
reduced value to reduce ageing is the primary reason for the slightly
worse resolution in the non-ME$\pm1/1$ chambers.

\begin{table}
\begin{center}
\caption{\label{tab:layer_res}
Resolution {\em per layer} for each chamber type, and the rms of the
pull distributions.}
\begin{tabular}{|cccccc|}
\hline
ring & \multicolumn{4}{c}{resolution} & pull rms \\
 & \multicolumn{2}{c}{{\sl fit to two Gaussians}} &
   \multicolumn{2}{c}{{\sl derived from $\chi^2_{0}$}} & \\
 & strip widths & $\mu m$ & strip widths & $\mu m$ & \\
\hline
ME$\pm 1/1b$ & $0.0214\pm0.0001$ & 129 & 0.020 & 119 & $1.80 \pm 0.06$ \\
ME$\pm 1/2$  &  $0.031\pm0.001$  & 265 & 0.033 & 278 & $1.40 \pm 0.01$ \\
ME$\pm 1/3$  &  $0.040\pm0.003$  & 513 & 0.046 & 606 & $1.73 \pm 0.01$ \\
ME$\pm 2/1$  &  $0.042\pm0.001$  & 474 & 0.051 & 571 & $1.41 \pm 0.02$ \\
ME$\pm 2/2$  &  $0.036\pm0.001$  & 447 & 0.045 & 551 & $1.47 \pm 0.01$ \\
ME$\pm 3/1$  &  $0.043\pm0.002$  & 503 & 0.053 & 619 & $1.44 \pm 0.03$ \\
ME$\pm 3/2$  &  $0.038\pm0.001$  & 461 & 0.046 & 569 & $1.44 \pm 0.01$ \\
ME$\pm 4/1$  &  $0.048\pm0.002$  & 579 & 0.057 & 693 & $1.43 \pm 0.03$ \\
\hline
\end{tabular}
\end{center}
\end{table}
\begin{table}
\begin{center}
\caption{\label{tab:chamber_res}
Resolution {\em per chamber} for each chamber type.}
\begin{tabular}{|cccc|}
\hline
ring & \multicolumn{3}{c|}{resolution ($\mu m$)} \\
 & design & {\em per layer} $/\sqrt{6}$ &  Eq.~(\ref{eq:res_segment_simple}) \\
\hline
ME$\pm 1/1b$ &  75 &  52 &  47 \\
ME$\pm 1/2$  &  75 & 116 & 104 \\
ME$\pm 1/3$  & 150 & 234 & 174 \\
ME$\pm 2/1$  & 150 & 208 & 159 \\
ME$\pm 2/2$  & 150 & 199 & 154 \\
ME$\pm 3/1$  & 150 & 258 & 193 \\
ME$\pm 3/2$  & 150 & 218 & 155 \\
ME$\pm 4/1$  & 150 & 264 & 243 \\
\hline
\end{tabular}
\end{center}
\end{table}

\subsection{Special Studies for ME1/1}
\par
The ME$\pm1/1$ chambers play a special role.  First, they provide
the key measurements for the high-momentum muon tracks expected
at high~$|\eta|$.  And second, they must operate in a very 
high magnetic field, which alters the drift of the electrons
inside the gas layers.  For these reasons, the gas gaps are 
smaller, the gas gain is higher, the strips are narrower, and the wires are tilted
with respect to wires in the other chambers~\cite{ME1}.
\par
\begin{figure}
\begin{center}
\includegraphics[width=0.49\textwidth]{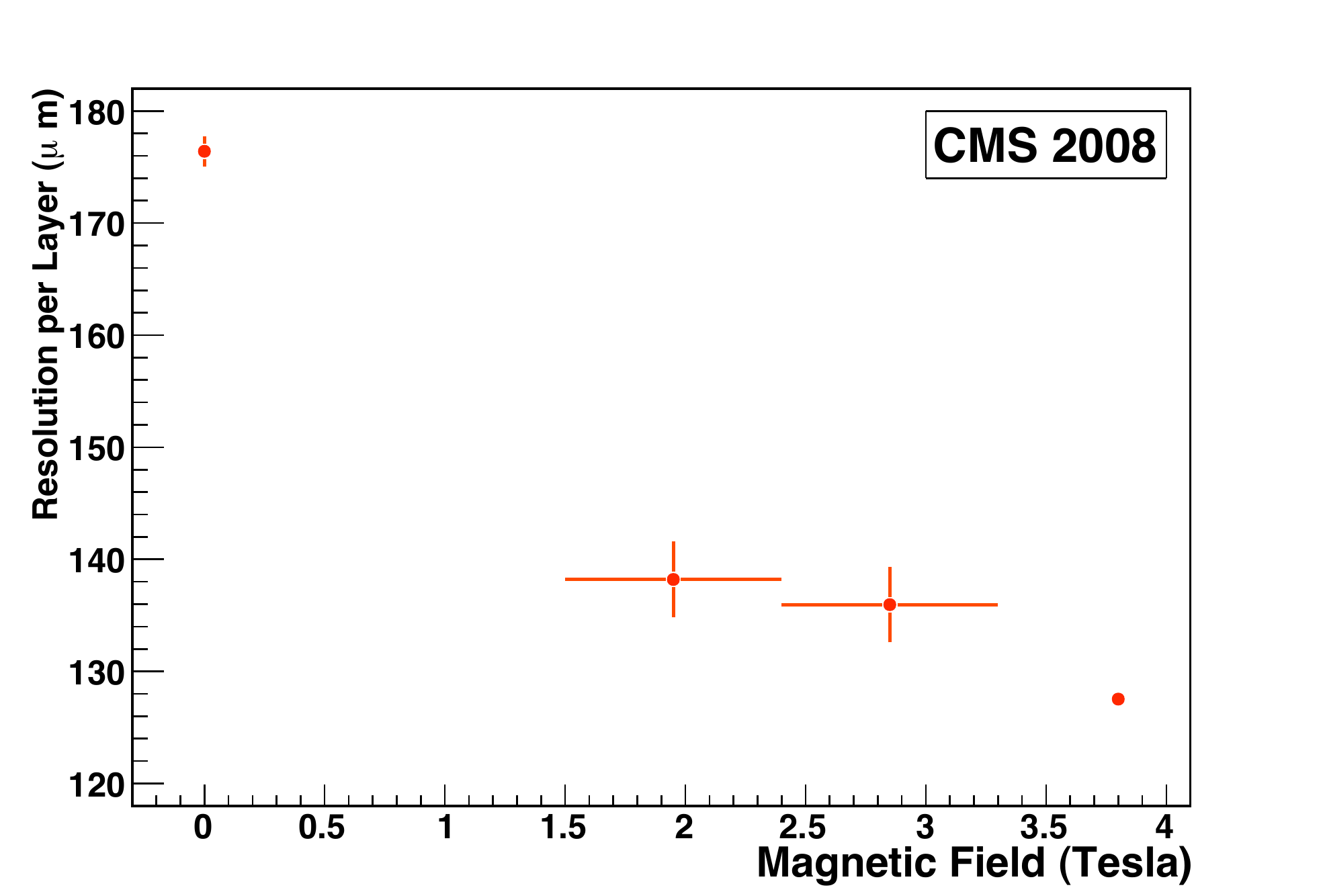}
\includegraphics[width=0.49\textwidth]{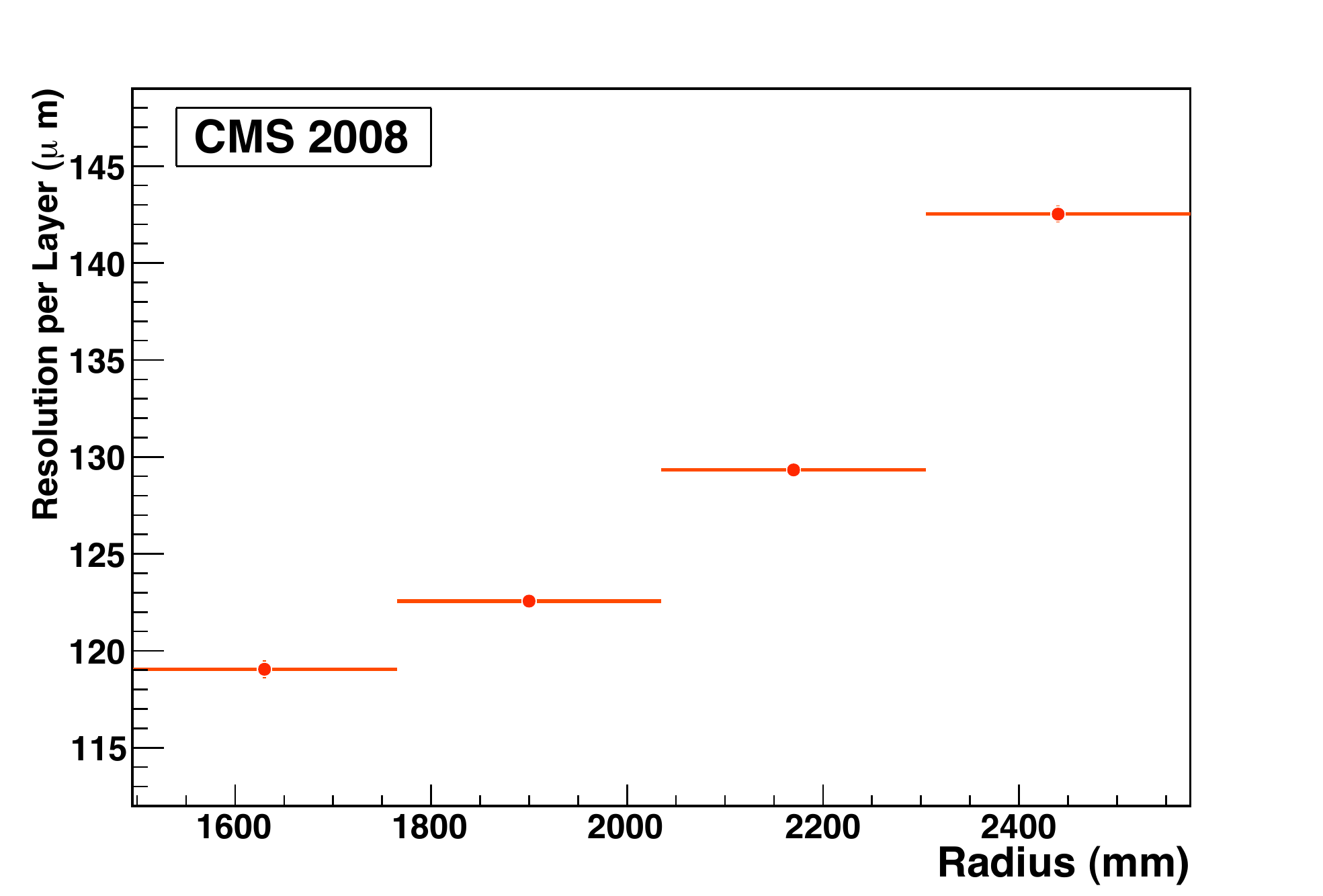}
\end{center}
\caption{\label{fig:dubna_bfield}
Left: Variation of the resolution in the ME$\pm1/1b$ chambers
as a function of magnetic field in Tesla.
The horizontal error bars on the center two points reflect the
changing value of the field for those data.
Right: Variation of the resolution as a function
of the radius (distance from the beam line).}
\end{figure}

The drift of the electrons perpendicular to the anode wires
depends sensitively on the magnetic field.  Most of the
CRAFT data were taken at full operating field, but some data were
taken with zero field, and with some intermediate values.
These data were analyzed to measure the resolution as a
function of the magnetic field, with the results shown in
Fig.~\ref{fig:dubna_bfield}~(left).  For the measurements
at $B \approx 2$~T and $2.9$~T, the field was changing, as
indicated by the horizontal error bars.  The resolution is
best at the maximum operating value of the field, confirming the 
details of the chamber design.
\par
The radial extent of the ME$\pm1/1b$ chambers was divided
into four regions in order to check the resolution at
different radii.  Figure~\ref{fig:dubna_bfield}~(right) shows that
the resolution is best near the beam line, where it is
most critical, and rises rapidly with radius.  A further
study of the resolution for different azimuthal regions
of the ME$\pm1/1b$ chambers shows a mild variation with
the angle of the anode wires, confirming the choices made
in the design of these chambers.

%

\typeout{=========== END OF RESOLUTION  ===========}

\typeout{=============START OF TIMING============}
\section{Timing}
\par
\begin{figure}
\begin{center}
\includegraphics[width=0.6\textwidth]{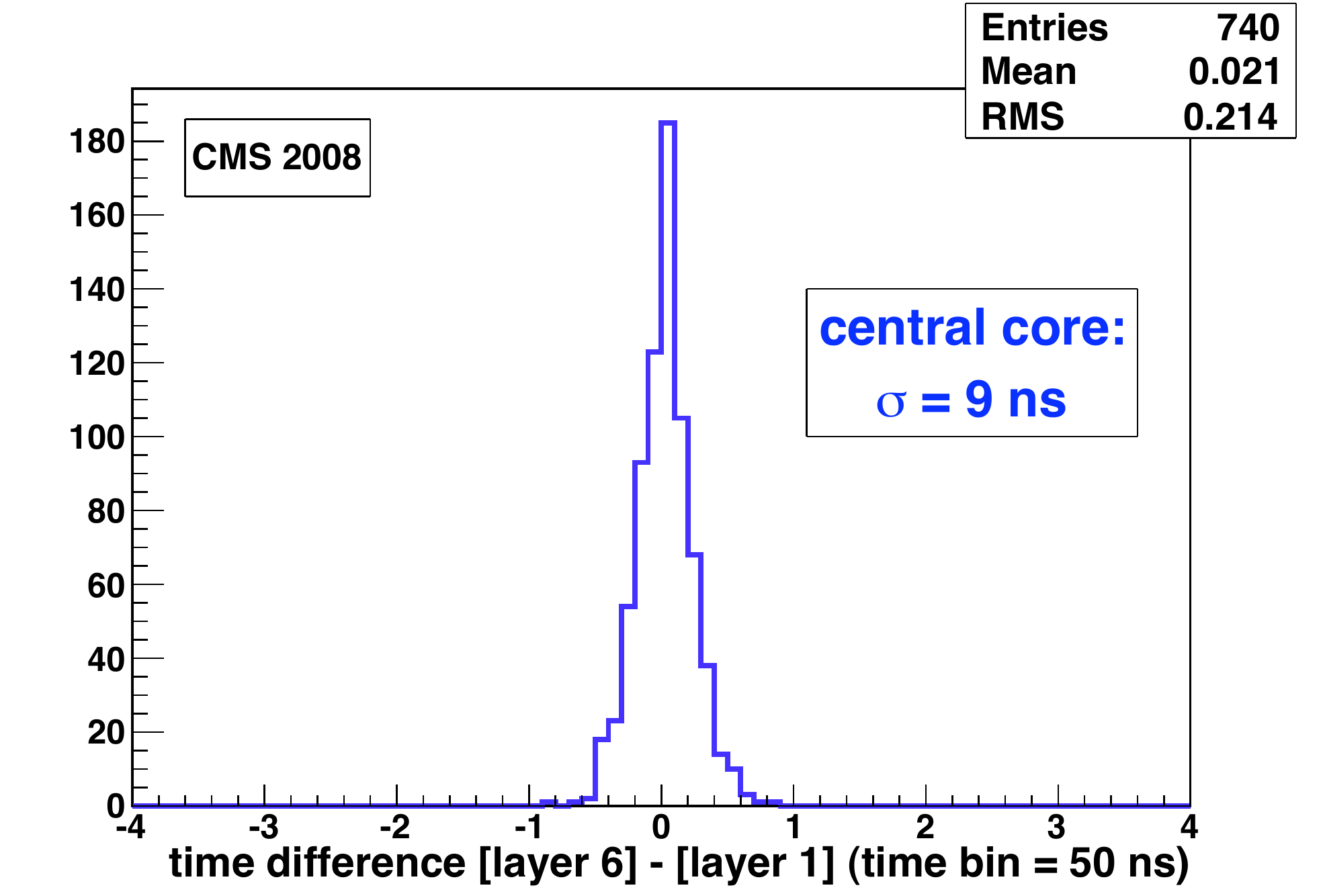}
\end{center}
\caption{\label{fig:dTime}
The difference in rechit times for layers~6 and~1 in chamber ME$+3/2/9$.
Units are 50~ns time bins.  A fit of the central core to a Gaussian
function gives a width of 9~ns.}
\end{figure}

We used the CRAFT data to make some simple tests
of the timing capabilities of the CSCs.  The
time of flight of a muon through a single chamber
is quite small, essentially zero compared to the
25~ns BX spacing.  Figure~\ref{fig:dTime} shows the
distribution of differences in measured times
for layers~6 and~1, in units of 50~ns time bins.
The mean is consistent with zero, and the rms
is $0.214$ time bins, which corresponds to 7.2~ns,
or 5~ns per layer.  Most segments have six
rechits (cf. Fig.~\ref{fig:AK_segment}), so a
single segment should have a time resolution of
about 2~ns.  This compares well with the
transit time of a muon from the interaction point
to the CSCs of roughly 30~ns, and of the beam
crossing time of 25~ns.
\par
Improvements in the use of the strip timing
information are foreseen, based on a more
detailed analysis of the subtle effects of
cross talk and noise correlations, as suggested
by pilot studies with test beam data. 
It is hoped to use this timing capability for
rejecting out-of-time hits and tagging the time
of the muon independently of the trigger system.

\typeout{=============END OF TIMING============}

\section{Summary}
\par
An assessment of the performance of the CSCs has been completed
using the large CRAFT data sample recorded in fall 2008.
More than 96\%  of the CSC muon detector system was in excellent 
working condition and participated in the bulk of this campaign.
The simulation reproduces well distributions of basic global 
quantities, such as the number of hits on track segments and 
the angular distributions of muon tracks, observed in the data.
The fraction of channels which provided no signal, or were
noisy, is less than 1\%.
All of the essential efficiencies have been measured, ranging
from the local charged tracks which trigger the chamber readout
through the reconstruction of segments.  These efficiencies
are all very high.
The position resolution has been studied, with variations observed
as a function of several relevant variables, such as the charge,
position within a strip, high voltage, track inclination, 
and in the case of the ME$\pm 1/1$ chambers, of the magnetic
field, radius and wire tilt.  The measured chamber resolutions
are not quite as good as design, due to an intentional reduction
of the high voltage, except for the ME$\pm 1/1$ chambers, which
surpass the design criterion.
Finally, the potential timing capabilities
of the CSCs was briefly investigated.
\par
The prospects for future studies are very good.
The operating conditions of the CSC subsystem have
been improved since the CRAFT data were taken, and one
can anticipate that the CSC subsystem will function
up to specifications once the LHC delivers collisions to CMS.


\section*{Acknowledgements}
We thank the technical and administrative staff at CERN and other CMS Institutes, and acknowledge support from: FMSR (Austria); FNRS and FWO (Belgium); CNPq, CAPES, FAPERJ, and FAPESP (Brazil); MES (Bulgaria); CERN; CAS, MoST, and NSFC (China); COLCIENCIAS (Colombia); MSES (Croatia); RPF (Cyprus); Academy of Sciences and NICPB (Estonia); Academy of Finland, ME, and HIP (Finland); CEA and CNRS/IN2P3 (France); BMBF, DFG, and HGF (Germany); GSRT (Greece); OTKA and NKTH (Hungary); DAE and DST (India); IPM (Iran); SFI (Ireland); INFN (Italy); NRF (Korea); LAS (Lithuania); CINVESTAV, CONACYT, SEP, and UASLP-FAI (Mexico); PAEC (Pakistan); SCSR (Poland); FCT (Portugal); JINR (Armenia, Belarus, Georgia, Ukraine, Uzbekistan); MST and MAE (Russia); MSTDS (Serbia); MICINN and CPAN (Spain); Swiss Funding Agencies (Switzerland); NSC (Taipei); TUBITAK and TAEK (Turkey); STFC (United Kingdom); DOE and NSF (USA). Individuals have received support from the Marie-Curie IEF program (European Union); the Leventis Foundation; the A. P. Sloan Foundation; and the Alexander von Humboldt Foundation.


\cleardoublepage\appendix\section{The CMS Collaboration \label{app:collab}}\begin{sloppypar}\hyphenpenalty=5000\widowpenalty=500\clubpenalty=5000\textbf{Yerevan Physics Institute,  Yerevan,  Armenia}\\*[0pt]
S.~Chatrchyan, V.~Khachatryan, A.M.~Sirunyan
\vskip\cmsinstskip
\textbf{Institut f\"{u}r Hochenergiephysik der OeAW,  Wien,  Austria}\\*[0pt]
W.~Adam, B.~Arnold, H.~Bergauer, T.~Bergauer, M.~Dragicevic, M.~Eichberger, J.~Er\"{o}
% ...
\vskip\cmsinstskip
% ...

\end{sloppypar}

\providecommand{\href}[2]{#2}\begingroup\raggedright\begin{thebibliography}{}

\end{thebibliography}\endgroup


\begin{thebibliography}{1}

\bibitem{CMS}
{\bf CMS} Collaboration, 
``The CMS Experiment at the CERN LHC,'' 
{\em JINST} {\bf 3} (2008)   S08004.\\
{\tt doi:10.1088/1748-0221/3/08/S08004}

\bibitem{LHC}
L. Evans and P. Bryant (eds.),
``LHC Machine,''
{\em JINST} {\bf 3} (2008) S08001.\\
{\tt doi:10.1088/1748-0221/3/08/S08001}


\bibitem{MAINCRAFT}
{\bf CMS} Collaboration,
``Commissioning of the CMS Experiment and the Cosmic Run at Four Tesla,''
{\em submitted to JINST} (2009).

\bibitem{MTDR}
{\bf CMS} Collaboration, 
``CMS -- The Muon Project,''
{\em CERN/LHCC} {\bf 97-32} (1997).
\vskip 5pt
D.~Acosta \etal,
``Large CMS Cathode Strip  Chambers: Design and Performance,''
{\em Nucl. Instrum. Meth.} {\bf A453} (2000) 182-187.


\bibitem{TDR} {\bf CMS} Collaboration, 
``CMS Physics Technical Design Report,''
{\em CERN/LHCC} {\bf 2006-001} (2006).

\bibitem{LCT} J. Hauser \etal, 
``Experience with Trigger Electronics for the CSC System of CMS,''
Proceedings of the 10th Workshop on Electronics for LHC Experiments and 
Future Experiments (2004).

\bibitem{CSCTF}
{\bf CMS} Collaboration,
``Performance of the CMS Level-1 Trigger during Commissioning with Cosmic Rays,''
{\em submitted to JINST} (2009).

\bibitem{MuonReco}
{\bf CMS} Collaboration,
``Performance of CMS Muon Reconstruction in Cosmic-Ray Events,''
{\em submitted to JINST} (2009).

\bibitem{CRMC}
P. Biallass, T. Hebbeker and K. Hoepfner, 
11Simulation of Cosmic Muons and Comparison with Data from 
the Cosmic Challenge using Drift Tube Chambers,''
{\bf CMS NOTE-2007/024} (2007).
\vskip 5pt
P. Biallass and T. Hebbeker, 
``Parametrization of the Cosmic Muon Flux for the Generator CMSCGEN,''
{\tt arXiv:0907.5514}

\bibitem{PDG} 
W.-M. Yao \etal,   {\em Journal of Physics} {\bf G33} (2006) 1.

\bibitem{MTCC_data}
D. Acosta \etal, 
``Measuring Muon Reconstruction Efficiency from Data,''
{\bf CMS NOTE-2006/060} (2006).
\vskip 5pt
D. Acosta \etal,
``Efficiency of Finding Muon Track Trigger Primitives
in CMS Cathode Strip Chambers,''
{\em Nucl. Instrum. Meth.} {\bf A592} (2008) 26.

\bibitem{Florida2}
V. Barashko \etal,
``Fast Algorithm for Track Segment and Hit Reconstruction
in the CMS Cathode Strip Chambers,''
{\em Nucl. Instrum. Meth.} {\bf A589/3} (2008) 26.

\bibitem{Gatti}
E. Gatti \etal,
``Optimum Geometry for Strip Cathodes or Grids in
MWPC for Avalanche Localization Along the Anode Wires,''
{\em Nucl. Instrum. Meth.} {\bf 163} (1979) 83.

\bibitem{Matheison}
E. Mathieson and J. Gordon,
``Cathode Charge Distributions in Multiwire Chambers:
I.~Measurement and Theory,''
{\em Nucl. Instrum. Meth.} {\bf 227} (1984) 267;
{\em op. cit.},
``II.~Approximate and Empirical Formulae,''
{\em Nucl. Instrum. Meth.}  {\bf 227} (1984) 277.

\bibitem{Moissenz}
I. Golutvin \etal,
``Cathode Strip Chambers Data Analysis,''
Proceedings of the {\sl{Seventh International Conference on 
Advanced Technology and Particle Physics}}, Como, Italy, 15-19 October 2001.


\bibitem {ME1} 
Yu V.~Erchov \etal, 
``Cathode Strip Chamber for CMS ME1/1 Endcap Muon Station,''
{\em Physics of Particles and Nuclei Letters,}
Vol. 3, no. 3 (2006) 183.

\end{thebibliography}
\end{document}